\documentclass[11pt]{article}
\usepackage{amsmath}
\usepackage{graphicx,psfrag,epsf}
\usepackage{enumerate}
\usepackage{natbib}
\usepackage{amssymb}
\usepackage[colorlinks,citecolor=blue,urlcolor=blue]{hyperref}
\usepackage{mathrsfs}
\usepackage{amsfonts,anysize,amsmath,indentfirst,geometry,xcolor}
\usepackage{bm}
\usepackage{amsthm}
\usepackage{multirow}
\usepackage{chngpage}
\usepackage{array}
\usepackage{rotating}
\usepackage{setspace}
\usepackage{graphicx}
\usepackage{algorithm}
\usepackage{algorithmic}
\usepackage{epsfig}
\usepackage{amsbsy}

\newtheorem{theorem}{Theorem}

\newtheorem{lemma}{Lemma}
\newcommand{\calF}{{\cal F}}

\newcommand{\0}{{\bf 0}}

\makeatletter
\def\singlespace{\def\baselinestretch{1}\@normalsize}

\@addtoreset{equation}{section}
\renewcommand{\baselinestretch}{1.412}

\begin{document}


\title{\bf Model Averaging based Semiparametric
\\
Modelling for Conditional
\\
Quantile Prediction
}
  \author{Chaohui Guo
\\
School of Mathematical Sciences
\\
Chongqing Normal University, China
    \and
Wenyang Zhang
\thanks{Corresponding author, email: wenyang.zhang@york.ac.uk, address:
Department of Mathematics, University of York, York YO10 5DD, UK.}
\\
Department of Mathematics
\\
The University of York, UK
    }
  \maketitle

\begin{abstract}

In real data analysis, the underlying model is usually unknown, modelling
strategy plays a key role in the success of data analysis.  Stimulated by
the idea of model averaging, we propose a novel semiparametric modelling
strategy for conditional quantile prediction, without assuming the underlying
model is any specific parametric or semiparametric model.  Thanks the
optimality of the selected weights by cross-validation, the proposed modelling
strategy results in a more accurate prediction than that based on some commonly
used semiparametric models, such as the varying coefficient models and additive
models. Asymptotic properties are established of the proposed modelling
strategy together with its estimation procedure.  Intensive simulation studies
are conducted to demonstrate how well the proposed method works, compared with
its alternatives under various circumstances.  The results show the proposed
method indeed leads to more accurate predictions than its alternatives.
Finally, the proposed modelling strategy together with its prediction procedure
are applied to the Boston housing data, which result in more accurate
predictions of the quantiles of the house prices than that based on some
commonly used alternative methods, therefore, present us a more accurate
picture of the housing market in Boston.

\end{abstract}

\noindent%

\section{Introduction}
\label{sec:1}

\subsection{Preamble}
\label{pre0}

Quantile prediction is an important topic in data analysis, it is widely used
in many scientific disciplines, which includes economics, finance, sociology,
engineering, medical science, to name but a few. Modelling strategy plays a
key role in the success of quantile prediction.  Traditional approach is mainly
based on linear models, see \cite{bc2011,wwl2012,dcl2012,lnp2014,FFB14,TXWF20,fbf2021}.
It is well know the linearity assumption may not hold in reality, some
nonlinear models are proposed in literature, see \cite{msw2006,jjs2012,zzlx2018,
u2019,zlx2021}, and the reference therein.  Like linear models, any parametric
models have their limitation, therefore, may suffer from misspecification and
lead to biased predictions.  Nonparametric modelling is flexible, however, it
is not applicable for multiple explanatory variables due to ``curse of
dimensionality".  Semiparametric modelling appears as a more promising
modelling strategy in quantile regression analysis, it makes use of model
information, yet incorporates the ingredients of nonparametric modelling into
the model building process, see \cite{fz1999,wzz2009,zfs2009,klz2010,ky2011,
syzl2014,CHLP14,FJR15,sw2016,CHL16,dlf2019,zwz2021}, and the reference therein.  However, to find
a suitable semiparametric model for a dataset of interest can be difficult.
Model averaging, as a modelling technique, takes all potential models into
account and assigns a weight to each model, which reduces model uncertainty and
may result in more accurate predictions, see \cite{HR12,LS15,GZWZ16,AL17,ZZLC20,
LLWL20,FLO20,FLYZ21}, and the reference therein.  Although plenty of models appear in
literature, which model should be used in reality depends on what the data is
like and which aspects are of interest.

\subsection{A motivating example}
\label{mot0}

The data which stimulates this work is the famous Boston housing data, see
detailed description of the data in Section \ref{sect.5}.  What we are
interested in is how some factors, which are commonly associated with house
price, affect the house prices in different parts of the market, such as
upper end of the market and lower end of the market, thereby predict the
market.  The factors of interest are:

\begin{enumerate}
\item[(1)] CRIM: per capita crime rate by town
\item[(2)]  ZN: proportion of residential land zoned for lots over $25,000$
sq.ft.
\item[(3)]  INDUS: proportion of non-retail business acres per town
\item[(4)]  CHAS: Charles River dummy variable (= 1 if tract bounds river; 0 otherwise)
\item[(5)]  NOX: nitric oxides concentration (parts per 10 million)
\item[(6)]  RM: average number of rooms per dwelling
\item[(7)]  AGE: proportion of owner-occupied units built prior to 1940
\item[(8)]  DIS: weighted distances to five Boston employment centres
\item[(9)]  RAD: index of accessibility to radial highways
\item[(10)] TAX: full-value property-tax rate per \$10,000
\item[(11)] PTRATIO: pupil-teacher ratio by town
\item[(12)] B: $1000(Bk - 0.63)^2$ where $Bk$ is the proportion of blacks by town
\item[(13)] LSTAT: \% lower status of the population.
\end{enumerate}

In the context of statistical modelling, what we are interested in can be
formulated as how the above listed factors affect the quantiles of the house
prices in Boston, and how to predict the conditional quantiles given these
factors.  This is basically a problem of conditional quantile prediction.

The analysis of this dataset has appeared in much literature.  One
frequently used approach is based on the varying coefficient models (VCMs), see
\cite{fh2005, wx2009,hx2012,fmd2014}. However, for this dataset, the selection
of the index in the varying coefficient models is not as obvious as the cases
where nonlinear dynamic is of interest.  Among the $13$ factors, which are the
explanatory variables, $10$ are continuous variables, therefore, any of them
can serve as the index.  Quite a few literature assumes LSTAT as the index,
for example \cite{fh2005,wx2009,hx2012}.  Whilst, their analysis has led to
some interesting findings, it is reasonable to question the rationale of
assuming LSTAT as the index.  In fact, regardless which variable is assumed as
the index, the same question can be asked.  To solve this problem, it would be
better to take a more data driven approach.  One promising data-driven-approach
is that we assume any of the $10$ continuous variables can be the index, and
let data assign a weight to the prediction obtained based on each varying
coefficient model resulted from assuming each of the $10$ continuous variables
to be the index, then take a model averaging approach to construct the final
prediction.  Indeed, we will see that this approach leads to a more accurate
and stable prediction for the conditional quantiles, and it stimulates us to
propose the semiparametric modelling strategy for conditional quantile
prediction in this paper.

Although the proposed modelling strategy and the resulting predictions are
stimulated by a particular dataset, apparently, they are widely applicable
for other datasets from many other disciplines.

This paper is organised as follows: in Section \ref{sect.2}, we will give a
detailed description about the proposed modelling strategy and the resulting
prediction.  Theoretical properties of the proposed prediction and estimation
involved are presented in Section \ref{sect.3}.  In Section \ref{sect.4}, we
conduct intensive simulation studies to demonstrate how well the proposed
prediction works.  In Section \ref{sect.5}, we apply the proposed prediction
method to the Boston house price data, and show the proposed method provides a
more accurate prediction to the market than its alternatives.  All theoretical
proofs are left in Section \ref{sec6}.

\section{A modelling strategy in quantile prediction}
\label{sect.2}

Although the proposed modelling strategy is motivated by the Boston house
price data, in this Section, rather than focusing on this specific data, we
introduce the methodology in a generic term.

Let $Y$ be the response variable, $\bm{X}$ the covariate.
$\bm{X}=(\bm X_{C}^\top, \ \bm X_{D}^\top)^\top$, $\bm X_{C}$ is a
$p$-dimensional continuous vector, $\bm X_{D}$ a $q$ dimensional discrete
vector.  We use $Q_{\tau}(Y|\bm{X})$ to denote the $\tau$th quantile of $Y$
given $\bm{X}$, which is what we would like to predict.

Let $\{(\bm{X}_i^{\top}, \ Y_i), \ 1 \le i \le n\} $ be a sample from
$(\bm{X}^{\top}, \ Y)$, we are going to construct a prediction of
$Q_{\tau}(Y|\bm{X})$ based on this sample.

\subsection{Individual varying coefficient model based quantile prediction}
\label{sect.2.1}

In this Sub-section, we are going to brief the quantile prediction based on a
varying coefficient model.

For each given $s$, $1 \leq s \leq p$, taking the $s$th component of
$\bm{X}_i$, $X_{is}$, as the index, we apply the following varying coefficient
model to fit the condition quantile $Q_{\tau}(Y|\bm{X}_i)$
\begin{equation}
\label{eq1}
\mathbb{M}_s:
{\mu _{\tau,s}}(\bm X_i)
=
\alpha_{\tau,s}\left( X_{is}\right)
+
\bm X_{i\setminus s}^\top\bm \beta_{\tau,s}\left( X_{is}\right),
\quad
i=1, \ \cdots, \ n, \ s=1, \ \cdots, \ p,
\end{equation}
where $\bm X_{i\setminus s}$ is $\bm X_{i}$ with the $s$th component
being removed.

For any given $x_s$, let $\dot \alpha _{\tau, s}(\cdot)$ and
$\dot {\bm \beta} _{\tau, s}(\cdot)$ be the first derivative of $\alpha_{\tau, s}(\cdot)$ and
$\bm \beta _{\tau, s}(\cdot)$, respectively, by the Taylor's expansion, we have
\[
{\alpha _{\tau, s}}\left( {{X_{is}}} \right)
\approx
{\alpha _{\tau,s}}\left( {{x_s}} \right)
+
{{\dot \alpha }_{\tau,s}}\left( {{x_s}} \right)
\left( {{X_{is}} - {x_s}} \right),
\quad
\bm \beta_{\tau,s}\left( X_{is}\right)
\approx
\bm \beta_{\tau,s}\left( x_s\right)
+
\dot{\bm \beta}_{\tau,s}\left( x_s\right) ( {{X_{is}} - {x_s}})
\]
when $X_{is}$ is in a small neighbourhood of $x_s$. This together with the $\rho$-loss
function for quantile regression lead to the following objective function for
estimating ${\alpha _{\tau,s}}\left( {{x_s}} \right)$ and
$\bm \beta_{\tau,s}\left( x_s\right)$:
\begin{equation}
\label{eq.2}
\sum\limits_{i = 1}^n
{{\rho _\tau }\left\{
{{Y_i} - a - b
\left( {{X_{is}} - {x_s}} \right) - \bm X_{i \setminus s}^\top
\left(
{{\bm c} + {\bm d}\left( {{X_{is}} - {x_s}}
\right)}
\right)}
\right\}}
{K_{{h_s}}}\left( {{X_{is}} - {x_s}} \right),
\end{equation}
where ${\rho _\tau }\left( u \right) = \tau u - uI\left( {u \leq 0} \right)$,
$K_{h_s}(\cdot) = K(\cdot/h_s)/h_s$, $K(\cdot)$ is a kernel function, usually
taken to be either the Gauss kernel or the Epanechnikov kernel, i.e.
$K(\cdot) = 0.75(1 - u^2)_+$, $h_s$ is a bandwidth.  In the numerical analysis
in this paper, we use the Gauss kernel.

Minimise (\ref{eq.2}) with respect to $a$, $b$, $\bm c$ and $\bm d$, and
denote the resulting minimiser by $(\hat{a}, \ \hat{b}, \ \hat{\bm c}^{\top},
\ \hat{\bm d}^{\top})$.  We use $\hat{a}$ and $\hat{\bm c}$ to estimate
${\alpha _{\tau,s}}\left( {{x_s}} \right)$ and
$\bm \beta_{\tau,s}\left( x_s\right)$, and denote them by
$\hat{\alpha} _{\tau,s}(x_s)$ and $\hat{\bm \beta}_{\tau,s}(x_s)$.

For a given $\bm X$, let $X_s$ be the $s$th component of $\bm X$ and
$\bm X_{\setminus s}$ be $\bm X$ with its $s$th component being removed.  Based
on model (\ref{eq1}), we use
\begin{equation}
\hat{\mu}_{\tau,s} (\bm X)
=
\hat{\alpha}_{\tau,s} (X_s)
+
\bm X_{\setminus s}^\top \hat{\bm \beta}_{\tau,s}(X_s),
\label{est0}
\end{equation}
to predict $Q_{\tau}(Y|\bm{X})$.

\subsection{Cross-validation based model averaging}
\label{sect.2.2}

As mentioned in Section \ref{mot0}, the selection of index in the varying
coefficient models can be difficult in some cases, and wrongly selected index
may lead to biased prediction due to misspecification, therefore, the
prediction (\ref{est0}) may not be a good one in practice when the index
variable is not obvious.  Rather than setting one particular continuous
explanatory variable as the index variable to construct a prediction, a
sensible approach would be applying the model averaging idea to construct the
prediction.  Specifically, for each continuous explanatory variable, we treat
it as the index variable to form a varying coefficient model and construct a
prediction based on this model, then use a weighted average of the obtained
predictions as the final prediction.  Namely, we use
\begin{equation}
\hat \mu_{\tau}^{[\bm w]}\left(\bm X\right)
=
\sum\limits_{s=1}^p w_s \hat{\mu}_{\tau, s} (\bm X),
\label{est1}
\end{equation}
as the prediction of $Q_{\tau}(Y|\bm{X})$, where
$$
w_s \geq 0,
\quad
\sum\limits_{s=1}^p w_s=1.
$$

The weights $w_s$s in (\ref{est1}) play a key role in the success of the
prediction.  We apply the cross-validation to select the weights.

Let $\hat{\mu}_{\tau, s}^{\setminus i} (\bm X_i)$ be the prediction of
$Q_{\tau}(Y|\bm{X}_i)$ constructed by (\ref{est0}) without using the $i$th
observation $(\bm{X}_i^{\top}, Y_i)$, $\bm w = (w_1, \ \cdots, \ w_p)^{\top}$.
Our cross-validation sum is defined as
\begin{equation}
\label{eq4}
\mbox{CV}_n (\bm w)
=
\frac{1}{n}
\sum\limits_{i = 1}^n
\rho_\tau
\left\{
Y_i - \sum\limits_{s = 1}^{p} w_s \hat{\mu}_{\tau, s}^{\setminus i} (\bm X_i)
\right\}.
\end{equation}
Let $\hat{\bm w}$, $\hat{\bm w} = (\hat{w}_1, \ \cdots, \ \hat{w}_p)^{\top}$,
minimise (\ref{eq4}).  Our proposed model averaging prediction of
$Q_{\tau}(Y|\bm{X})$ is
\begin{equation}
\label{eq.5}
\hat \mu_{\tau}^{[\hat{\bm w}]}\left(\bm X\right)
=
\sum\limits_{s=1}^p \hat{w}_s \hat{\mu}_{\tau, s} (\bm X),
\end{equation}
and termed jackknife varying coefficient quantile model averaging prediction,
denoted by {\sf JVCQMA}.  Although the idea used here is leave-one-out
cross-validation, there are no much ingredients of jackknife, to conform with
the conventional names used in this regard, such as that in \cite{HR12}, we
still name the proposed method jackknife varying coefficient quantile model
averaging prediction.

\subsection{Measurement of accuracy and computational issues}
\label{fpe0}

Let ${\mathcal{D}_n} = \left\{ \left(\bm X_i^{\top}, \ Y_i \right)
\right\}_{i = 1}^n$, and $(\bm X^{\top}, \ Y)$ be independent of
$\mathcal{D}_n$, and shares the same distribution with $(\bm X_i^{\top}, \ Y_i)$
.  A natural measurement of the accuracy of a model averaging prediction
defined by (\ref{est1}) is
\begin{equation}
{\rm{FPE}}_n(\bm w)
=
E\left\{
\rho _\tau\left(
Y - \sum\limits_{s=1}^p w_s \hat{\mu}_{\tau, s} (\bm X)
\right)|\mathcal{D}_n
\right\}.
\label{fpe}
\end{equation}
We will show, in Section \ref{sect.3}, that the proposed {\sf JVCQMA}
prediction is the optimal one among all model averaging predictions
defined by (\ref{est1}), in terms of ${\rm{FPE}}_n(\bm w)$.

There are two optimization problems involved in the implementation of the
proposed prediction, i.e. minimization of (\ref{eq.2}) and of (\ref{eq4}).
However, they are not difficult to solve numerically.  In fact, the
minimization of (\ref{eq.2}) can be easily solved by the R package
{\tt quantreg}, and the minimization of (\ref{eq4}) can be done by the R
package {\tt Rsolnp}.  Indeed, our numerical analysis conducted in Sections
\ref{sect.4} and \ref{sect.5} shows these two packages work very well in the
implementation of the proposed prediction.

\section{Asymptotic properties}
\label{sect.3}

In this Section, we are going to show the proposed prediction, {\sf JVCQMA}, is
the optimal one among the varying coefficient models based model averaging
prediction.

In the asymptotic properties of the proposed method, because we do not assume
the underlying model is any specific model, there are no true functions or true
parameters for the estimators, involved in the proposed prediction, to converge
to.

For the response variable $Y$ and covariate $\bm X$, let $X_s$ be the $s$th
component of $\bm X$ and $\bm X_{\setminus s}$ be $\bm X$ with its $s$th
component being removed. We define
\begin{equation}
\label{eq.6}
\left(
\alpha_{\tau,s}(\cdot), \ \bm \beta_{\tau,s}(\cdot)^{\top}
\right)^{\top}
=
\mathop {\arg \min }\limits_{a(\cdot), \bm c (\cdot) \in \calF}
E\left\{
\rho _\tau\left(
Y -
a(x_s)
-
\bm X_{\setminus s}^\top\bm c(x_s)
\right)
\right\},
\end{equation}
where $\calF$ is the set of all functions with bounded second derivatives.  We
will show the estimators $\hat{\alpha} _{\tau,s}(x_s)$ and
$\hat{\bm \beta}_{\tau,s}(x_s)$ in Section \ref{sect.2.1} converge to
$\alpha_{\tau,s}(x_s)$ and ${\bm \beta}_{\tau,s}(x_s)$ uniformly.

\subsection{Technical conditions}\label{sect.3.1}

Before presenting our theoretical results, we first state the technical
conditions needed.  We start with the introduction of some notations.  Let
$\varepsilon_i =  Y_i-Q_{\tau}(Y_i|\bm{X}_i)$, and
$f\left( { \cdot \left| {{\bm X_i}} \right.} \right)$ and
$F\left( { \cdot \left| {{\bm X_i}} \right.} \right)$ be the conditional
probability density function (PDF) and cumulative distribution
function (CDF) of $\varepsilon_i$ given $\bm X_i$, respectively.  For each $s$,
$s=1, \ \cdots, \ p$, we use $f_{X_s}(\cdot)$ to denote the marginal density
function of $X_s$, the $s$th component of $\bm X$, and define
${\bm \Sigma_s}
=
E\left[ {{{\left( {1,\bm X_{\setminus s}^\top} \right)}^\top}
\left( {1,\bm X_{\setminus s}^\top} \right) } \right]$.  For any matrix $\bm A$,
we use $\lambda _{\min }(\bm A)$ and $\lambda _{\max }(\bm A)$ to denote its
smallest and largest eigenvalue, respectively, and for any vector
$\bm a=(a_1, \ \cdots, \ a_s)^\top$, we define
$\left\| \bm a \right\| = {\left( {a_1^2 +  \cdots  + a_s^2} \right)^{1/2}}$.

The technical conditions needed in order to derive the asymptotic properties
of the proposed methods are as follows:

\begin{enumerate}
\item[(C1)] $f_{X_s}(\cdot)$ has a compact support and continuous derivatives
up to the second order, and $0 < c_1 \leq f_{X_s}(x) \leq c_2 < \infty$.

\item[(C2)] $\alpha_{\tau,s}\left( x_s\right) $ and
$\bm \beta_{\tau,s}\left( x_s\right) $ have continuous second derivatives.

\item[(C3)] $K\left( \cdot\right) $ is a symmetric density function with
bounded support and satisfies a Lipschitz condition.

\item[(C4)] $f\left( { \cdot \left| {{\bm X_i}} \right.} \right)$ is
differentiable, and both $f\left( { \cdot \left| {{\bm X_i}} \right.} \right)$
and its derivative are bounded by a constant $c_f$.

\item[(C5)] There exist two constants $C_1$ and $C_2$ such that
$0<C_1\leq {\lambda _{\min }}\left({\bm \Sigma_s} \right)\leq {\lambda _{\max }}\left({\bm \Sigma_s}\right)\leq C_2<\infty$.
\end{enumerate}

Conditions (C1) and (C2) impose some smoothness restrictions on the marginal
density functions and marginal regression functions, which are similar to the
conditions (A2) and (A4) in \cite{CLLL18}.  Condition (C1) could be relaxed by
slightly modifying our methodology.  For example, if the marginal density
function of $X_{is}$ is the standard normal density which does not have a
compact support, we can truncate the tail of $X_{is}$ by replacing $X_{is}$
with $X_{is}I(|X_{is}|\leq \varsigma _n)$ and $\varsigma _n$ divergent to
infinity at a slow rate.  Condition (C2) is necessary for local linear
estimation.  Condition (C3) is a commonly-used condition for a kernel
function.  Condition (C4) specifies the quantile restriction, which is similar
to the condition (A.2) (i) in \cite{LS15} and (C5)(i) in \cite{KLZ11}.
Condition (C5) imposes some regularity conditions on the eigenvalues of the
positive definite matrix involved in the asymptotic theory.

\subsection{Main theoretical results}\label{sect.3.2}

Let
$$
\underline{h} = {\min _{1 \le s \le p}}{h_s},
\quad
{\bar h }= {\max _{1 \le s \le p}}{h_s},
\quad
{\bm{ \theta }_{\tau ,s}}\left( {{x_s}} \right)
=
{\left( {{{ \alpha }_{\tau ,s}}\left( {{x_s}} \right),
\bm \beta _{\tau ,s}^{\top}\left( {{x_s}} \right)} \right)^\top},
$$
and
$$
{\bm{\hat \theta }_{\tau ,s}}\left( {{x_s}} \right)
=
{\left( {{{\hat \alpha }_{\tau ,s}}\left( {{x_s}} \right),
\bm{\hat \beta} _{\tau ,s}^{\top}\left( {{x_s}} \right)} \right)^\top},
\quad
{\bm{\hat \theta }_{\tau ,s}^{\setminus i}}\left( {{x_s}} \right)
=
{\left( {{{\hat \alpha }_{\tau ,s}^{\setminus i}}\left( {{x_s}} \right),
\left(\bm{\hat \beta} _{\tau ,s}^{\setminus i}\left( {{x_s}} \right)\right)^\top} \right)^\top},
$$
where ${{\hat \alpha }_{\tau ,s}^{\setminus i}}\left( {{x_s}} \right)$ and
$\bm{\hat \beta} _{\tau ,s}^{\setminus i}\left( {{x_s}} \right)$ are the
estimators of ${\alpha _{\tau,s}}\left( {{x_s}} \right)$ and
$\bm \beta_{\tau,s}\left( x_s\right)$ without using the $i$th observation
$(\bm{X}_i^{\top}, Y_i)$.

\begin{theorem}\label{th.1}  For any compact sets $\mathcal{C}_s$,
$s=1, \ \cdots, \ p$, under conditions (C1)--(C5), if ${\bar h }\rightarrow0$
and $n\underline{h}\rightarrow\infty$ as $n\rightarrow\infty$, we have
$$
\mathop {\max }\limits_{1 \le s \le p} \mathop {\sup }
\limits_{x_s \in \mathcal{C}_s} \left\|{\bm{\hat \theta }_{\tau ,s}}
\left( {{x_s}} \right)-{\bm{ \theta }_{\tau ,s}}\left( {{x_s}} \right) \right\|
=
{O_p}\left({\log }^{1/2}\left( {1/\underline{h}} \right)/
\sqrt {n\underline{h}} +\bar{h}^2\right),
$$
and for each $i$, $i=1, \ \cdots, \ n$,
$$
\mathop {\max }\limits_{1 \le s \le p} \mathop {\sup }
\limits_{x_s \in \mathcal{C}_s}
\left\|{\bm{\hat \theta }_{\tau ,s}^{\setminus i}}\left( {{x_s}}
\right)-{\bm{ \theta }_{\tau ,s}}\left( {{x_s}} \right) \right\|
=
{O_p}\left({\log }^{1/2}\left( {1/\underline{h}} \right)
/\sqrt {n\underline{h}} +\bar{h}^2\right).
$$
 \end{theorem}

\bigskip

Theorem \ref{th.1} gives the uniform convergence rates of
${\bm{\hat \theta }_{\tau ,s}}\left( {{x_s}} \right)$ and
${\bm{\hat \theta }_{\tau ,s}^{\setminus i}}\left( {{x_s}} \right)$ to
${\bm{ \theta }_{\tau ,s}}\left( {{x_s}} \right)$.  These results are building
blocks for the asymptotic optimality of our proposed model averaging estimator,
which is presented in Theorem \ref{th.2}.

\begin{theorem}\label{th.2}
Under the same conditions as Theorem \ref{th.1}.
If ${\log }\left( {1/\underline{h}} \right)/ {n\underline{h}}
\rightarrow 0$ as $n\rightarrow\infty$, we have
\begin{equation}
\label{eq.7}
 \frac{ {\rm{FPE}}_n\left( \hat{\bm w} \right)}{\mathop {\inf}\limits_{w\in \mathbb{W}}{{\rm{FPE}}_n\left( {\bm w} \right)}} =1+o_p(1).
 \end{equation}
 \end{theorem}

\bigskip

Theorem \ref{th.2} shows the proposed prediction {\sf JVCQMA} is asymptotically
optimal among all varying coefficient models based model averaging
predictions.  As a consequence, the proposed prediction {\sf JVCQMA} is more
accurate than the prediction based on either the varying coefficient models or
additive models.

\section{ Simulation Studies}
\label{sect.4}

In this section, we are going to use simulated examples to demonstrate the
superiority of the proposed {\sf JVCQMA} over its alternatives.

In all numerical analysis in this paper, the kernel function involved in the
proposed {\sf JVCQMA} is taken to be the Gauss kernel, and the bandwidth
involved is selected in the same way as that in \cite{KLZ11}, namely,
$$
h_s
=
\tilde{h}_{s}\times\left\{ {\tau(1-\tau)/\phi^2(\Phi^{-1}(\tau))} \right\} ^{1/5},
\quad
s=1, \ \cdots, \ p,
$$
where $\phi(\cdot)$ and $\Phi(\cdot)$ are the density function and distribution
function of standard normal distribution, respectively, and $\tilde{h}_{s}$ is
selected by the cross-validation criterion based on the local least squares
estimation for $\alpha_{\tau,s}(\cdot)$ and $\beta_{\tau,s}(\cdot)$, pretending
(\ref{eq1}) is the regression function of $Y_i$ on $\bm X_i$.

The methods we are going to compare with are as follows:

\begin{enumerate}
\item[(1)] The linear quantile regression model, denoted by {\sf LQR}.  It is
implemented by the R function {\tt rq} in the R package {\tt quantreg}.

\item[(2)] The penalized linear quantile regression model with SCAD proposed
by \cite{WL09}, denoted by {\sf PLQR}.  It is implemented by the R function
{\tt cv.rq.pen} in the R package {\tt rqPen}.

\item[(3)] The linear quantile model averaging proposed by \cite{LS15},
denoted by {\sf LQMA}.

\item[(4)] The varying coefficient quantile regression model, denoted by
{\sf VCQR}.

\item[(5)] The partial linear additive quantile regression model proposed by
\cite{L12}, denoted by {\sf AQR}.

\item[(6)] The varying coefficient quantile model averaging prediction with equal weights, denoted by {\sf VCQMA1}.

\item[(7)] The varying coefficient quantile model averaging prediction
with weights being selected by the smoothed BIC criterion \citep{BBA97},
namely,
$$
{{\hat w}_s}
=
\exp \left( { - 0.5{\rm {BIC}}_{(s)}} \right)/
\sum\limits_{j = 1}^p {\exp \left( { - 0.5{\rm {BIC}}_{(j)}} \right)},
\quad
s=1, \ \cdots, \ p
$$
with
\[
{\rm {BIC}}_{(s)}
=
2n\ln \left[ {\frac{1}{n}\sum\limits_{i = 1}^n
{{\rho _\tau }\left( {Y_i} -\hat{\mu}_{\tau, s} (\bm X_i)\right)} } \right]
+
\left( {p+q - 1} \right)\ln \left( n \right).
\]
This method is denoted by {\sf VCQMA2}.

\item[(8)] The proposed {\sf JVCQMA}.
\end{enumerate}

We are going to examine the performances of the above eight methods by four
simulated examples, which are detailed as follows:

\bigskip

\noindent
\textbf{Example 1}.  In this example, we consider an unconventional varying
coefficient model and generate the random samples from the following model
\begin{equation}\label{ex1}
Y_i
=
\alpha\left( X_{i1}\right)
+
\bm X_{i \setminus 1}^\top\bm \beta\left( X_{i1}\right)
+
\bm X_{i \setminus 2}^\top\bm \beta\left( X_{i2}\right)
+
\varepsilon_i,
\quad
i=1, \ \cdots, \ n,
\end{equation}
where, we set $\alpha\left( u\right) = u$,
$$
\bm \beta\left( u\right)
=
(u(1-0.5u), \ {\rm exp}(u/2-0.5), \ {\rm sin}\left(2\pi u\right)-u, \
2{\rm exp}(-0.5u^2)/({\rm exp}(-0.5u^2)+1), \ \0_{p-5})^\top,
$$
where $\0_k$ is a $k$-dimensional row vector with each component being $0$.

The covariate $\bm X_i=\left(X_{i1}, \ \cdots, \ X_{ip} \right)^\top$s are
generated from a normal distributions with mean zero and
$Cov\left( {{X_{ij}},{X_{il}}} \right) = {0.5^{\left| {j - l} \right|}}$ for
$1\leq j, l \leq p$.

We note that this true model involves two types of varying coefficients induced
by two index variables.  In order to examine the robustness of the proposed
procedure, we consider the following three different error distributions for
$\varepsilon_i$:  standard normal distribution (\emph{case1}),
$t$-distribution with three degrees of freedom (\emph{case2}) and a mixture of
two normal distributions (\emph{case3}), which is a mixture of $N(0,1)$ and
$N(0,25)$ with the weights 95\% and 5\%, respectively.

\bigskip

\noindent
\textbf{Example 2}. In this example, we generate the random samples from the following heteroscedastic model
$$
{Y_i}
=
\alpha\left( X_{i1}\right)
+
\bm X_{i \setminus 3}^\top\bm \beta_1\left( X_{i3}\right)
+
\bm X_{i \setminus 4}^\top\bm \beta_2\left( X_{i4}\right)
+
{\varepsilon _i}, \quad i=1, \ \cdots, \ n,
$$
where, we set $\alpha\left( u\right)=u(1-u)$,
$$
\bm \beta_1\left( u\right)
=
\left((2+u^2)/(1+u^2), \ u, \ \0_{p-3} \right)^\top,
$$
$$
\bm \beta_2\left( u\right)
=\left(
2{\rm sin}\left(2\pi u\right)/(2-{\rm cos}\left(2\pi u\right)),\
{\rm exp}(-0.5u^2),\ 1,\ -1,\ \0_{p-5} \right)^\top.
$$

$\bm X_i=\left(X_{i1}, \ \cdots, \ X_{ip}\right)^\top$s are independently
generated from $U(-2, 2)$, $\varepsilon_i$s are generated through
$$
\varepsilon_i
=
0.5\left({\rm sin}^2(X_{i1})+{\rm cos}^2(X_{i2})
+0.5 \right) e_i.
$$
To examine the performance of the proposed method when the error distribution
is asymmetric, we generate $e_i$s from one of the following three asymmetric
distributions: chi-square distribution with one degree of freedom (\emph{case4}),
gamma-distribution $G(1,1)$ (\emph{case5}), and log normal distribution
(\emph{case6}) with mean and standard deviation being 0.5 and 0.5,
respectively, on the log scale.  $\bm X_i$ and $e_i$ are generated
independently.

\bigskip

\noindent
\textbf{Example 3}.  To demonstrate the flexibility of the proposed method,
similar to \cite{CLLL18}, we generate the random samples from the following
partially linear additive model
\begin{align}
{Y_i}
=&
\sum\limits_{j = 1}^6 {{f_j}} \left( {{X_{ij}}} \right) + {\beta _1}{X_{i7}}
+ {\beta _2}{X_{i8}} + {\beta _3}{X_{i9}} + {\beta _4}{X_{i10}}
+{\varepsilon _i}, \quad i=1, \ \cdots, \ n,
\nonumber
\end{align}
 where, we set
 $$
 f_1(u)={\rm{-sin}}(2u),\quad f_2(u)=0.5(u^2-25/12), \quad f_3(u)=u,
 $$
$$
f_4(u)={\rm{exp}} (-u)-\frac{2}{5}{{\rm{sinh}}(5/2)}, \quad f_5(u)=u, \quad f_6(u)=-2u,
 $$
 $$
 {\beta _1}={\beta _3}=1, \quad {\beta _2}={\beta _4}=-2.
 $$

The covariate $\bm X_i=\left(X_{i1},\cdots,X_{i6} \right)^\top$ is
generated from a normal distribution with mean zero and
$Cov\left( {{X_{ij}},{X_{il}}} \right) = {0.5^{\left| {j - l} \right|}}$ for
$1\leq j, l \leq 6$, $X_{i7}$ from $Binomial(2, \ 0.5)$, $X_{i8}$ from
$Binomial(2, \ 0.5)$, $X_{i9}$ from $Binomial(3, \ 0.5)$, and $X_{i10}$ from
$Binomial(3,0.5)$.  $\varepsilon_i$ is generated through
$$
\varepsilon_i
=
\left( {\left| {0.5{X_{i1}} - 0.5{X_{i2}}} \right| + 0.5} \right) e_i,
$$
and $e_i$ is generated from one of the six error distributions, marked as
$case1, \ \cdots, \ case6$, in Examples 1 and 2.  We generate $\bm X_i$,
$X_{i7}$, $X_{i8}$, $X_{i9}$, $X_{i10}$, and $e_i$ independently.

\bigskip

\noindent
\textbf{Example 4}.  In examples 1--3, we generate data from the models with
varying coefficient and additive structure.  To demonstrate the flexibility of
the proposed method in model specification and the advantage over commonly used
methods, we consider the following multivariate regression model, where
the commonly assumed structures do not hold,
$$
{Y_i}
=
\alpha(X_{i1}, \ \cdots, \ X_{i10})
+
\varepsilon_i,
\quad
i=1, \ \cdots, \ n,
$$
where, we set
$$
\alpha(X_{i1}, \ \cdots, \ X_{i10})
=
4{\rm{cos}}(X_{i1}X_{i2}X_{i3}X_{i4})
-
X_{i5}X_{i6}+X_{i7}X_{i8}X_{i9}X_{i10}.
$$
The covariate $\bm X_i=\left(X_{i1}, \ \cdots, \ X_{i6} \right)^\top$ is
generated from a normal distribution with mean zero and
$Cov\left( {{X_{ij}},{X_{il}}} \right) = {0.5^{\left| {j - l} \right|}}$ for
$1\leq j, l \leq 6$, $X_{i7}$ from $Binomial(1,0.5)$, $X_{i8}$ from
$Binomial(1,0.5)$, $X_{i9}$ from $Binomial(2,0.5)$, and $X_{i10}$ from
$Binomial(2,0.5)$.  $\varepsilon_i$ is generated through
$$
\varepsilon_i
=
\left( {\left| {0.5{X_{i1}} - 0.5{X_{i2}}} \right| + 0.5} \right) e_i
$$
and $e_i$ is generated from one of the six error distributions, marked as
$case1, \ \cdots, \ case6$, in Examples 1 and 2.  We generate $\bm X_i$,
$X_{i7}$, $X_{i8}$, $X_{i9}$, $X_{i10}$, and $e_i$ independently.

\bigskip

For Examples 1 and 2, we set sample size $n$ to be either $200$ or $400$, and
consider $p=5$ or $10$ to reflect different sparsity levels.  For Examples 3
and 4, we only consider the cases when $n=400$ to save space.

For each simulation in each simulated example, we generate a training data set
of sample size $n$ to estimate unknown parameters, functions and model weights,
then generate another $100$ observations (a testing set) to calculate
the FPE, defined in (\ref{fpe}), for each of the eight methods under
comparison.  The FPE is used to measure the accuracy of the out-of-sample
quantile prediction, and is calculated through
$$
\mbox{FPE}
=
\frac{1}{|\mathcal{I}|}
\sum\limits_{(Y_i, \bm X_i) \in \mathcal{I}} {{\rho _\tau }
\left(
{{Y_i} - \bar \mu_{\tau}\left(\bm X_i\right)}
\right)}
$$
where $\bar \mu_{\tau }\left(\bm X_i\right)$ is a prediction of the
conditional quantile $Q_{\tau }\left(Y|\bm X_i\right)$, $\mathcal{I}$ stands
for the testing set, and $|\mathcal{I}|$ is the size of $\mathcal{I}$.

\begin{figure}\center
\includegraphics[scale=0.4]{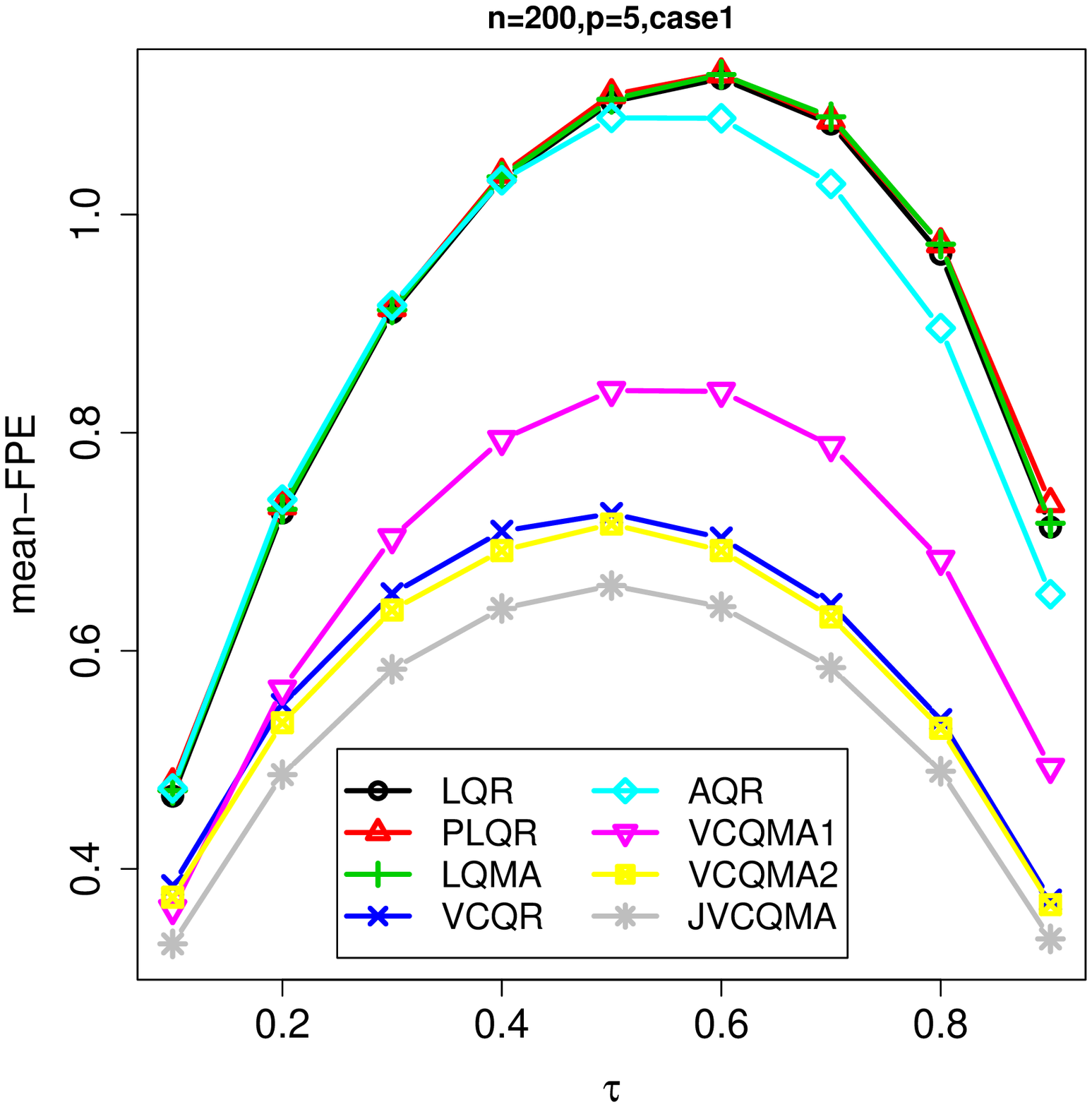}
\includegraphics[scale=0.4]{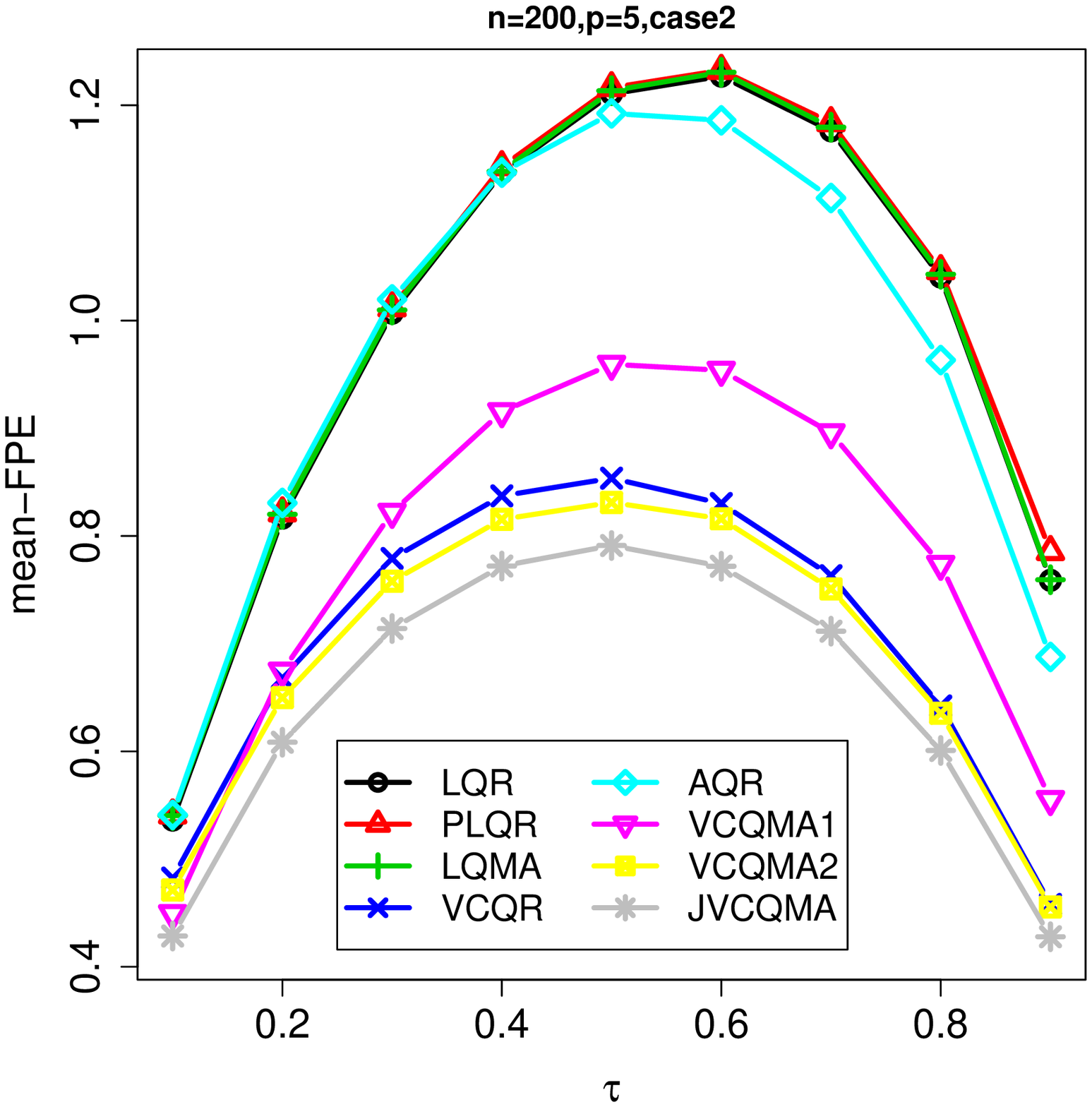}
\includegraphics[scale=0.4]{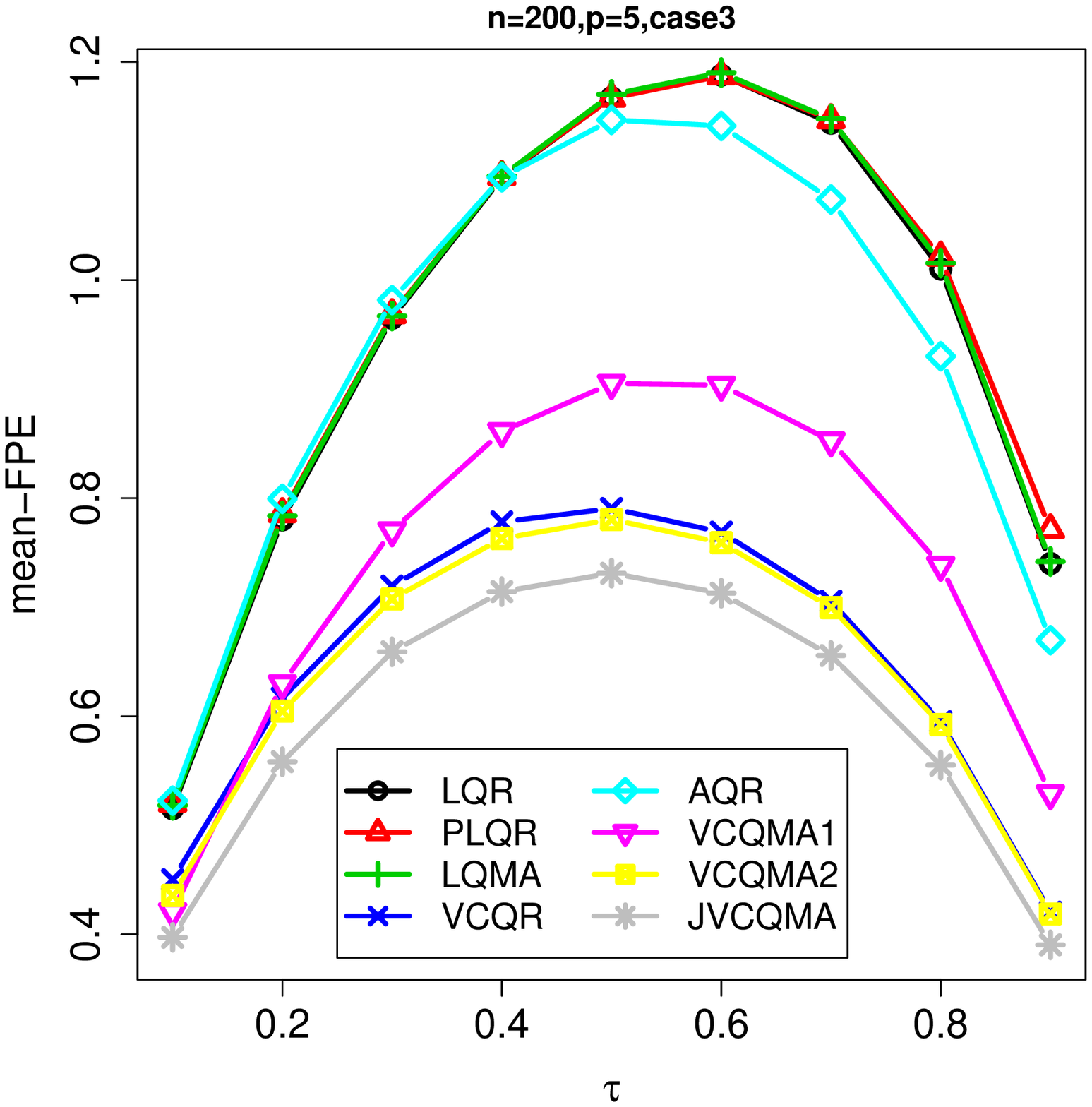}
\includegraphics[scale=0.4]{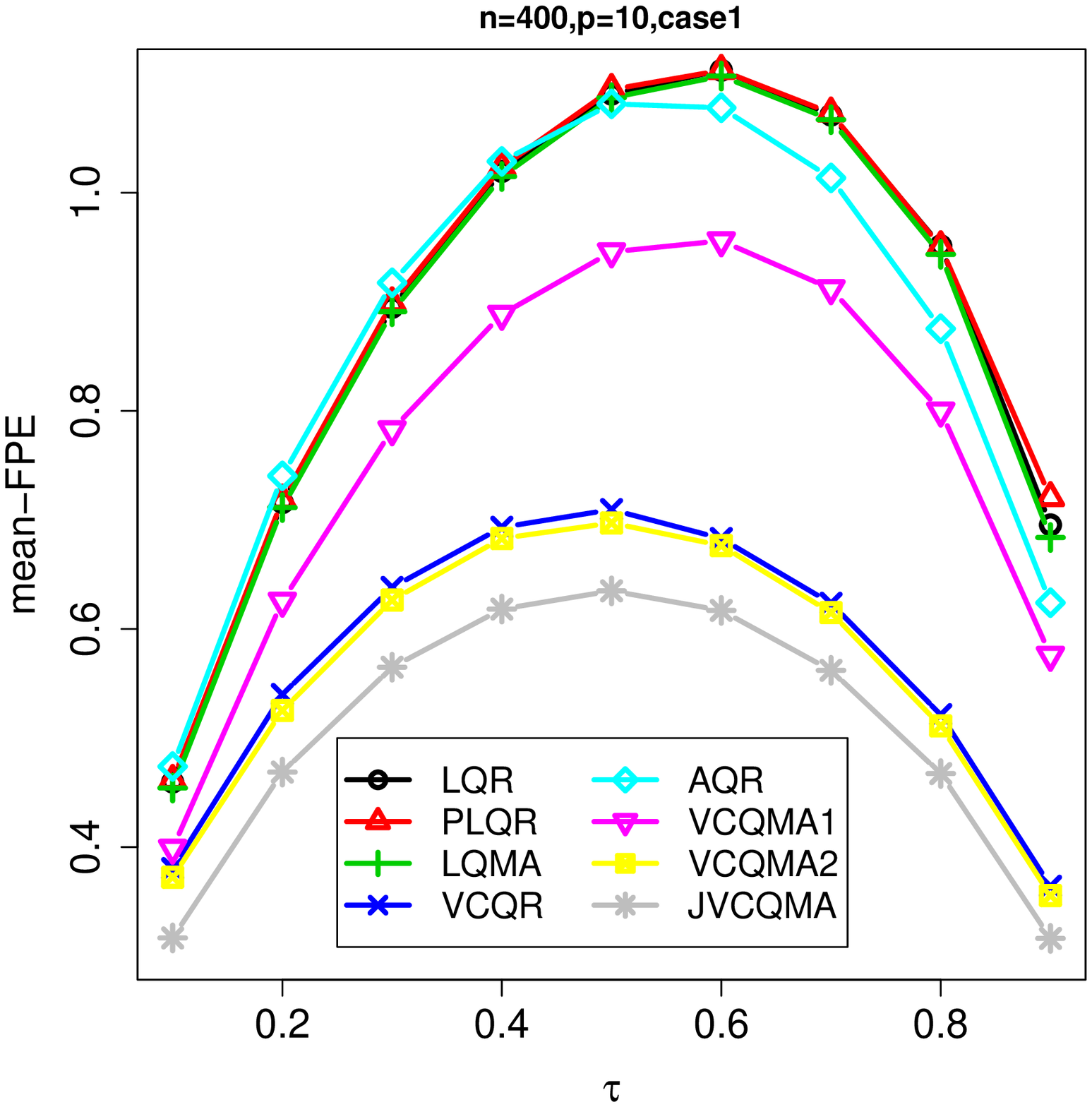}
\includegraphics[scale=0.4]{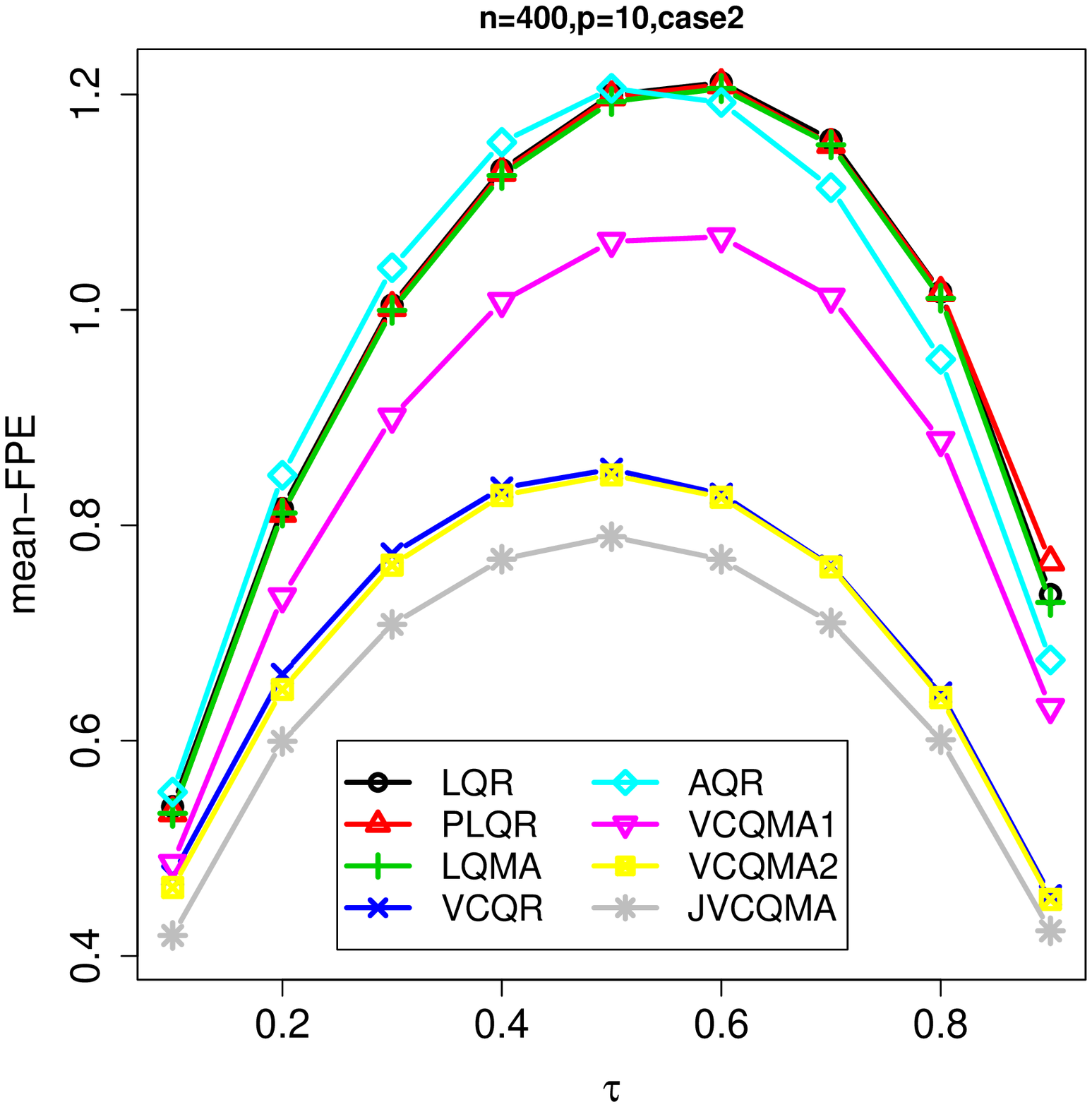}
\includegraphics[scale=0.4]{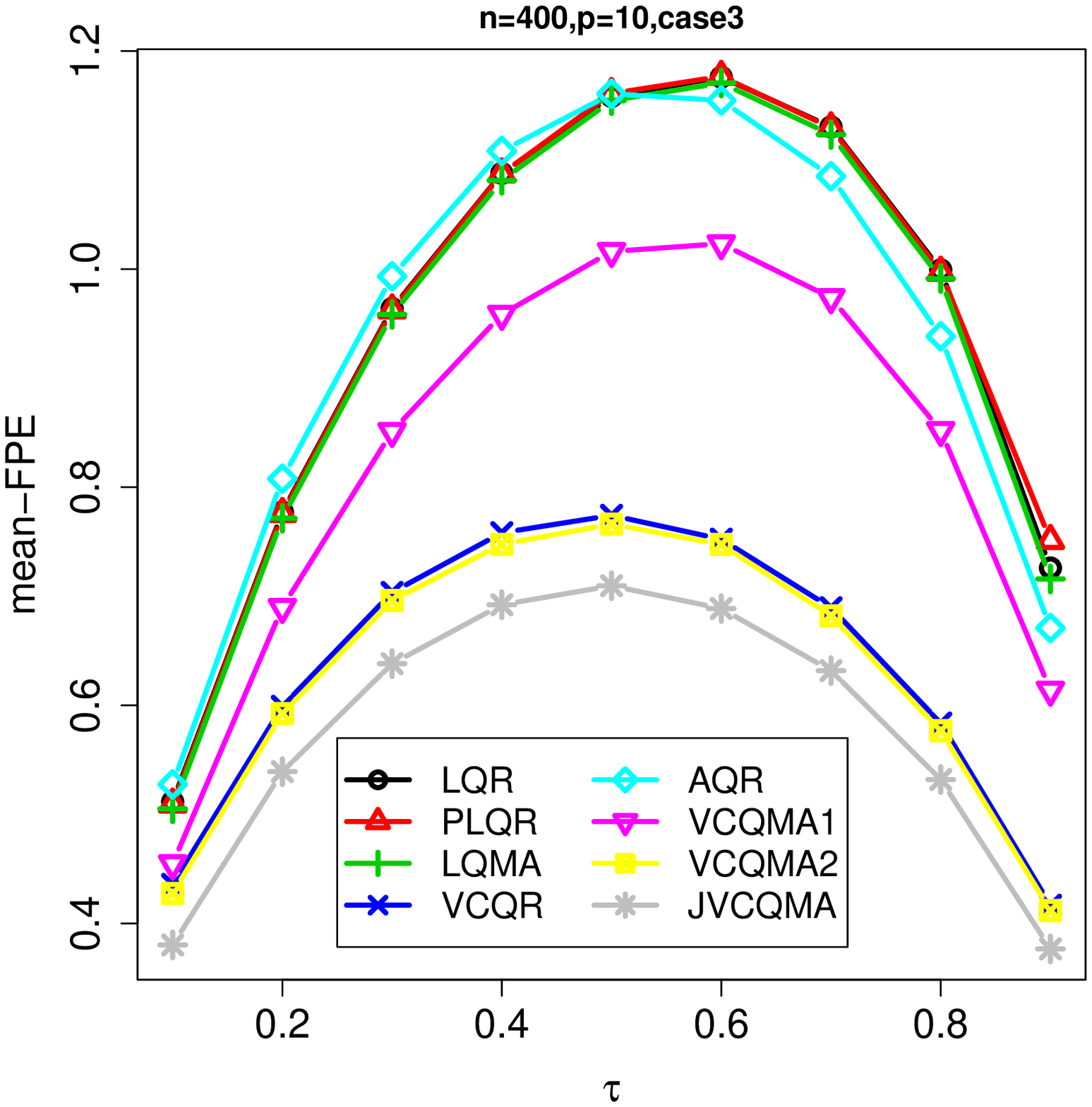}
\caption{Mean-FPEs at different quantiles with $\tau$ ranging from 0.1 to 0.9
for example 1.}
\label{figure1}
\end{figure}

\begin{figure}\center
\includegraphics[scale=0.4]{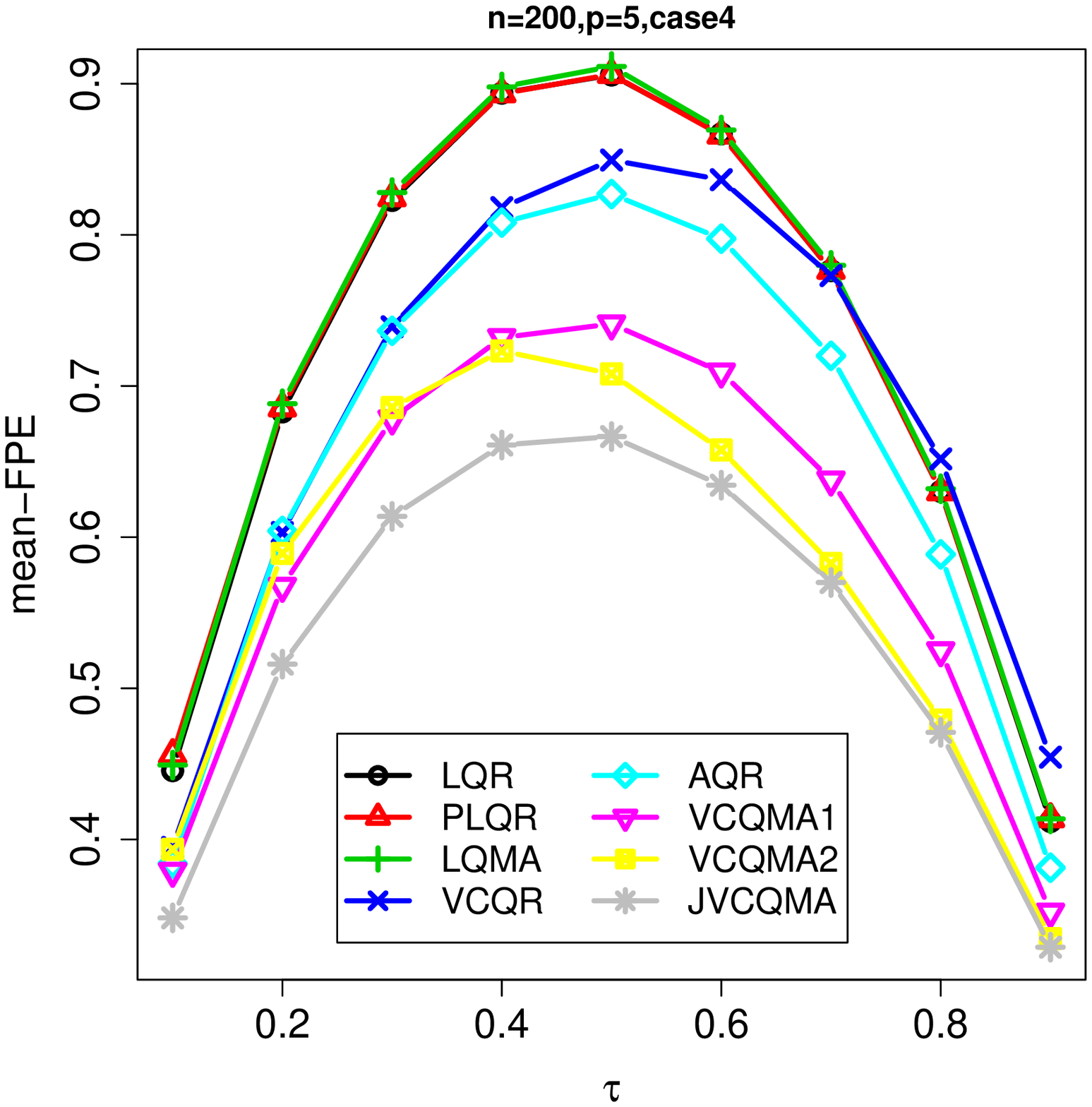}
\includegraphics[scale=0.4]{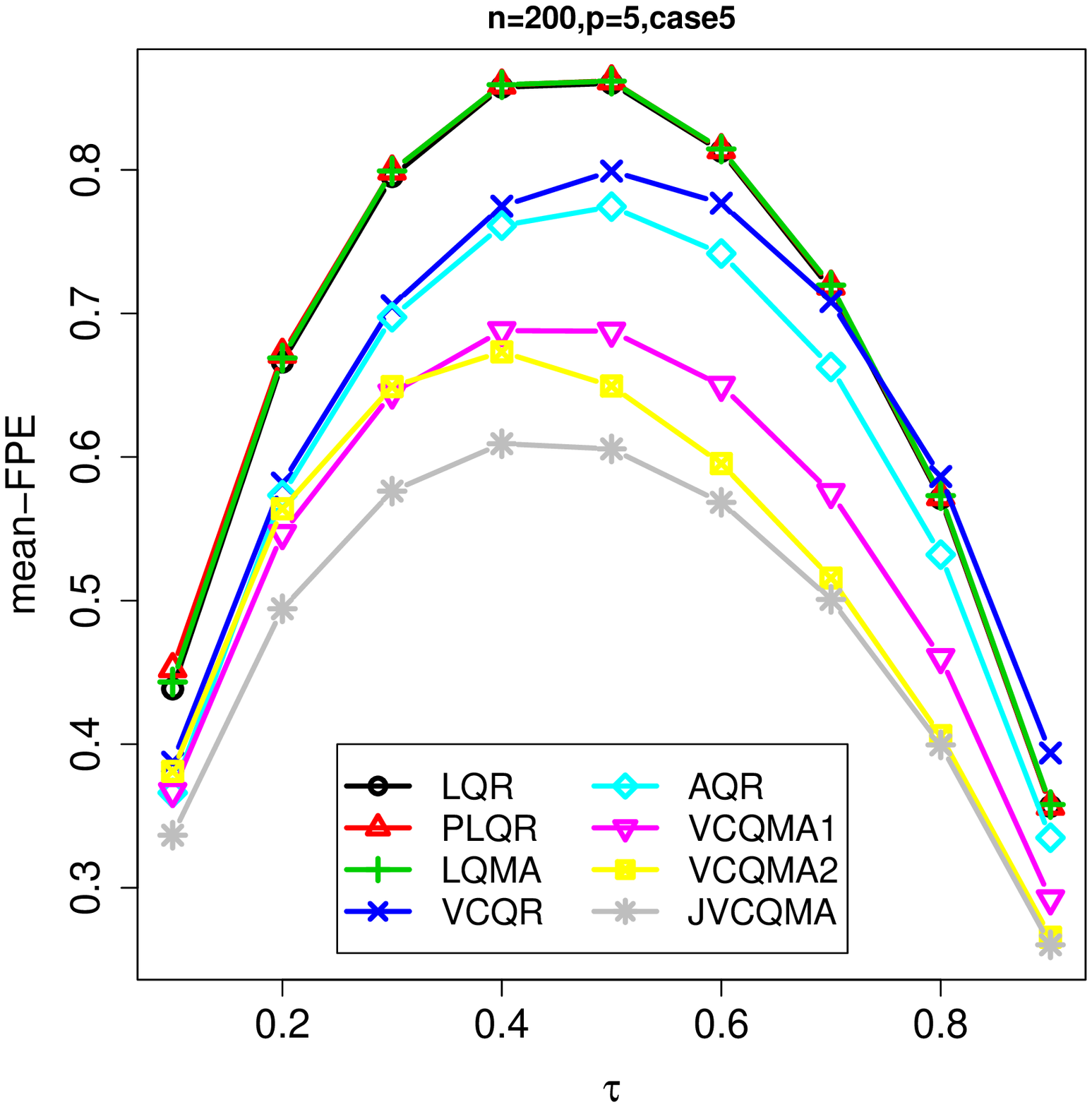}
\includegraphics[scale=0.4]{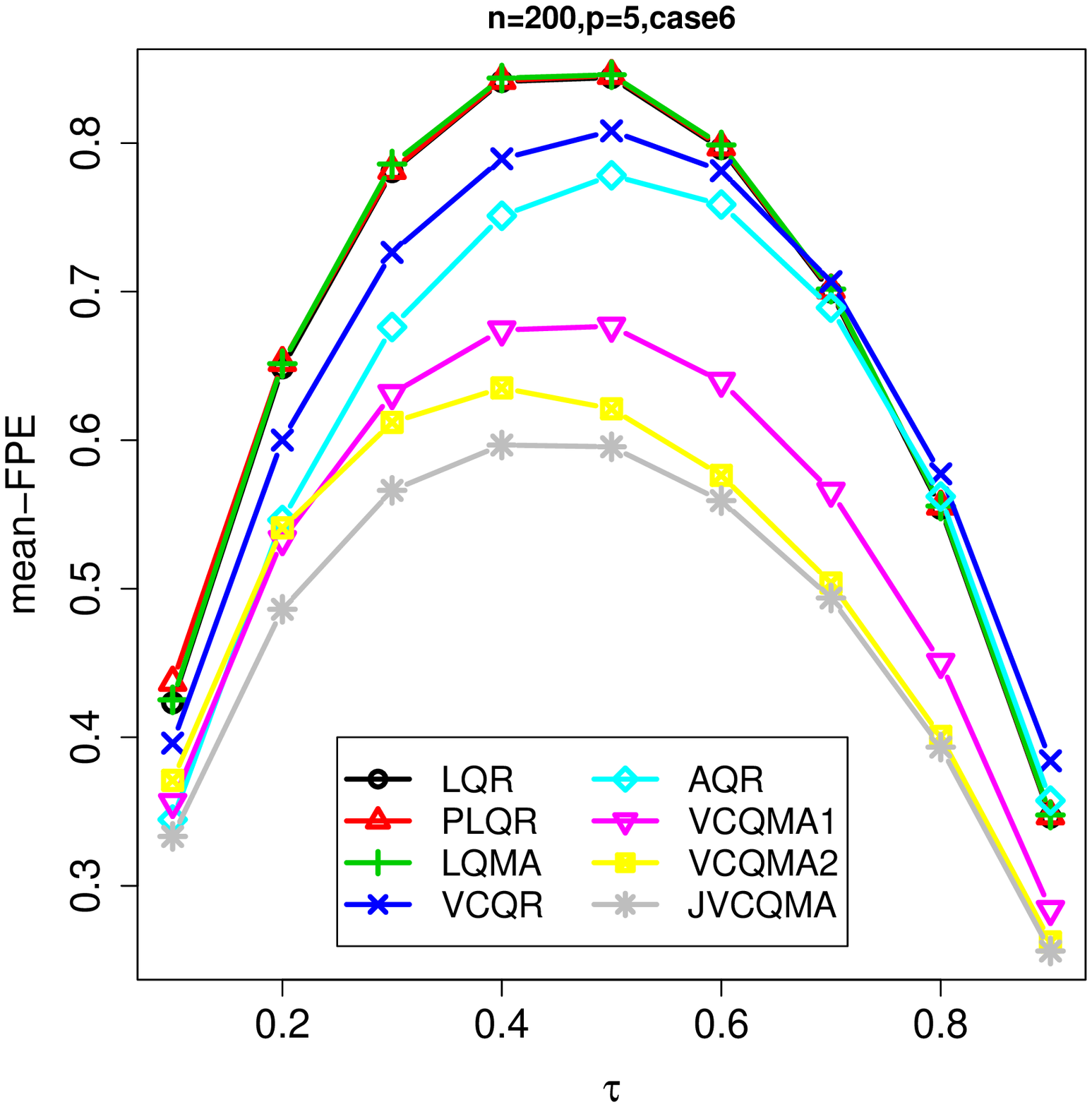}
\includegraphics[scale=0.4]{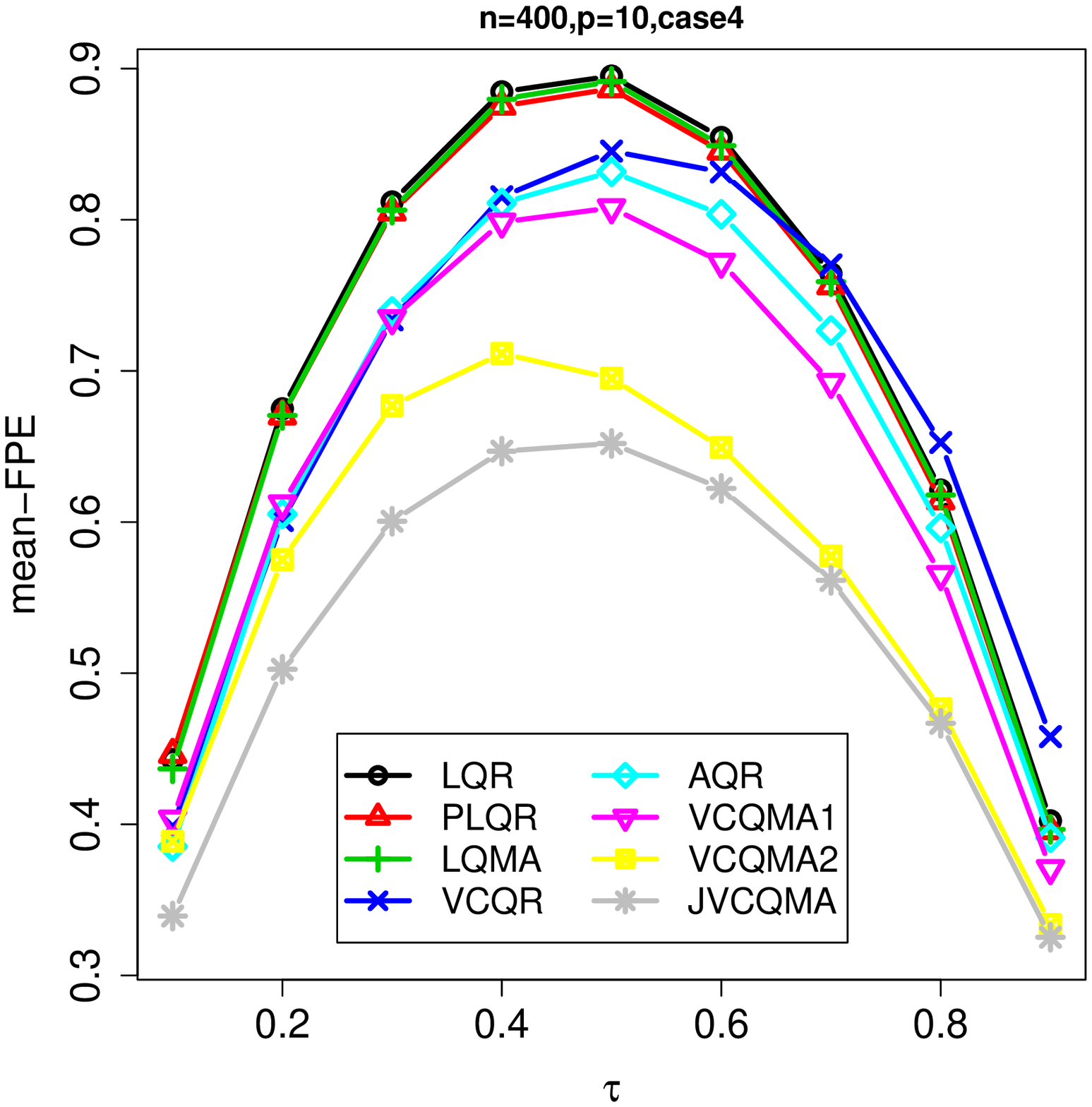}
\includegraphics[scale=0.4]{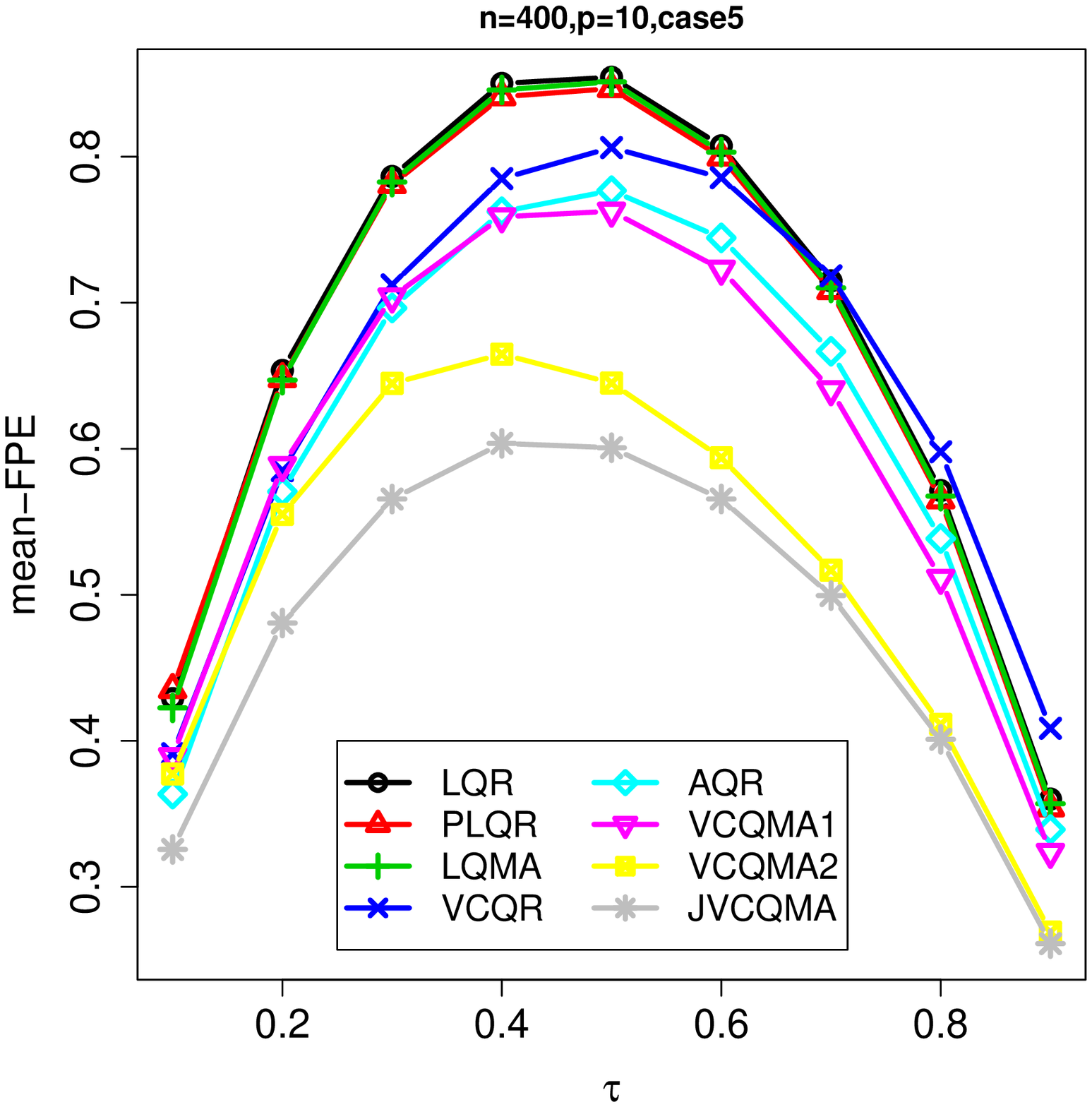}
\includegraphics[scale=0.4]{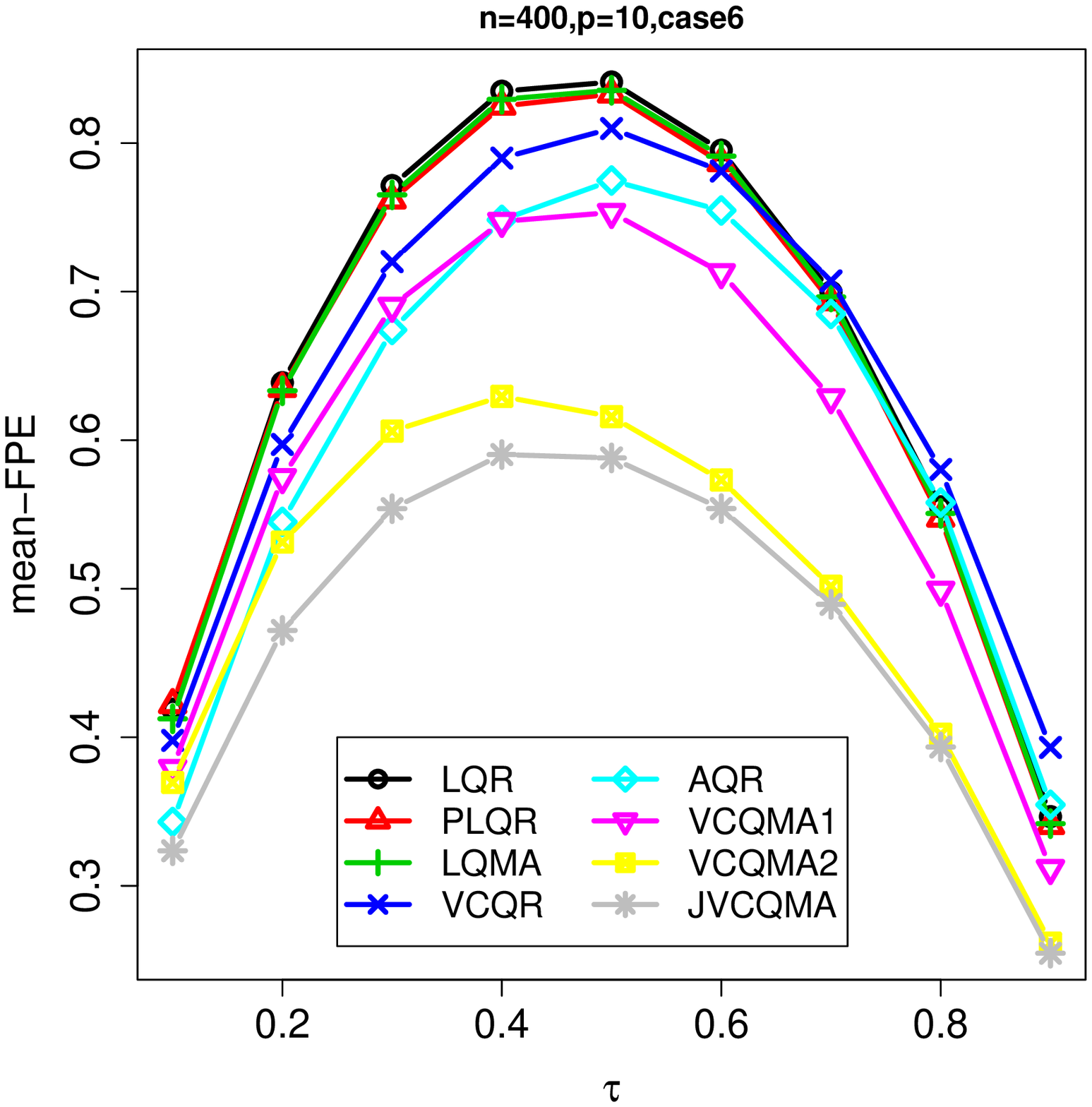}
\caption{Mean-FPEs at different quantiles with $\tau$ ranging from 0.1 to 0.9
for example 2.}
\label{figure2}
\end{figure}

\begin{figure}\center
\includegraphics[scale=0.4]{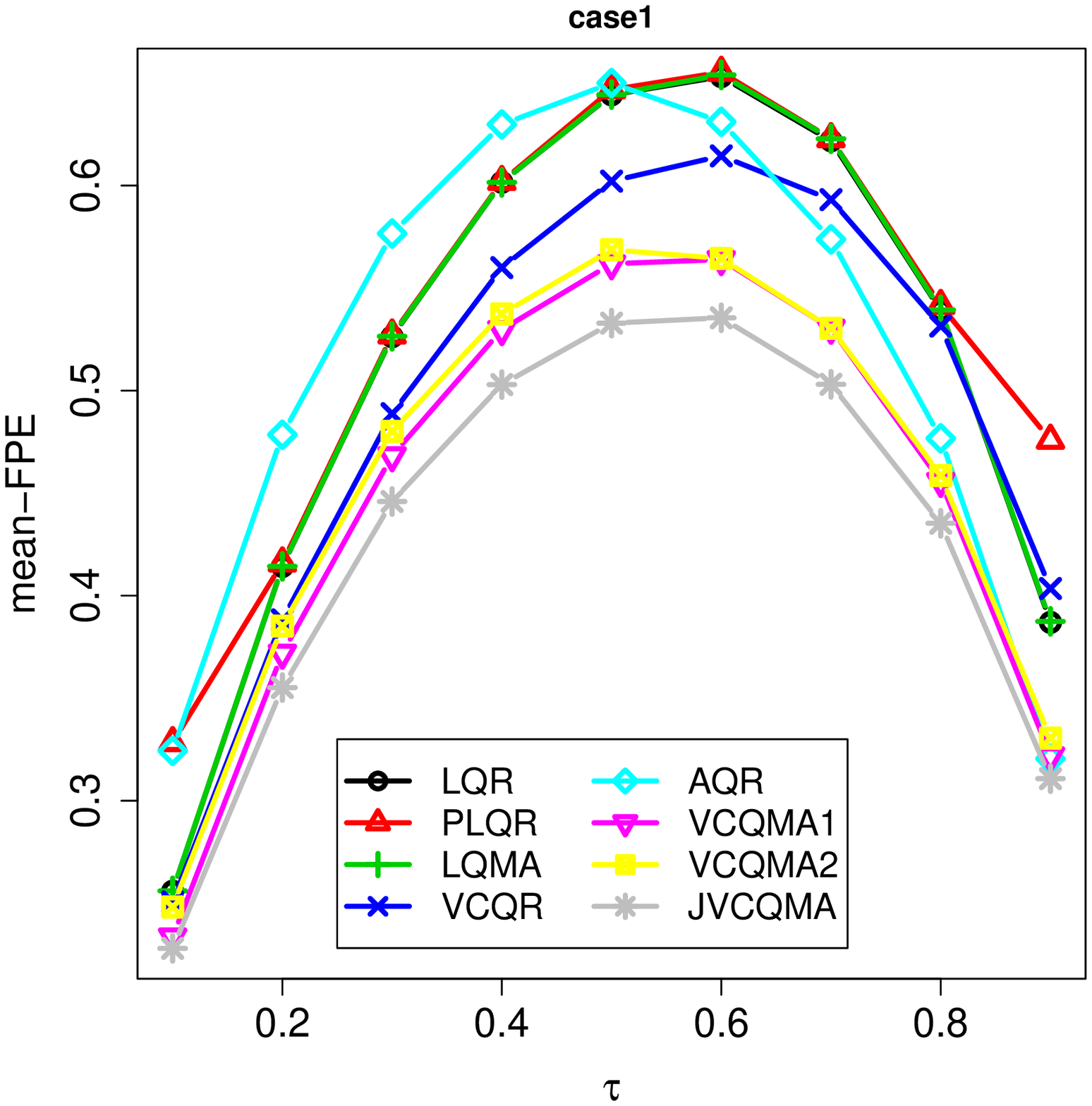}
\includegraphics[scale=0.4]{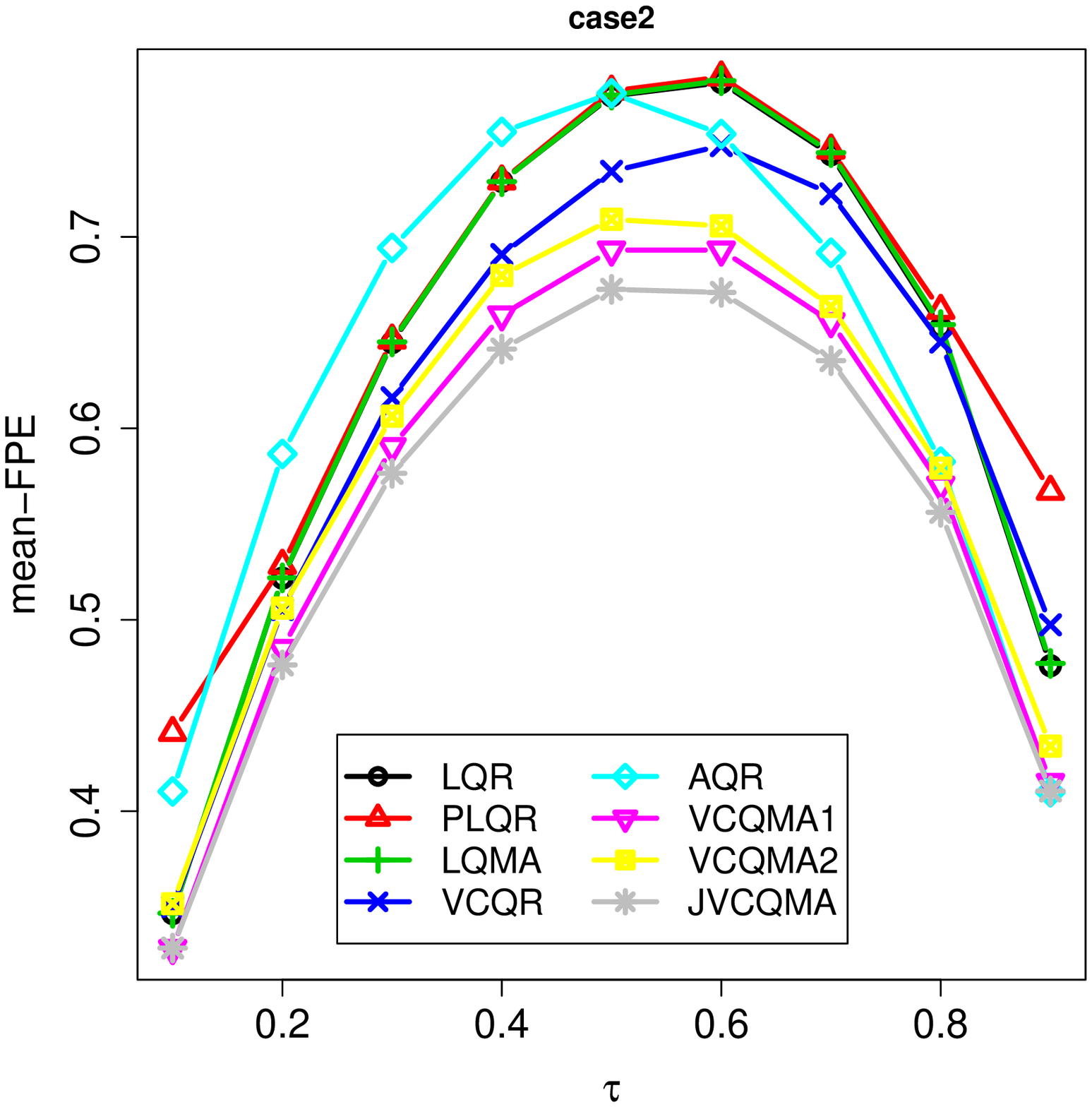}
\includegraphics[scale=0.4]{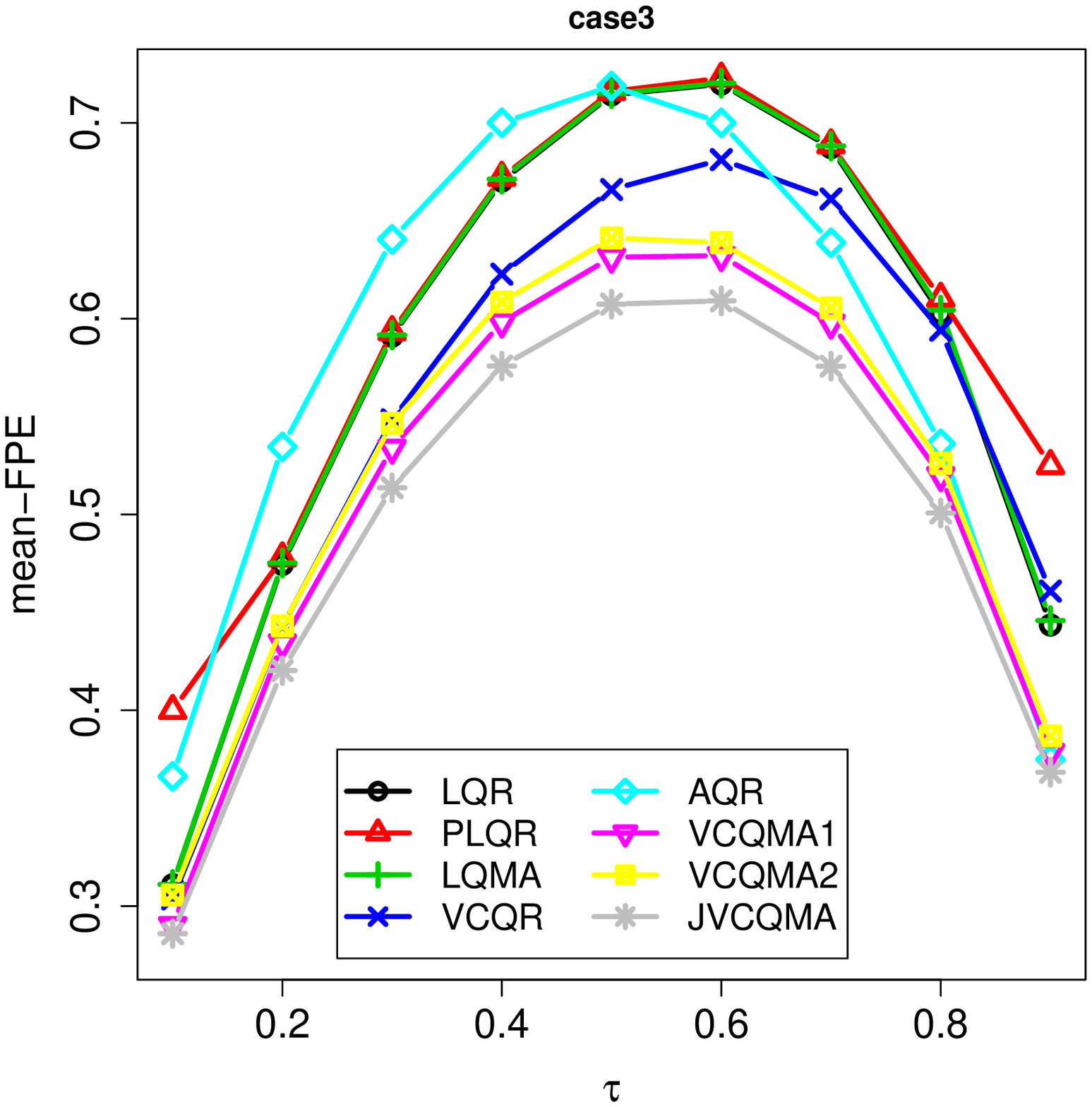}
\includegraphics[scale=0.4]{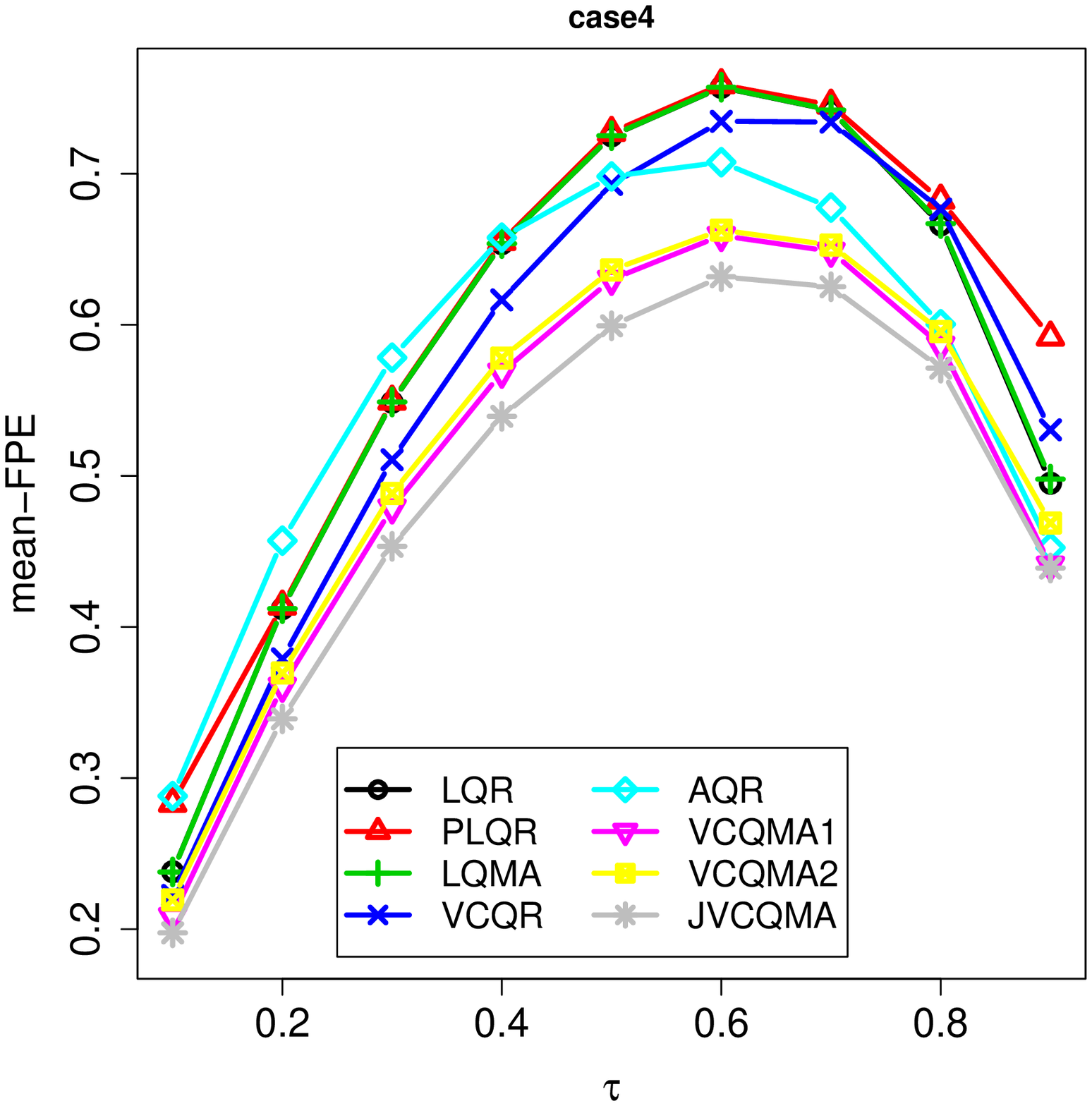}
\includegraphics[scale=0.4]{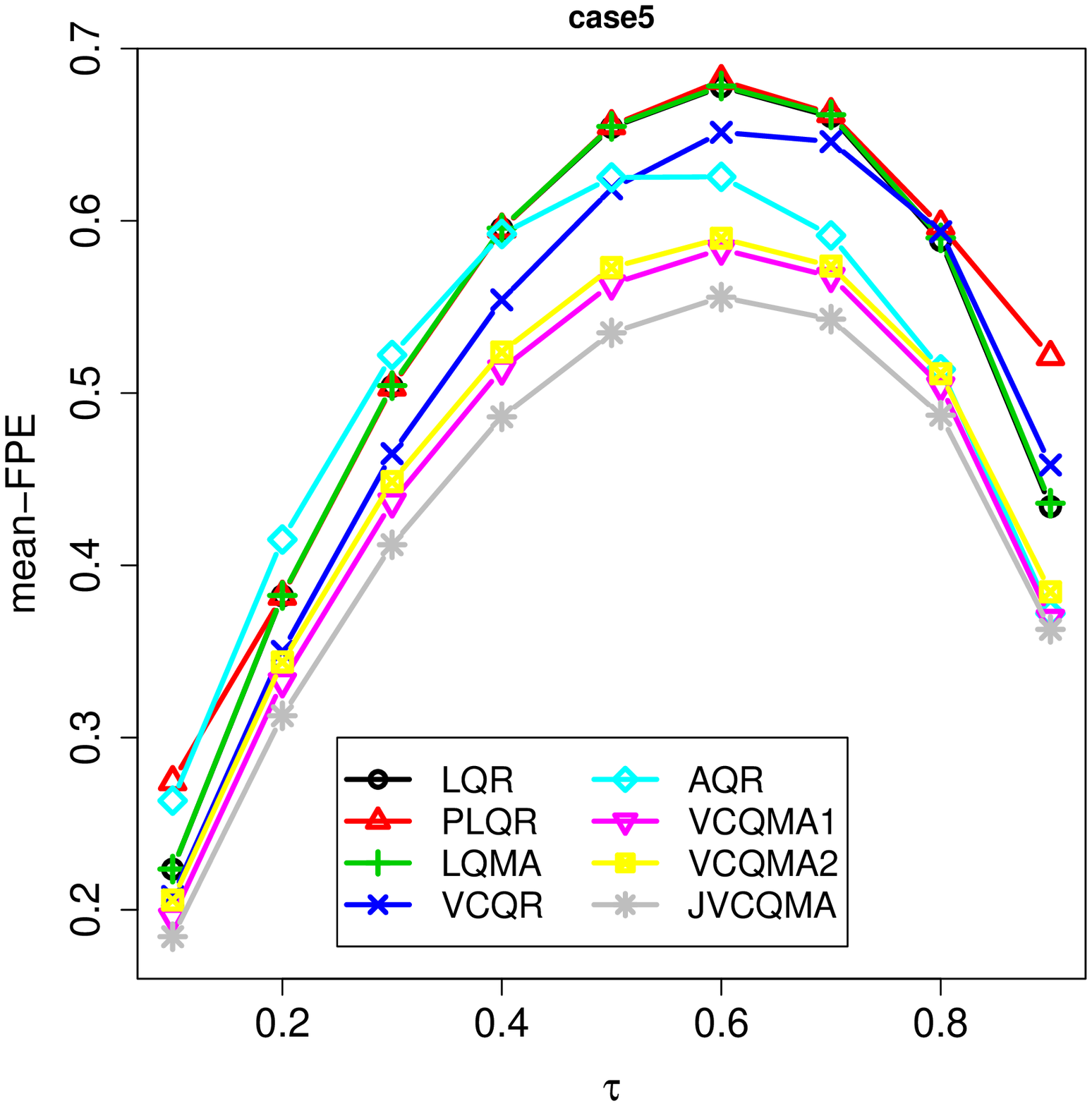}
\includegraphics[scale=0.4]{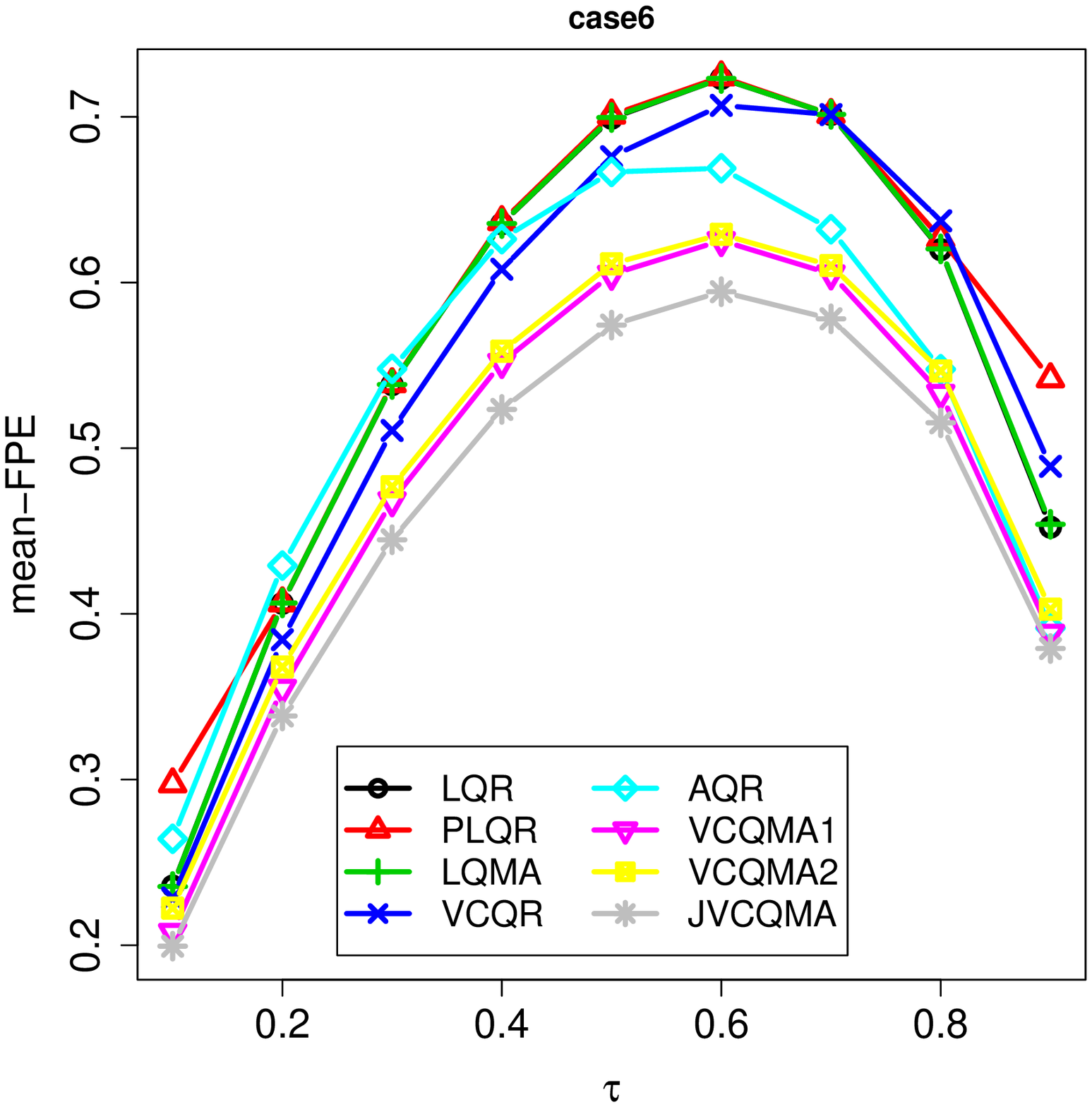}
\caption{Mean-FPEs at different quantiles with $\tau$ ranging from 0.1 to 0.9
for example 3.}
\label{figure3}
\end{figure}

\begin{figure}\center
\includegraphics[scale=0.4]{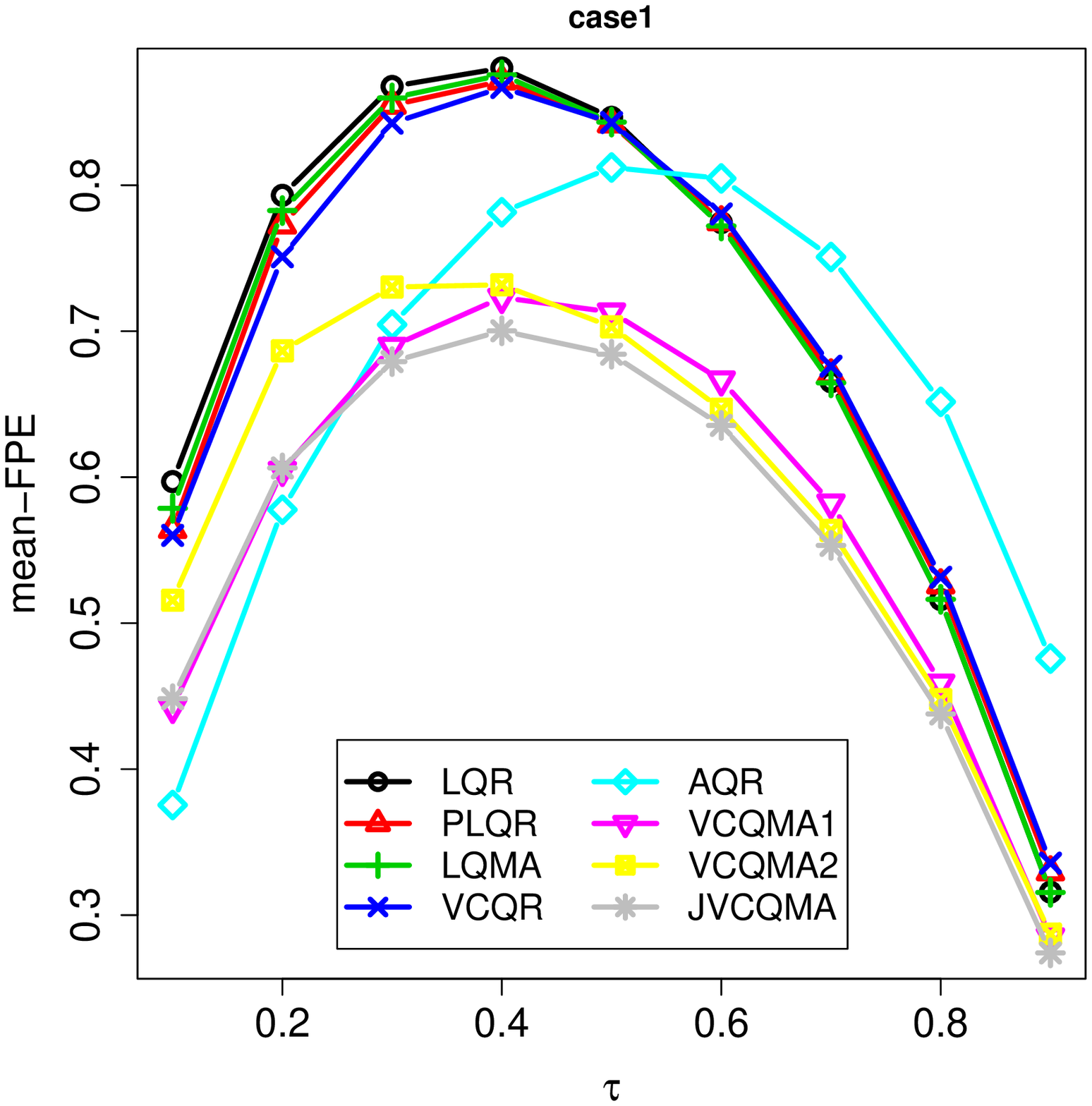}
\includegraphics[scale=0.4]{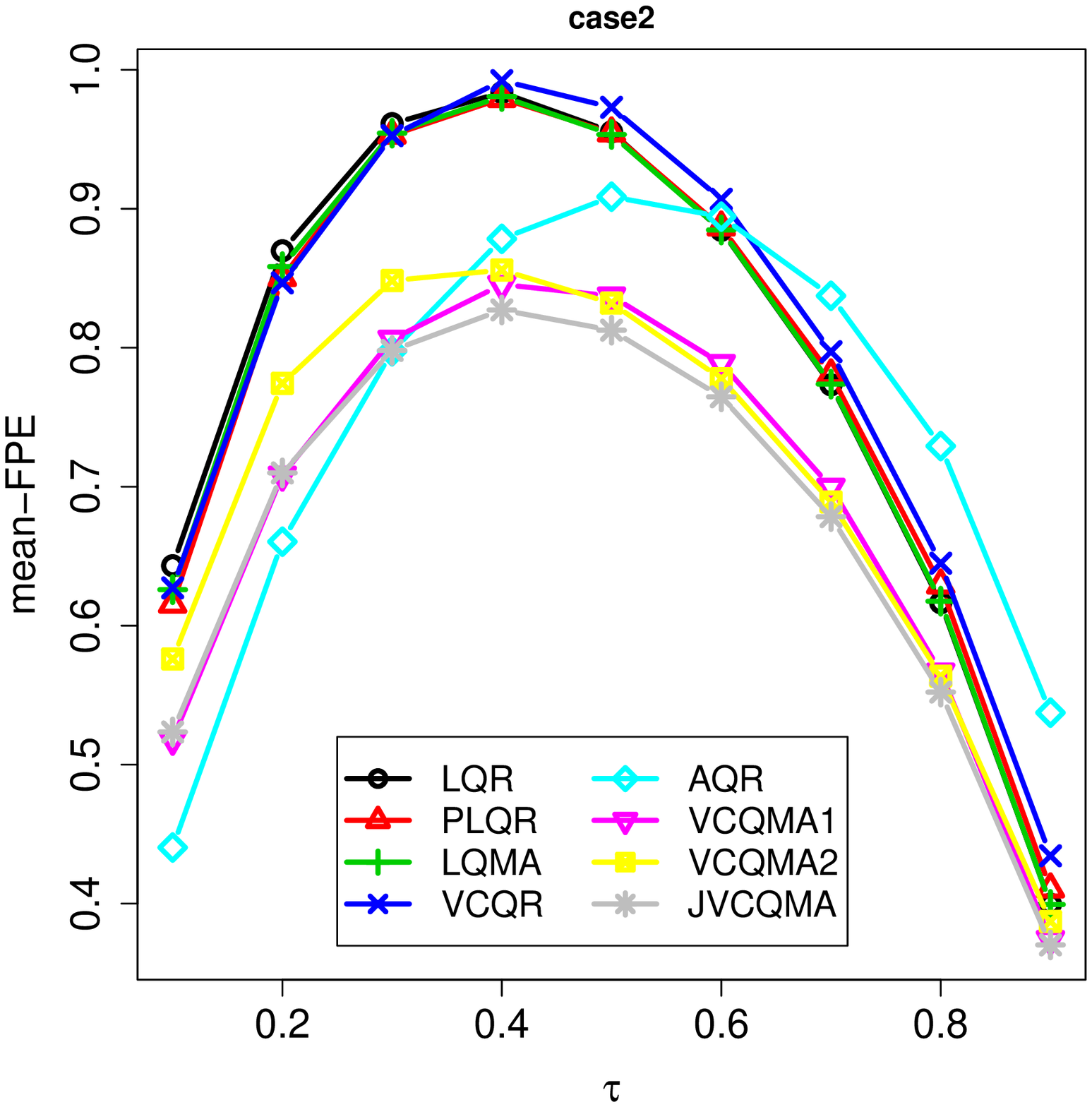}
\includegraphics[scale=0.4]{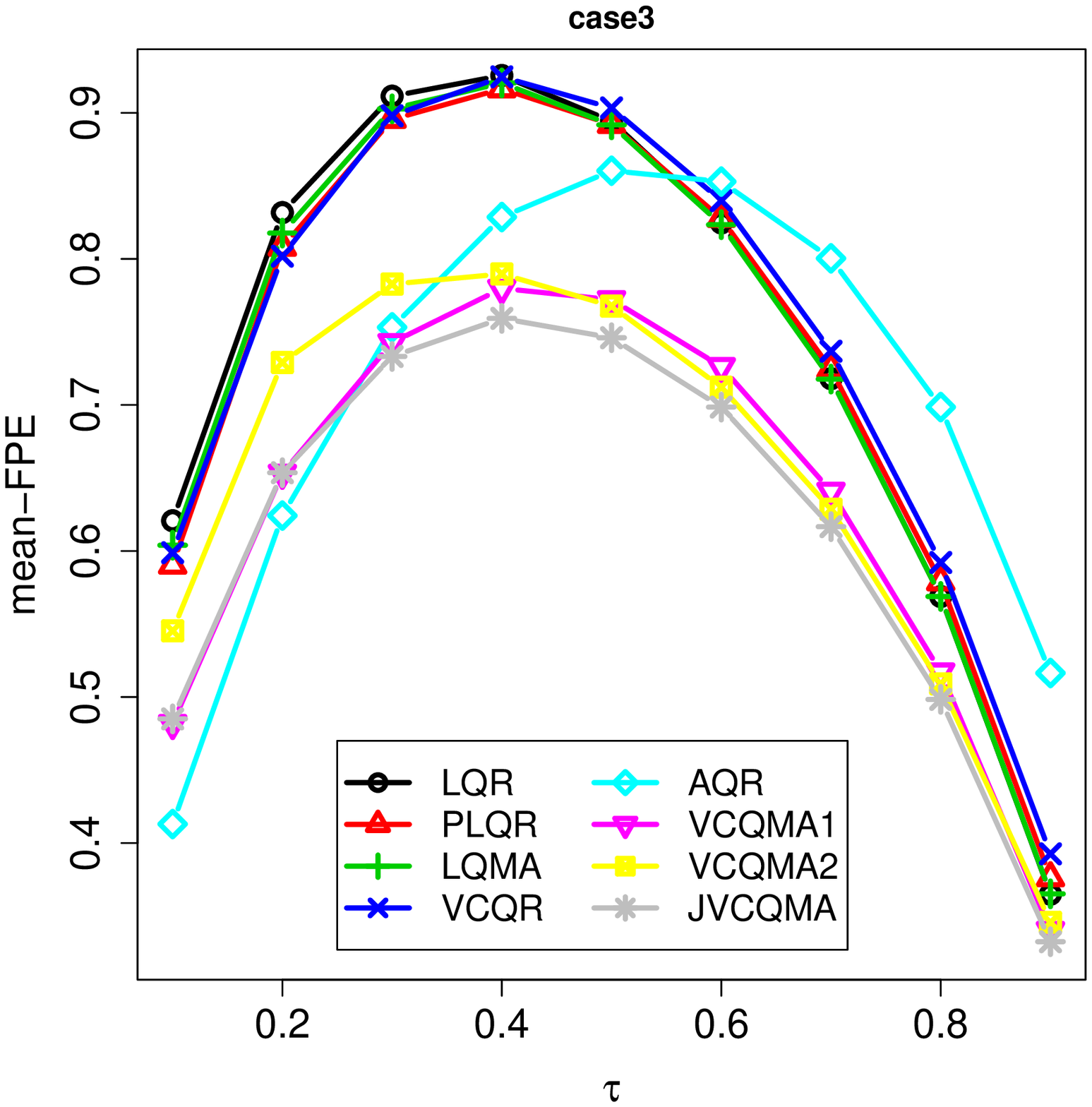}
\includegraphics[scale=0.4]{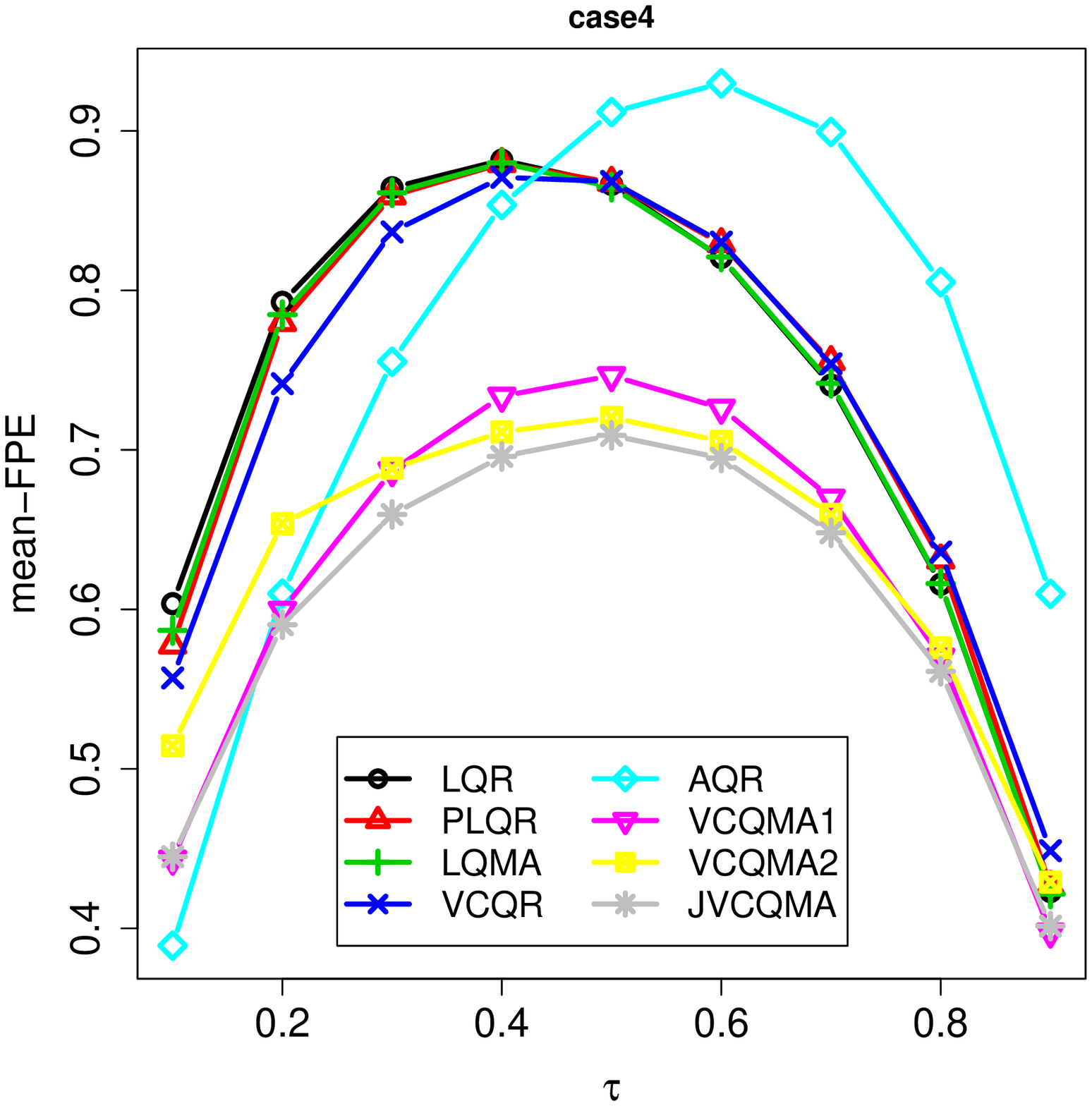}
\includegraphics[scale=0.4]{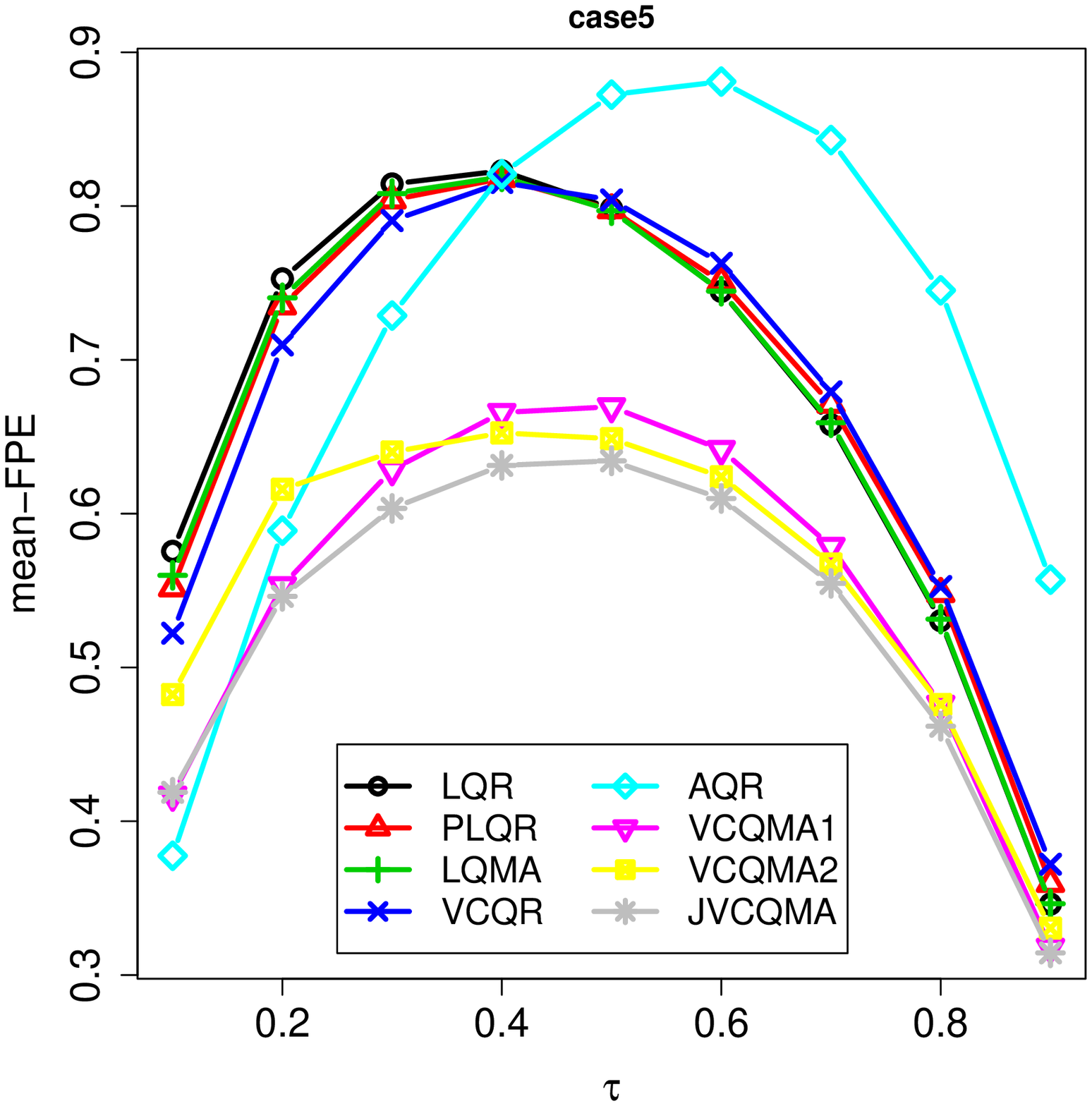}
\includegraphics[scale=0.4]{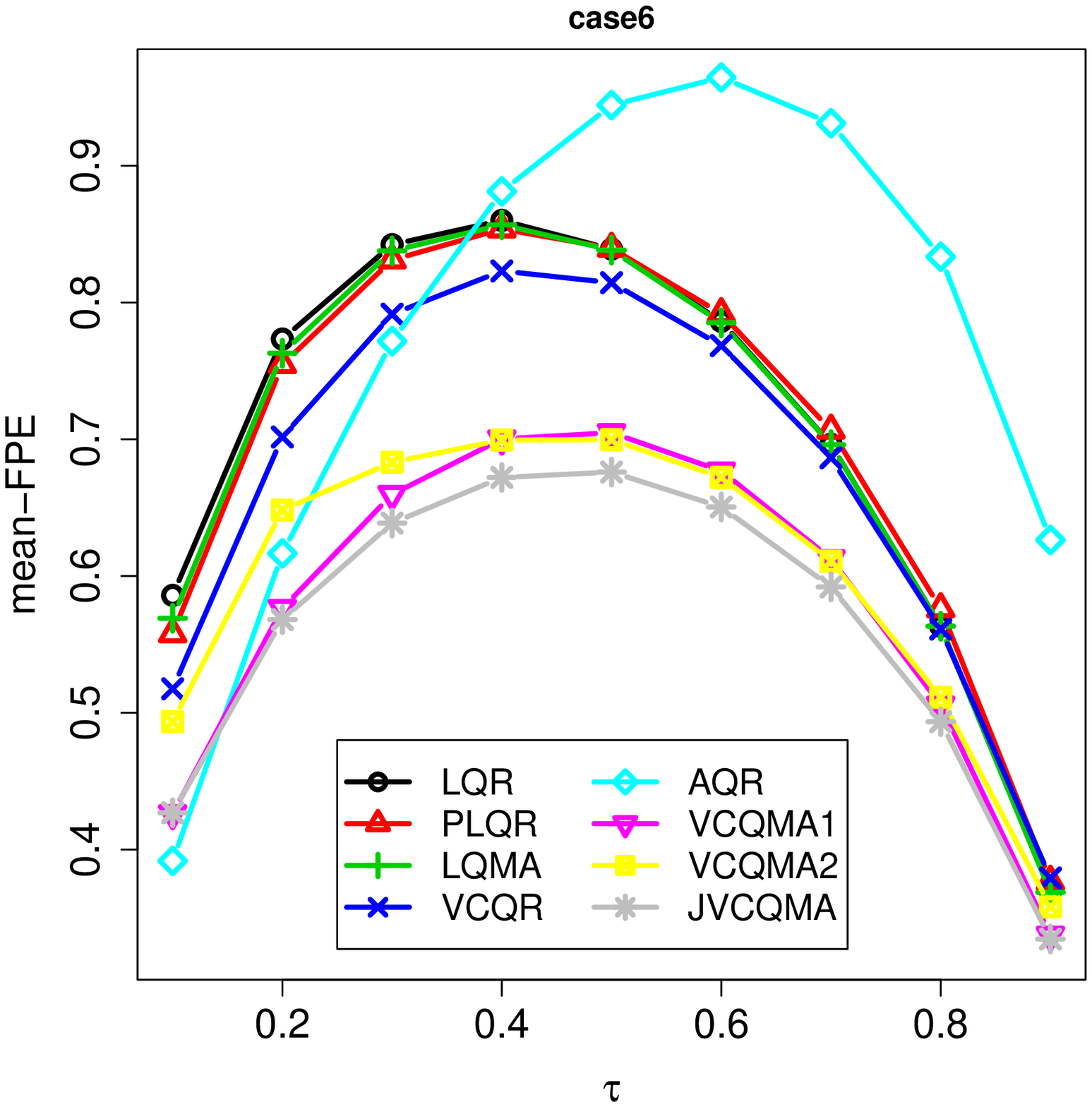}
\caption{Mean-FPEs at different quantiles with $\tau$ ranging from 0.1 to 0.9
for example 4.}
\label{figure4}
\end{figure}

For each setting in the $4$ simulated examples, we repeat the simulation 200
times and report, in Figures \ref{figure1} to \ref{figure4}, the
mean of the obtained $200$ FPEs for each method under comparison for
different quantiles with $\tau$ ranging from 0.1 to 0.9 with increment 0.1.
Obviously, the smaller the mean of the FPE, the better the method in terms of
the out-of-sample quantile prediction error.

We also report, in Tables \ref{table1} to \ref{table4}, the means and standard
deviations of the estimated model weights, obtained by the proposed method,
over the 200 simulations for different settings and different quantiles.
$$
[\mbox{Tables \ 1 \ to \ 4\ about\ here}.]
$$

We first examine the performance of the estimators of the weights in the
proposed {\sf JVCQMA}.  For Example 1, Table \ref{table1} shows the first two
sub-models carry almost all the weights, and the combination of the two models
is indeed the true model, which indicates the proposed cross-validation based
method works very well for selection of the weights in the model averaging
prediction.  This implies the proposed {\sf JVCQMA} has the best prediction
ability by optimally combining the most suitable candidate models.
Furthermore, we can also see from Tables \ref{table1}--\ref{table2} that
different quantiles may lead to different estimators of model weights, and
they are becoming more stable when sample size is increasing.
$$
[\mbox{Figures \ 1 \ to \ 4\ about\ here}.]
$$
After having examined the performance of the estimated model weights, we now
evaluate the prediction accuracy.  The results about the mean FPEs for each
method under consideration for different sample sizes, sparsity levels and error
distributions are presented in Figures \ref{figure1}--\ref{figure4}, which
clearly show the proposed {\sf JVCQMA} has the best out-of-sample prediction
performance in the sense that it has the smallest mean FPE, except for the lower
quartile in example 4, compared to its alternatives.  {\sf LQR},{\sf PLQR} and
{\sf LQMA} apply linear models for the prediction, which completely ignores the
nonlinear functional relationship between the response and covariates,
therefore, results in very poor predictions.  {\sf AQR} and {\sf VCQR} employ
misspecified model structures for predictions, hence, also perform poorly.

Although {\sf JVCQMA} may also use misspecified candidate models, it combines
useful information from different candidate models in a sensible way, thus
leads to more accurate predictions.  It is clear the proposed model averaging
prediction is more flexible in terms of model specification and has clear
advantages over the other methods considered in the simulations.  Furthermore,
it is also worth mentioning that {\sf JVCQMA} clearly outperforms {\sf VCQMA1}
and {\sf VCQMA2}, suggesting the weights in model averaging play a key role
and the proposed cross-validation based selection of weights works very well,
which is in line with Theorem \ref{th.2} in Section \ref{sect.3}.

\bigskip

\noindent
{\bf Remark:} {\it When applying {\sf VCQR} to the $4$ simulated examples, we
only report the results for the cases where the index variable is set to be
$X_{i1}$.  When the index variable is set to be any other continuous variable,
the performance of {\sf VCQR} is also poor.  To save space, we don't report
those results.  When applying {\sf AQR}, the variables for nonparametric
additive part are set to be all continuous variables, and the variables for
linear part are set to be all discrete variables.}

\section{Real data analysis}
\label{sect.5}

In this Section, we are going to analyse the Boston housing data by using
each method mentioned in simulation study and the commonly used varying
coefficient model based method.  Specifically, we are going to examine the
prediction power of each method for different parts of the housing market in
Boston, that is to predict different quantiles of the house price there by each method and assess the accuracy of the resulting prediction.

The Boston housing data is freely available in the R package {\tt mlbench},
(http://cran.r-project.org/).  It has been analysed in much literature, see
\cite{fh2005}, \cite{hx2012} and \cite{syzl2014}.  The dataset consists of the
median value of owner price in 1970 of owner-occupied houses in 506 census
tracts within the Boston metropolitan area, together with several variables
which are commonly believed to be associated with housing values, detailed in
Section \ref{mot0}.  Like \cite{fh2005} and \cite{hx2012}, we take
MEDV (median value of owner-occupied homes in 1,000 United States dollar (USD))
as the response, denoted by $Y$, and the variables CRIM, ZN, INDUS, CHAS, NOX,
RM, AGE, DIS, RAD, TAX, PTRATIO, B and LSTAT as covariates.  Among the
covariates, CRIM, INDUS, NOX, RM, AGE, DIS, TAX, PTRATIO, B and LSTAT are
continuous variables, we denote them by $X_1, \ X_2, \ \cdots, \ X_{10}$,
respectively.  ZN, CHAS and RAD are discrete variables, we denote them by
$X_{11}$, $X_{12}$, and $X_{13}$, respectively.  Each continuous
covariate is standardized such that it has mean zero and variance $1$, before
any analysis is carried out.

This dataset has $10$ continuous covariates, i.e. $X_i$, $i=1, \ \cdots, \ 10$,
if a varying coefficient model based approach is employed for the quantile
prediction, any of the $10$ continuous covariates could be the index variable
in the varying coefficient model used.  We denote the varying coefficient model
based approach with $X_i$ being the index variable by {\sf VCQR$_i$}.

\begin{figure}\center
\includegraphics[scale=0.25]{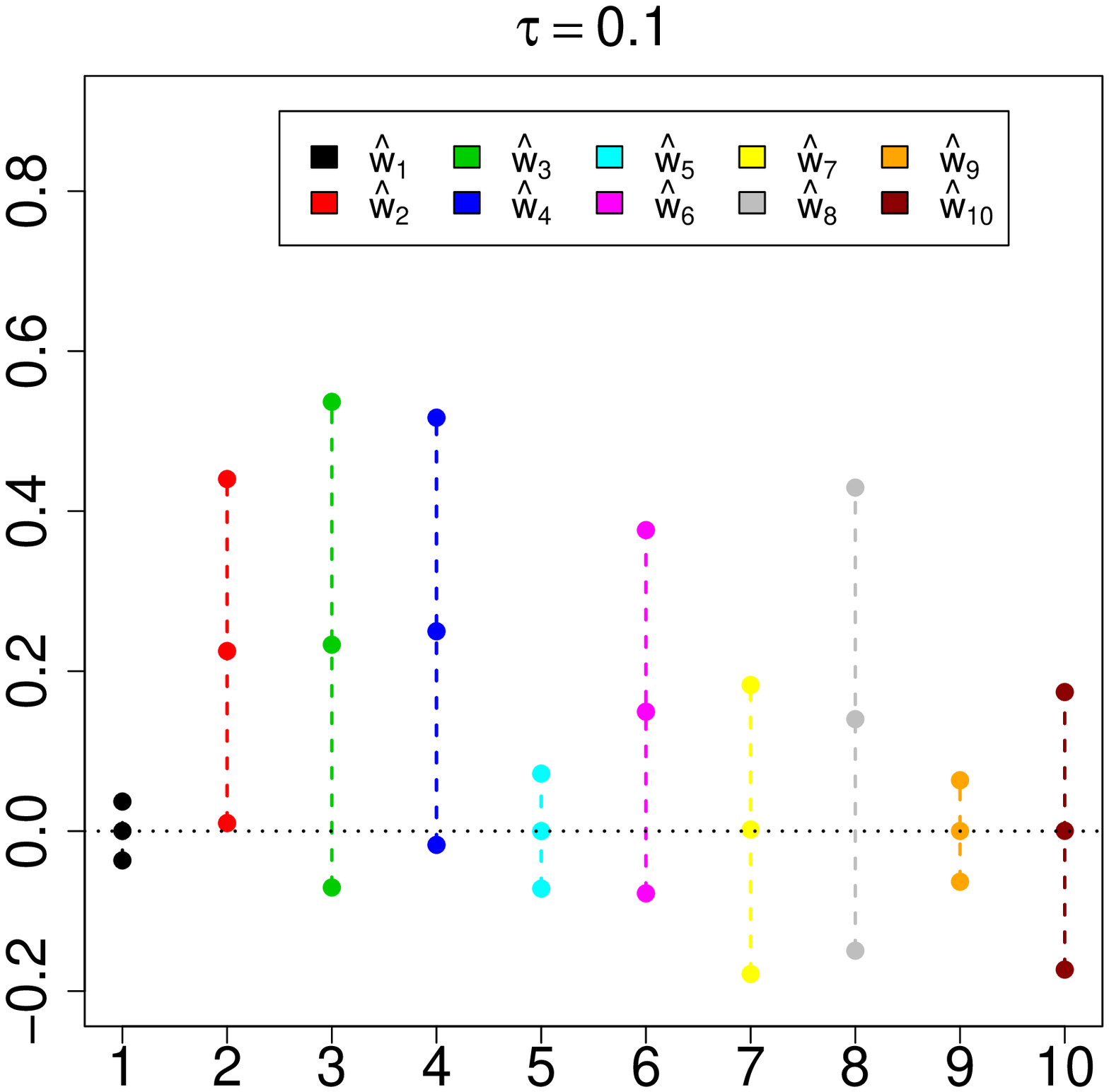}
\includegraphics[scale=0.25]{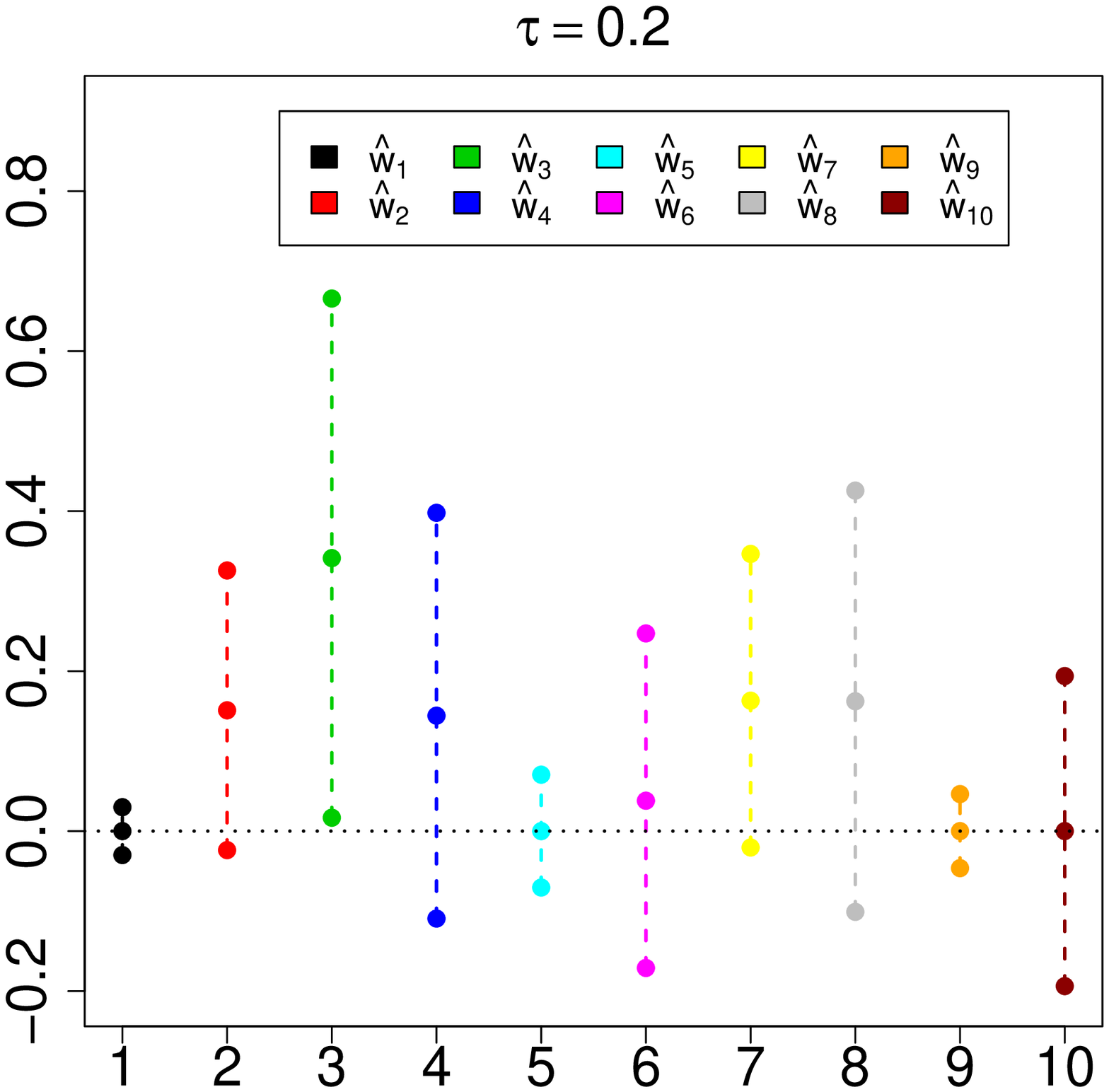}
\includegraphics[scale=0.25]{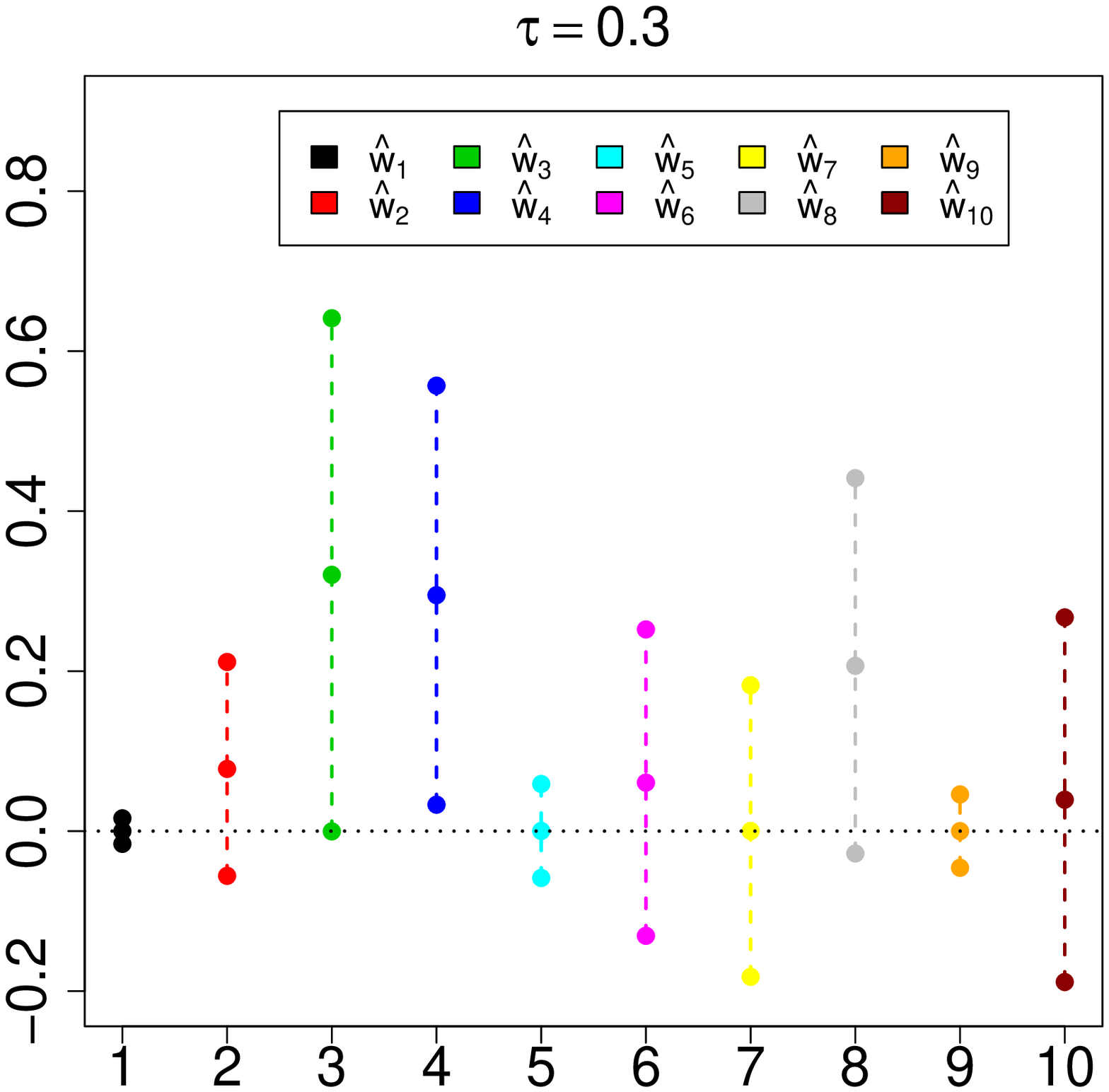}
\includegraphics[scale=0.25]{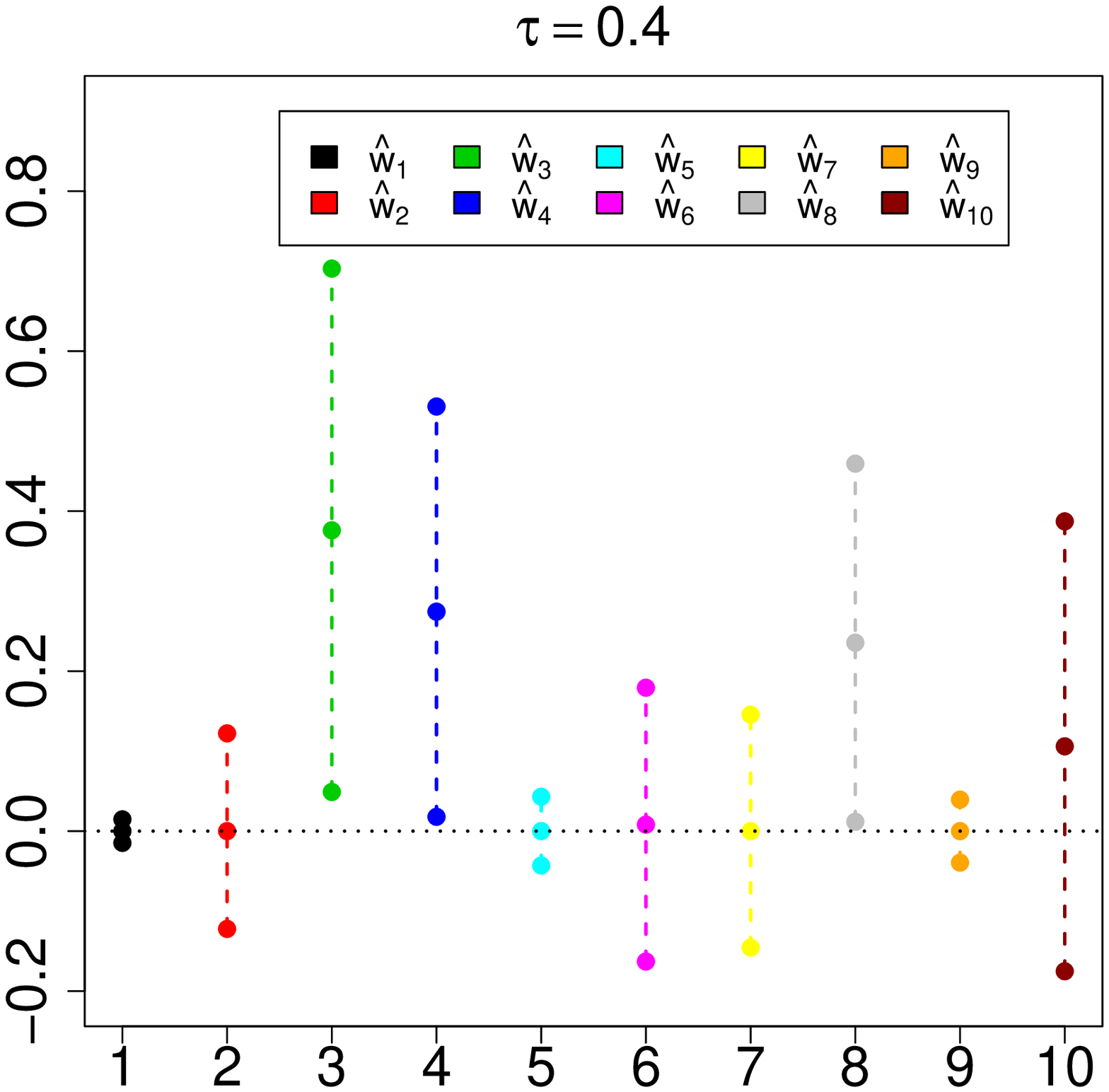}
\includegraphics[scale=0.25]{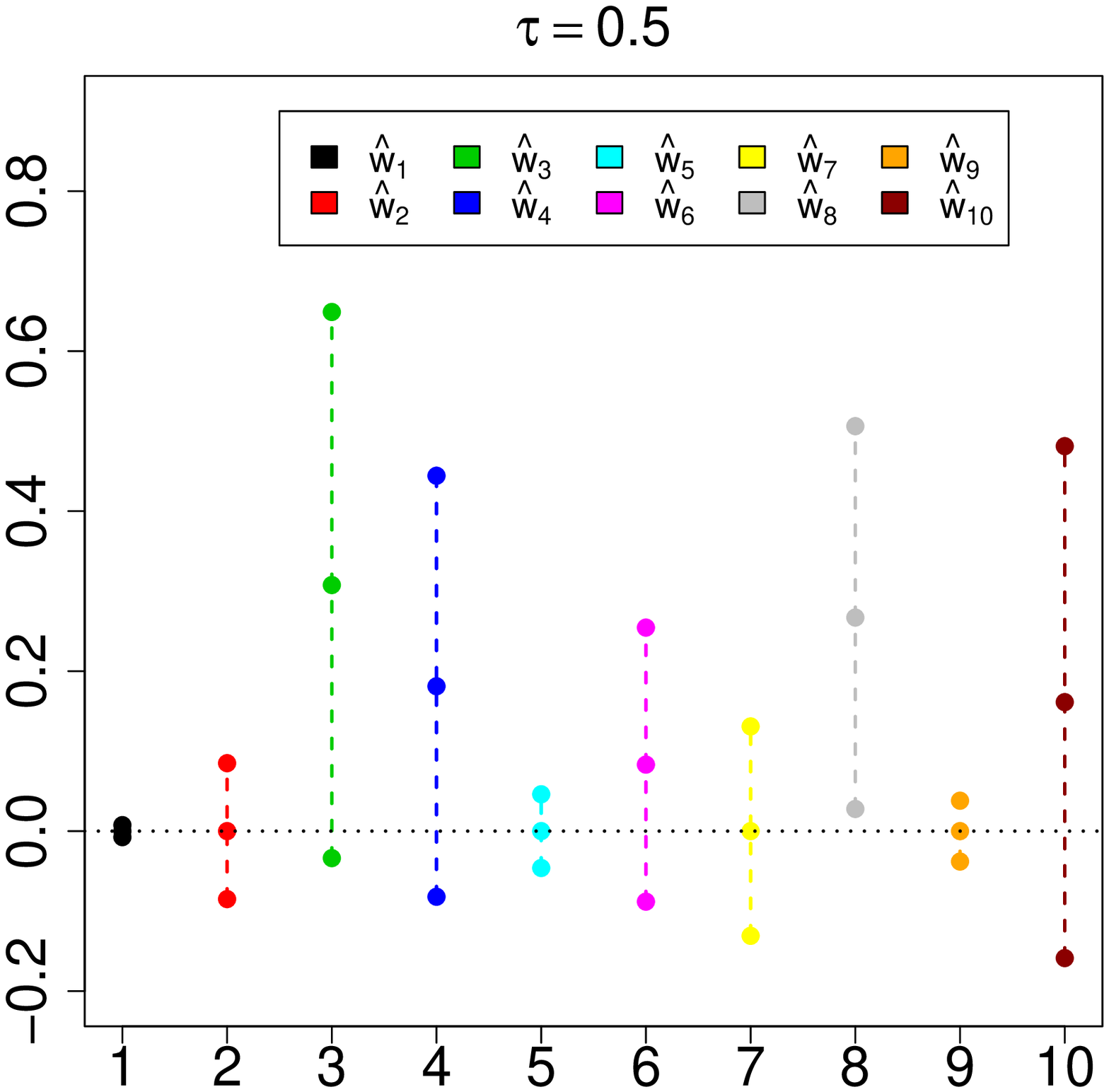}
\includegraphics[scale=0.25]{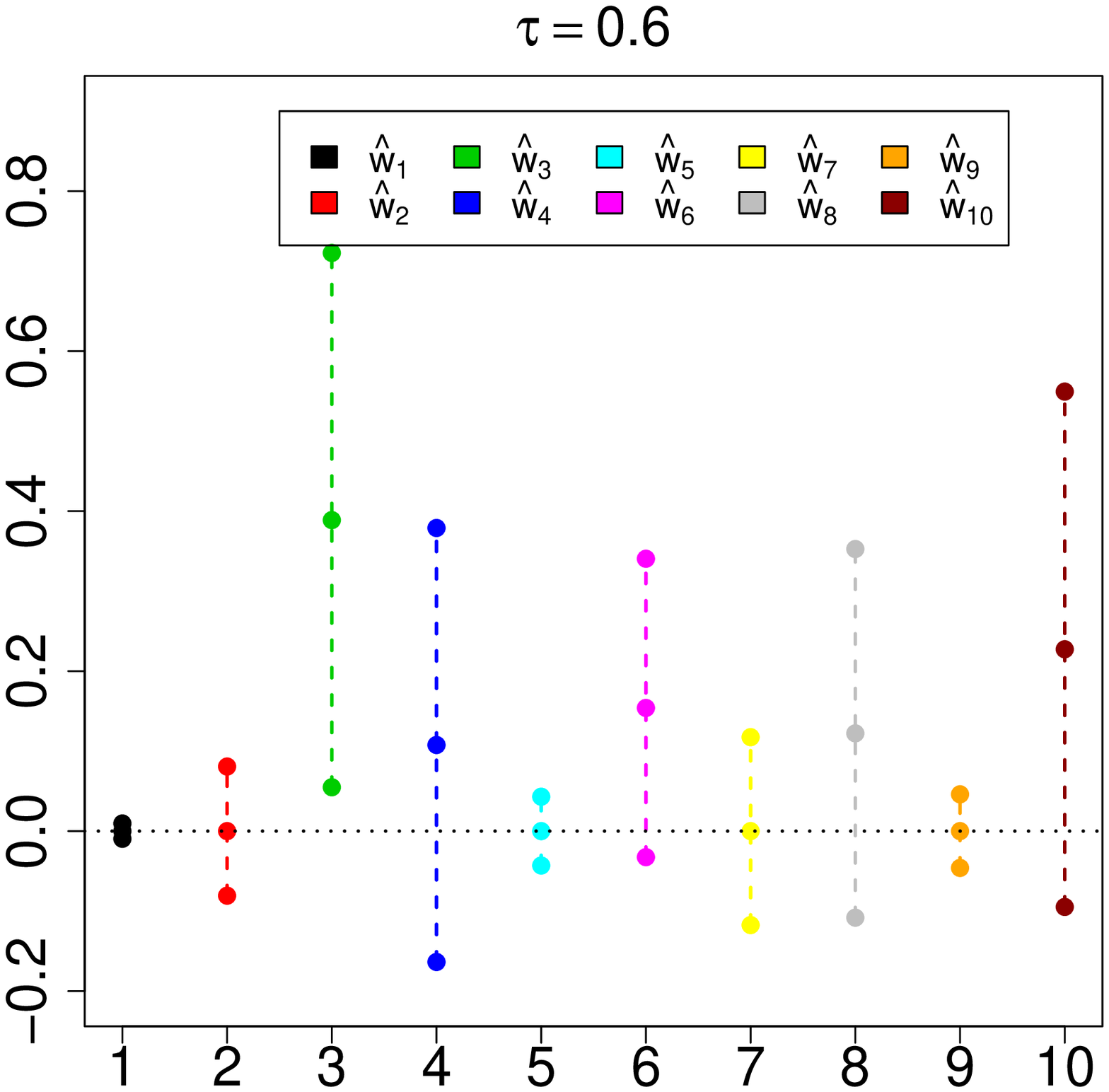}
\includegraphics[scale=0.25]{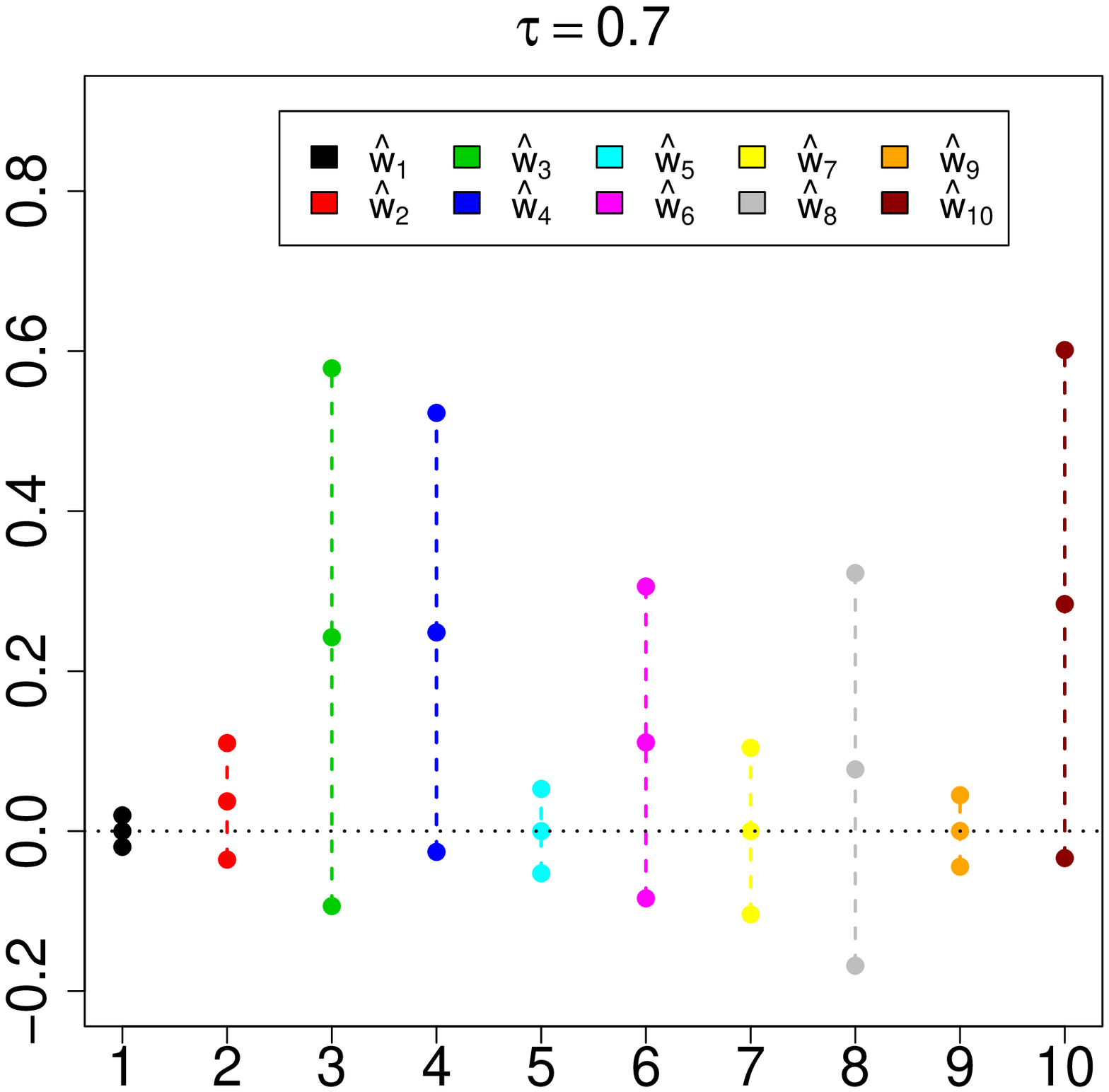}
\includegraphics[scale=0.25]{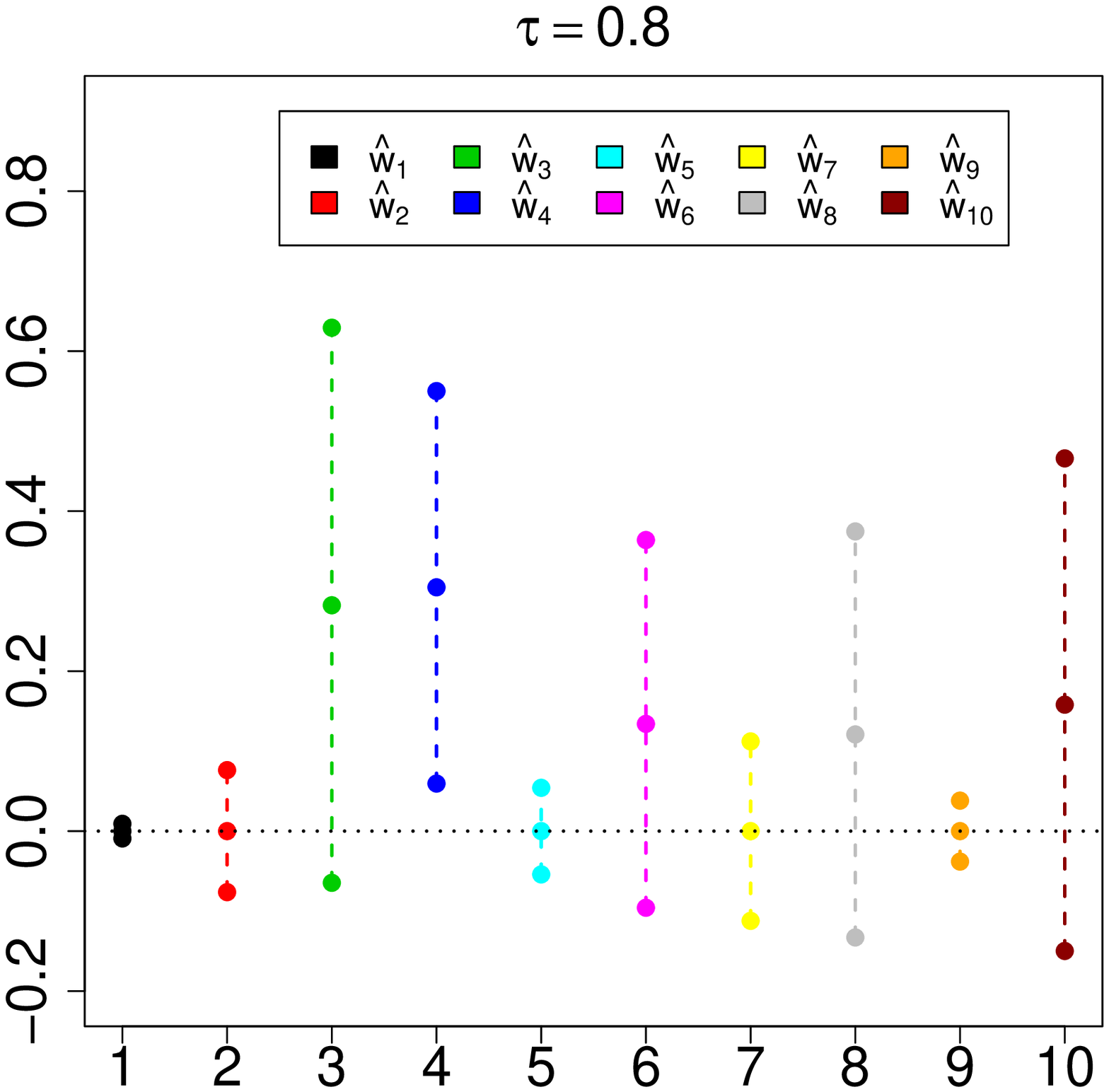}
\includegraphics[scale=0.25]{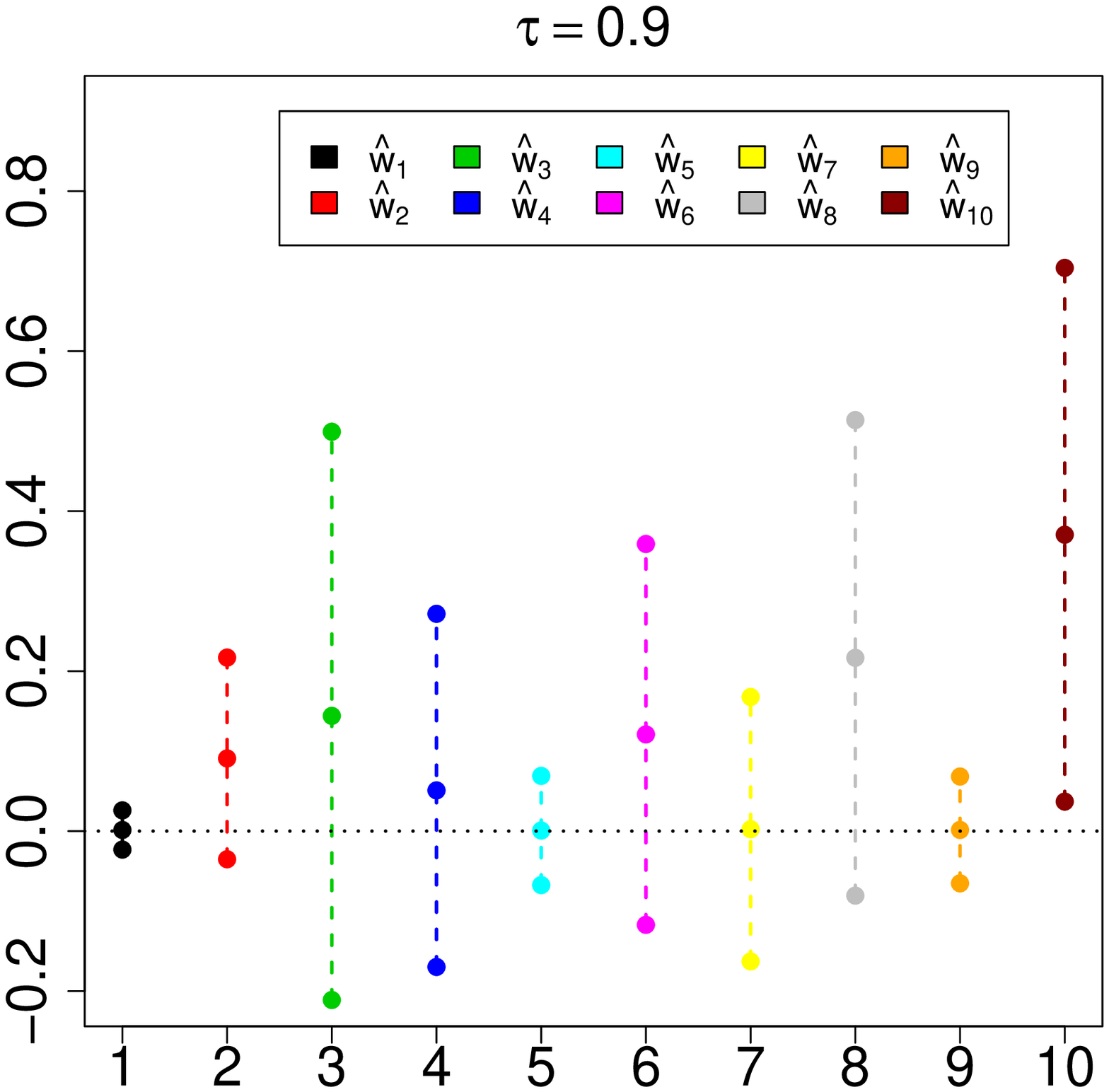}
\caption{Plots of the estimated model weights $\bm w=(w_1,\cdots, w_{10})$ and their 95\% confidence limits, where the standard errors are computed based on 500 bootstrap samples.}\label{figure5}
\end{figure}

When applying {\sf AQR} to this datset, the variables for nonparametric
additive part are set to be all continuous variables, i.e. $X_i$, $i=1, \
\cdots, \ 10$, and the variables for linear part are set to be all discrete
variables, i.e. $X_i$, $i=11, \ 12, \ 13$.
$$
[\mbox{Figure \ 5 \ about\ here}.]
$$
To have a visible idea about how the weights are assigned, by the proposed
cross-validation, to the $10$ varying coefficient models, formed by setting
each $X_i$, $i=1, \ \cdots, \ 10$, to be the index variable, involved in the
proposed {\sf JVCQMA}, we apply bootstrap to compute the standard errors of the
estimated weights, and present in Figure \ref{figure5} the $95\%$ confidence
intervals of the weights for different quantiles with $\tau$ ranging from 0.1
to 0.9 with increment 0.1.  Figure \ref{figure5} shows the allocation of the
weights is different for different quantiles.

To examine the prediction power of each method and make a comparison between
different methods, we randomly split the dataset to training set of size
$n_{train}$ and testing set of size $n_{test}$.  We apply each method under
comparison to the training set to form quantile predictions, and use the
testing set to compute the out-of-sample quantile prediction error of this
method.  We repeat the random splitting procedure $200$ times, and compute the
mean of the obtained $200$ FPEs for each method.
$$
[\mbox{Table \ 5 \ about\ here}.]
$$
We set the size $n_{test}$ of the test set to be either $50$, $100$ or $200$,
and compute the mean FPE for each method for different quantiles with $\tau$
ranging from 0.1 to 0.9 with increment 0.1, and present the results in Table
\ref{tablereal}.  Table \ref{tablereal} shows the proposed {\sf JVCQMA} gives
the most accurate prediction under any circumstance.

\section{Theoretical proofs}
\label{sec6}

 Lemma \ref{lemma:1}
below, which is a direct result of \cite{MS82}, will be used repeatedly.
\begin{lemma}\label{lemma:1}
Let $(X_1,Y_1),\cdots,(X_n,Y_n)$ be i.i.d. random vectors, where $Y_i$s are scalar random variables. Assume further that $E{\left| Y \right|^r}<\infty$ and that
${\sup _{x}}\int {{{\left| y \right|}^r}f\left( {x,y} \right)dy < \infty } $, where $f$ denotes the joint density of $(X,Y)$. Let $K$
be a bounded positive function with bounded support, satisfying a Lipschitz condition.
Then,
\[{\sup _{ x \in \mathcal{D}}}\left| {{n^{ - 1}}\sum\limits_{i = 1}^n {\left\{ {{K_{{h}}}\left( {{X_i} -x} \right){Y_i} - E\left[ {{K_{{h}}}\left( {{X_i} - x} \right){Y_i}} \right]} \right\}} } \right| = {O_p}\left( {{{\log }^{1/2}}\left( {1/{h}} \right)/\sqrt {n{h}} } \right),\]
provided that $0<h\rightarrow0$ and $n^{2\epsilon-1}h \rightarrow\infty $ for some $\epsilon<1-r^{-1}$, where $h$ is a bandwidth and $\mathcal{D}$ is some closed set.
\end{lemma}

Let $\varepsilon _{i,s}=Y_i-{\mu _{\tau,s}}(\bm X_i)$, and $f_s\left( { \cdot \left| {{\bm X_i}} \right.} \right)$ and $F_s\left( { \cdot \left| {{\bm X_i}} \right.} \right)$ be the conditional probability density function and cumulative distribution
function of $\varepsilon_{i,s}$ given $\bm X_i$, respectively. Furthermore, denote
 ${u_{i,s}}\left( {{x_s}} \right) = {\mu _{\tau,s}}(\bm X_i) - {\alpha _{\tau ,s}}\left( {{x_s}} \right) - {{\dot \alpha }_{\tau ,s}}\left( {{x_s}} \right)\left( {{X_{is}} - {x_s}} \right) -\bm X_{i \setminus s}^ \top {\bm\beta _{\tau ,s}}\left( {{x_s}} \right) - \bm X_{i \setminus s}^ \top {\bm{\dot \beta }_{\tau ,s}}\left( {{x_s}} \right)\left( {{X_{is}} - {x_s}} \right)$, ${ \textbf{X}_{i,s}}\left( {{x_s}} \right) = {\left\{ {1,\bm X_{i \setminus s}^ \top ,\left( {{X_{is}} - {x_s}} \right)/{h_s},\bm X_{i \setminus s}^ \top \left( {{X_{is}} - {x_s}} \right)/{h_s}} \right\}^ \top }$,
${\eta _{i,s}}\left( {{x_s}} \right) = I\left( {{\varepsilon _{i,s}} \le  - {u_{i,s}}\left( {{x_s}} \right)} \right) - \tau $ and $\bm{\hat \vartheta}  = \sqrt {n{h_s}} {\left\{ {\hat a - {\alpha _{\tau ,s}}\left( {{x_s}} \right),{{\left( {\bm{\hat c} - {\bm\beta _{\tau ,s}}\left( {{x_s}} \right)} \right)}^ \top },{h_s}\left( {\hat b - {{\dot \alpha }_{\tau ,s}}\left( {{x_s}} \right)} \right),{h_s}{{\left( {\bm{\hat d} - {\bm{\dot \beta }_{\tau ,s}}\left( {{x_s}} \right)} \right)}^ \top }} \right\}^ \top }$.

\begin{lemma}\label{lemma:2}
For $s=1,\cdots,p$, under conditions (C1)--(C4), if $h_s\rightarrow0$
and $nh_s\rightarrow\infty$, as $n\rightarrow\infty$, we have
\begin{align}
 &\sqrt {n{h_s}} \left( {{\bm{\hat \theta }_{\tau ,s}}\left( {{x_s}} \right) - {\bm\theta _{\tau ,s}}\left( {{x_s}} \right)} \right)   \nonumber\\
 &= - f_{{X_s}}^{ - 1}\left( {{x_s}} \right){\left( {{\bm D_s}\left( {{x_s}} \right)} \right)^{ - 1}}\bm W_{n,s}^{(1)}\left( {{x_s}} \right)+ {O_p}\left( {h_s^2 + {{\log }^{1/2}}\left( {1/{h_s}} \right)/\sqrt {n{h_s}} } \right), \nonumber
 \end{align}
where
\[{\bm D_s}\left( {{x_s}} \right) = E\left\{ {f_s\left( { - {u_{i,s}}\left| {{\bm X_i}} \right.} \right){{\left( {1,\bm X_{i \setminus s}^ \top } \right)}^ \top }\left( {1,\bm X_{i \setminus s}^ \top } \right)\left| {{X_{is}} = {x_s}} \right.} \right\},\]
and
\[\bm W_{n,s}^{(1)}\left( {{x_s}} \right) = \frac{1}{{\sqrt {n{h_s}} }}\sum\limits_{i = 1}^n {K\left\{ {\left( {{X_{is}} - {x_s}} \right)/{h_s}} \right\}{\eta _{i,s}}\left( {{x_s}} \right){{\left( {1,\bm X_{i \setminus s}^ \top } \right)}^ \top }} .\]

\end{lemma}

\noindent \textbf{Proof of Lemma \ref{lemma:2}.} Recall that $\left\{ {\hat a,{\bm{\hat c}^ \top },\hat b,{\bm{\hat d}^ \top }} \right\}^ \top $ minimizes
\begin{align}
\sum\limits_{i = 1}^n
{{\rho _\tau }\left\{
{{Y_i} - a - b
\left( {{X_{is}} - {x_s}} \right) - \bm X_{i \setminus s}^\top
\left(
{{\bm c} + {\bm d}\left( {{X_{is}} - {x_s}}
\right)}
\right)}
\right\}}
{K_{{h_s}}}\left( {{X_{is}} - {x_s}} \right).\nonumber
 \end{align}
We write ${Y_i} - a - b\left( {{X_{is}} - {x_s}} \right) -\bm X_{i \setminus s}^ \top \left( {\bm c +\bm d\left( {{X_{is}} - {x_s}} \right)} \right) = {\varepsilon _{i,s}} + {u_{i,s}}\left( {{x_s}} \right) - {\Delta _{i,s}}$, where ${\Delta _{i,s}} = { \textbf{X}}_{i,s}^ \top \left( {{x_s}} \right)\bm\vartheta /\sqrt {n{h_s}} $. Then, $\bm{\hat\vartheta}$ is also the minimizer of
\[{L_{n,s}}\left( \bm\vartheta  \right) = \sum\limits_{i = 1}^n {\left[ {{\rho _\tau }\left\{ {{\varepsilon _{i,s}} + {u_{i,s}}\left( {{x_s}} \right) - {\Delta _{i,s}}} \right\} - {\rho _\tau }\left\{ {{\varepsilon _{i,s}} + {u_{i,s}}\left( {{x_s}} \right)} \right\}} \right]} {K_i}\left( {{x_s}} \right),\]
where ${K_i}\left( {{x_s}} \right) = K\left\{ {\left( {{X_{is}} - {x_s}} \right)/{h_s}} \right\}$. By applying the identity in \cite{K98}
\begin{gather}
{\rho _\tau }\left( {u - v} \right) - {\rho _\tau }\left( u \right) = v \left\{I\left( {u \le 0} \right)-\tau\right\}+ \int_0^{  v} {\left\{ {I\left( {u \le t} \right) - I\left( {u \le 0} \right)} \right\}dt},
  \tag{A.1}\label{A.1}
\end{gather}
we have
\begin{align}
 {L_{n,s}}\left( \bm \vartheta  \right) = &\sum\limits_{i = 1}^n {{K_i}\left( {{x_s}} \right){\Delta _{i,s}}\left[ {I\left\{ {{\varepsilon _{i,s}} \le  - {u_{i,s}}\left( {{x_s}} \right)} \right\} - \tau } \right]}  \nonumber\\
  &+ \sum\limits_{i = 1}^n {{K_i}\left( {{x_s}} \right)\int_0^{{\Delta _{i,s}}} {\left[ {I\left\{ {{\varepsilon _{i,s}} \le  - {u_{i,s}}\left( {{x_s}} \right) + t} \right\} - I\left\{ {{\varepsilon _{i,s}} \le  - {u_{i,s}}\left( {{x_s}} \right)} \right\}} \right]dt} } \nonumber \\
  \buildrel \Delta \over = &{\left[ {{\bm W_{n,s}}\left( {{x_s}} \right)} \right]^ \top }\bm\vartheta  +  {{B_{n,s}}\left( \bm\vartheta  \right)},  \nonumber
 \end{align}
where
\[{\bm W_{n,s}}\left( {{x_s}} \right) = \frac{1}{{\sqrt {n{h_s}} }}\sum\limits_{i = 1}^n {{K_i}\left( {{x_s}} \right)\left[ {I\left\{ {{\varepsilon _{i,s}} \le  - {u_{i,s}}\left( {{x_s}} \right)} \right\} - \tau } \right]{\textbf{X}_{i,s}}\left( {{x_s}} \right)}, \]
and
\[{B_{n,s}}\left( \bm\vartheta  \right) = \sum\limits_{i = 1}^n {{K_i}\left( {{x_s}} \right)\int_0^{{\Delta _{i,s}}} {\left[ {I\left\{ {{\varepsilon _{i,s}} \le  - {u_{i,s}}\left( {{x_s}} \right) + t} \right\} - I\left\{ {{\varepsilon _{i,s}} \le  - {u_{i,s}}\left( {{x_s}} \right)} \right\}} \right]dt} } .\]
Since ${B_{n,s}}\left( \bm\vartheta  \right)$ is a summation of i.i.d. random variables of the kernel form, it
follows, by Lemma \ref{lemma:1}, that
\[{B_{n,s}}\left( \bm\vartheta  \right) = E\left[ {{B_{n,s}}\left( \bm\vartheta  \right)} \right] + {O_p}\left( {{{\log }^{1/2}}\left( {1/{h_s}} \right)/\sqrt {n{h_s}} } \right).\]
Denote by $\mathcal{X}$ the observed covariates vector, namely $\mathcal{X}=(X_{11}, \cdots,X_{1(p+q)},\cdots,X_{n1},\cdots,X_{n(p+q)})^\top$. The conditional expectation of ${B_{n,s}}\left( \bm\vartheta  \right)$ can be calculated as
\begin{align}
 &E\left[ {{B_{n,s}}\left( \bm\vartheta  \right)\left| \mathcal{X} \right.} \right] \nonumber\\
 = &\sum\limits_{i = 1}^n {{K_i}\left( {{x_s}} \right)\int_0^{{\Delta _{i,s}}} {\left[ {F_s\left\{ { - {u_{i,s}}\left( {{x_s}} \right) + t\left| {{\bm X_i}} \right.} \right\} - F_s\left\{ { - {u_{i,s}}\left( {{x_s}} \right)\left| {{\bm X_i}} \right.} \right\}} \right]dt} }  \nonumber\\
  = &\frac{1}{2}{\bm\vartheta ^ \top }\left[ {\frac{1}{{n{h_s}}}\sum\limits_{i = 1}^n {{K_i}\left( {{x_s}} \right)f_s\left\{ { - {u_{i,s}}\left( {{x_s}} \right)\left| {{\bm X_i}} \right.} \right\}{ \textbf{X}_{i,s}}\left( {{x_s}} \right)\textbf{X}_{i,s}^ \top \left( {{x_s}} \right)} } \right]\bm\vartheta  \left\{1+o_p\left( 1\right)\right\}\nonumber\\
  \buildrel \Delta \over = &\frac{1}{2}{\bm\vartheta ^ \top }{\bm D_{n,s}}\left( {{x_s}} \right)\bm\vartheta \left\{1+o_p\left( 1\right)\right\} .\nonumber
 \end{align}
Then,
\begin{align}
 {L_{n,s}}\left( \bm\vartheta  \right) = &{\left[ {{\bm W_{n,s}}\left( {{x_s}} \right)} \right]^ \top }\bm\vartheta  + E\left[ {{B_{n,s}}\left(\bm \vartheta  \right)} \right] + {O_p}\left( {{{\log }^{1/2}}\left( {1/{h_s}} \right)/\sqrt {n{h_s}} } \right) \nonumber\\
  =& {\left[ {{\bm W_{n,s}}\left( {{x_s}} \right)} \right]^ \top }\bm\vartheta  + E\left\{ {E\left[ {{B_{n,s}}\left( \bm\vartheta  \right)\left| \mathcal{X} \right.} \right]} \right\} + {O_p}\left( {{{\log }^{1/2}}\left( {1/{h_s}} \right)/\sqrt {n{h_s}} } \right) \nonumber\\
  = &{\left[ {{\bm W_{n,s}}\left( {{x_s}} \right)} \right]^ \top }\bm\vartheta  + \frac{1}{2}{\bm\vartheta ^ \top }E\left[ {{\bm D_{n,s}}\left( {{x_s}} \right)} \right]\bm\vartheta  + {O_p}\left( {{{\log }^{1/2}}\left( {1/{h_s}} \right)/\sqrt {n{h_s}} } \right) .\nonumber
 \end{align}
It can be shown that
\[E\left[ {{\bm D_{n,s}}\left( {{x_s}} \right)} \right] = {f_{{X_s}}}\left( {{x_s}} \right)\bm D_s^*\left( {{x_s}} \right) + {O_p}\left( {h_s^2} \right), \]
where $\bm D_s^*\left( {{x_s}} \right)=diag\left\{ {{\bm D_s}\left( {{x_s}} \right),{\mu _2}{\bm D_s}\left( {{x_s}} \right)} \right\}$.

Therefore, we can write $ {L_{n,s}}\left(\bm \vartheta  \right)$ as
\begin{align}
 {L_{n,s}}\left(\bm \vartheta  \right) =& {\left[ {{\bm W_{n,s}}\left( {{x_s}} \right)} \right]^ \top }\bm\vartheta  + \frac{{{f_{{X_s}}}\left( {{x_s}} \right)}}{2}{\bm\vartheta ^ \top }\bm D_s^*\left( {{x_s}} \right)\bm\vartheta  + {O_p}\left( {h_s^2 + {{\log }^{1/2}}\left( {1/{h_s}} \right)/\sqrt {n{h_s}} } \right). \nonumber
 \end{align}
By applying the convexity lemma of \cite{P91} and the quadratic approximation lemma of \cite{FG96},
the minimizer of $ {L_{n,s}}\left( \bm\vartheta  \right) $ can be expressed as
\[{\bm{\hat \vartheta}  =  - f_{{X_s}}^{ - 1}\left( {{x_s}} \right){{\left( {\bm D_s^*\left( {{x_s}} \right)} \right)}^{ - 1}}{\bm W_{n,s}}\left( {{x_s}} \right) + {O_p}\left( {h_s^2 + {{\log }^{1/2}}\left( {1/{h_s}} \right)/\sqrt {n{h_s}} } \right)},\]
which holds uniformly for $x_s\in \mathcal{C}_s$. Note that $\bm D_s^*\left( {{x_s}} \right)=diag\left\{ {{\bm D_s}\left( {{x_s}} \right),{\mu _2}{\bm D_s}\left( {{x_s}} \right)} \right\}$ is a quasi-diagonal matrix. So,
\begin{align}
& \sqrt {n{h_s}} \left( {{\bm{\hat \theta }_{\tau ,s}}\left( {{x_s}} \right) - {\bm\theta _{\tau ,s}}\left( {{x_s}} \right)} \right)  \nonumber\\
 & = - f_{{X_s}}^{ - 1}\left( {{x_s}} \right){\left( {{\bm D_s}\left( {{x_s}} \right)} \right)^{ - 1}}\bm W_{n,s}^{(1)}\left( {{x_s}} \right)+ {O_p}\left( {h_s^2 + {{\log }^{1/2}}\left( {1/{h_s}} \right)/\sqrt {n{h_s}} } \right),\nonumber
 \end{align}
uniformly for $x_s\in \mathcal{C}_s$. We complete the proof of Lemma \ref{lemma:2}. $\square$\\

\noindent \textbf{Proof of Theorem \ref{th.1}.} We only prove (i) as the proof of (ii) is analogous.
By Lemma \ref{lemma:2}, we have
\begin{align}
 &{\bm{\hat \theta }_{\tau ,s}}\left( {{x_s}} \right) - {\bm\theta _{\tau ,s}}\left( {{x_s}} \right) \nonumber\\
 =&  - f_{{X_s}}^{ - 1}\left( {{x_s}} \right){\left( {{\bm D_s}\left( {{x_s}} \right)} \right)^{ - 1}}\frac{1}{n}\sum\limits_{i = 1}^n {{K_{{h_s}}}\left( {{X_{is}} - {x_s}} \right){\eta _{i,s}}\left( {{x_s}} \right){{\left( {1,\bm X_{i \setminus s}^ \top } \right)}^ \top }}  \nonumber\\
  &+ {O_p}\left( {n^{-1/2}h_s^{3/2} + {{\log }^{1/2}}\left( {1/{h_s}} \right)/ {n{h_s}} } \right),
    \tag{A.2}\label{A.2}
 \end{align}
uniformly for $x_s\in \mathcal{C}_s$. By lemma \ref{lemma:1}, it can be shown that
\begin{align}
\frac{1}{n}\sum\limits_{i = 1}^n {{K_{{h_s}}}\left( {{X_{is}} - {x_s}} \right){\eta _{i,s}}\left( {{x_s}} \right){{\left( {1,\bm X_{i \setminus s}^ \top } \right)}^ \top }}  = {O_p}\left( {h_s^2 + {{\log }^{1/2}}\left( {1/{h_s}} \right)/\sqrt {n{h_s}} } \right),
  \tag{A.3}\label{A.3}
 \end{align}
uniformly for $x_s\in \mathcal{C}_s$. By \eqref{A.2} and \eqref{A.3}, we have
\begin{align}
\mathop {\max }\limits_{1 \le s \le p} \mathop {\sup }\limits_{x_s \in \mathcal{C}_s} \left\|{\bm{\hat \theta }_{\tau ,s}}\left( {{x_s}} \right)-{\bm{ \theta }_{\tau ,s}}\left( {{x_s}} \right) \right\| ={O_p}\left({\log }^{1/2}\left( {1/\underline{h}} \right)/\sqrt {n\underline{h}} +\bar{h}^2\right).\nonumber
 \end{align}
We complete the proof of Theorem 1 (i). $\square$\\

\noindent \textbf{Proof of Theorem 2.}
Following the proof of Theorem 3.3 in \cite{LS15}, if we can show
that the difference ${\rm{CV}}_n\left( w \right)-{\rm{FPE}}_n\left( \bm w \right)$ is negligible compared with ${\rm{FPE}}_n\left( \bm w \right)$ uniformly for any $\bm w\in \mathbb{W}$, then the optimality property is established for $\bm{\hat w}$. More precisely, it suffices to show that
\begin{align}
\mathop {\sup }\limits_{\bm w\in \mathbb{W}} \left| {\frac{{{\rm{CV}}_n\left(\bm w \right)-{\rm{FPE}}_n\left(\bm w \right)}}{{{\rm{FPE}}_n\left(\bm w \right)}}} \right|=o_p(1).
 \tag{A.4}\label{A.4}
\end{align}
 By Knight's identity \eqref{A.1}, we have
\begin{align}
  &{{\rm{CV}}_n\left( \bm w \right)-{\rm{FPE}}_n\left(\bm w \right)} \nonumber\\
  = &\left\{ {\frac{1}{n}\sum\limits_{i = 1}^n {\left[ { {{\rho _\tau }} \left\{ {{Y_i} - \sum\limits_{s = 1}^p {{w_s}} \hat \mu _{\tau ,s}^{ \setminus i}({\bm X_i})} \right\}- {\rho _\tau }\left( {{\varepsilon _i}} \right)} \right]} } \right\} \nonumber\\
  &- \left\{ {{\rm{FPE}}_n\left(\bm w \right) - E\left[ {{\rho _\tau }\left( {{\varepsilon }} \right)} \right]} \right\} + \frac{1}{n}\sum\limits_{i = 1}^n {\left\{ {{\rho _\tau }\left( {{\varepsilon _i}} \right) - E\left[ {{\rho _\tau }\left( {{\varepsilon }} \right)} \right]} \right\}}  \nonumber\\
  = &\frac{1}{n}\sum\limits_{i = 1}^n {\left[ {{Q_\tau }\left( {{Y_i}|{\bm X_i}} \right) - \sum\limits_{s = 1}^p {{w_s}} \hat \mu _{\tau ,s}^{ \setminus i}\left( {{\bm X_i}} \right)} \right]{\psi _\tau }\left( {{\varepsilon _i}} \right)}  \nonumber\\
  &+ \frac{1}{n}\sum\limits_{i = 1}^n {\int_0^{\sum\limits_{s = 1}^p {{w_s}} \hat \mu _{\tau ,s}^{ \setminus i}\left( {{\bm X_i}} \right) - {Q_\tau }\left( {{Y_i}|{\bm X_i}} \right)} {\left[ {I\left( {{\varepsilon _i} \le t} \right) - I\left( {{\varepsilon _i} \le 0} \right)} \right]dt} }  \nonumber\\
  &- E\left[ {\int_0^{\sum\limits_{s = 1}^p {{w_s}} {{\hat \mu }_{\tau ,s}}\left( \bm X \right) - {Q_\tau }\left( {Y|\bm X} \right)} {\left[ {I\left( {\varepsilon  \le t} \right) - I\left( {\varepsilon  \le 0} \right)} \right]dt\left| {{{\cal D}_n}} \right.} } \right] \nonumber\\
  &+ \frac{1}{n}\sum\limits_{i = 1}^n {\left\{ {{\rho _\tau }\left( {{\varepsilon _i}} \right) - E\left[ {{\rho _\tau }\left( {{\varepsilon _i}} \right)} \right]} \right\}}  \nonumber\\
\triangleq &{\Lambda _{n1}}\left(\bm w \right) + {\Lambda _{n2}}\left( \bm w \right) + {\Lambda _{n3}}\left(\bm w \right) + {\Lambda _{n4}}\left(\bm w \right) + {\Lambda _{n5}},\nonumber
\end{align}
where ${\psi _\tau }\left( {{\varepsilon _i}} \right)=\tau-I\left( {\varepsilon _i \leq 0} \right)$,
\begin{align}
 {\Lambda _{n1}}\left( \bm w \right) = &\frac{1}{n}\sum\limits_{i = 1}^n {\left[ {{Q_\tau }\left( {{Y_i}|{\bm X_i}} \right) - \sum\limits_{s = 1}^p {{w_s}} \hat \mu _{\tau ,s}^{ \setminus i}\left( {{\bm X_i}} \right)} \right]{\psi _\tau }\left( {{\varepsilon _i}} \right)},\nonumber
   \end{align}
\begin{align}
 {\Lambda _{n2}}\left(\bm  w \right) = &\frac{1}{n}\sum\limits_{i = 1}^n \int_0^{\sum\limits_{s = 1}^p {{w_s}} \hat \mu _{\tau ,s}^{ \setminus i}\left( {{\bm X_i}} \right) - {Q_\tau }\left( {{Y_i}|{\bm X_i}} \right) }
 {\left\{ {\left[ {I\left( {{\varepsilon _i} \le t} \right) - I\left( {{\varepsilon _i} \le 0} \right)} \right]} \right.}\nonumber\\
 &- \left. {\left[ {F\left( {t\left| {{\bm X_i}} \right.} \right) - F\left( {0\left| {{\bm X_i}} \right.} \right)} \right]} \right\}dt\nonumber
  \end{align}
  \begin{align}
 {\Lambda _{n3}}\left(\bm  w \right) = &\frac{1}{n}\sum\limits_{i = 1}^n {\left\{ {\int_0^{\sum\limits_{s = 1}^p {{w_s}} \hat \mu _{\tau ,s}^{ \setminus i}\left( {{\bm X_i}} \right) - {Q_\tau }\left( {{Y_i}|{\bm X_i}} \right)} {\left[ {F\left( {t\left| {{\bm X_i}} \right.} \right) - F\left( {0\left| {{\bm X_i}} \right.} \right)} \right]dt} } \right.} , \nonumber\\
 &\left. { - {E_{{\bm X_i}}}\left[ {\int_0^{\sum\limits_{s = 1}^p {{w_s}} \hat \mu _{\tau ,s}^{ \setminus i}\left( {{\bm X_i}} \right) - {Q_\tau }\left( {{Y_i}|{\bm X_i}} \right) } {\left[ {F\left( {t\left| {{\bm X_i}} \right.} \right) - F\left( {0\left| {{\bm X_i}} \right.} \right)} \right]dt} } \right]} \right\} , \nonumber
 \end{align}
\begin{align}
 {\Lambda _{n4}}\left(\bm  w \right) = &\frac{1}{n}\sum\limits_{i = 1}^n {\left\{ {{E_{{\bm X_i}}}\left[ {\int_0^{\sum\limits_{s = 1}^p {{w_s}} \hat \mu _{\tau ,s}^{ \setminus i}\left( {{\bm X_i}} \right) - {Q_\tau }\left( {{Y_i}|{\bm X_i}} \right)} {\left[ {F\left( {t\left| {{\bm X_i}} \right.} \right) - F\left( {0\left| {{\bm X_i}} \right.} \right)} \right]dt} } \right]} \right.}  \nonumber\\
  &- {E_{{\bm X_i}}}\left[ {\int_0^{\sum\limits_{s = 1}^p {{w_s}} \hat \mu _{\tau ,s}\left( {{\bm X_i}} \right) - {Q_\tau }\left( {{Y_i}|{\bm X_i}} \right)} {\left[ {F\left( {t\left| {{\bm X_i}} \right.} \right) - F\left( {0\left| {{\bm X_i}} \right.} \right)} \right]dt} } \right] ,\nonumber
   \end{align}
   \begin{align}
 {\Lambda _{n5}} =& \frac{1}{n}\sum\limits_{i = 1}^n {\left\{ {{\rho _\tau }\left( {{\varepsilon _i}} \right) - E\left[ {{\rho _\tau }\left( {{\varepsilon _i}} \right)} \right]} \right\}} . \nonumber
 \end{align}
Hence
\begin{align}
&\mathop {\sup }\limits_{\bm w\in \mathbb{W}} \left| {\frac{{{\rm{CV}}_n\left(\bm w \right)-{\rm{FPE}}_n\left(\bm w \right)}}{{{\rm{FPE}}_n\left(\bm w \right)}}} \right|\nonumber\\
=&\mathop {\sup }\limits_{\bm w\in \mathbb{W}} \left| {\frac{ {\Lambda _{n1}}\left( \bm w \right)+ {\Lambda _{n2}}\left( \bm w \right)+ {\Lambda _{n3}}\left( \bm w \right)+ {\Lambda _{n4}}\left( \bm w \right)+ {\Lambda _{n5}}    }{{{\rm{FPE}}_n\left(\bm w \right)}}} \right|\nonumber\\
\leq &\frac{\mathop {\sup }\limits_{\bm w\in \mathbb{W}} \left| {\Lambda _{n1}}\left( \bm w \right)+ {\Lambda _{n2}}\left( \bm w \right)+ {\Lambda _{n3}}\left( \bm w \right)+ {\Lambda _{n4}}\left( \bm w \right)+ {\Lambda _{n5}}  \right|  }   {\mathop {\min }\limits_{\bm w\in \mathbb{W}}\left| {\rm{FPE}}_n\left(\bm w \right)\right| }\nonumber\\
\leq &\frac{\mathop {\sup }\limits_{\bm w\in \mathbb{W}} \left| {\Lambda _{n1}}\left( \bm w \right) \right| +\mathop {\sup }\limits_{\bm w\in \mathbb{W}} \left| {\Lambda _{n2}}\left( \bm w \right) \right|+\mathop {\sup }\limits_{\bm w\in \mathbb{W}} \left| {\Lambda _{n3}}\left( \bm w \right) \right|+\mathop {\sup }\limits_{\bm w\in \mathbb{W}} \left| {\Lambda _{n4}}\left( \bm w \right) \right|+ \left| {\Lambda _{n5}} \right| }   {\mathop {\min }\limits_{\bm w\in \mathbb{W}}\left| {\rm{FPE}}_n\left(\bm w \right)\right| }.\nonumber
\end{align}
Therefore, to prove \eqref{A.4}, we will prove (i) $\mathop {\min }\limits_{\bm w\in \mathbb{W}} {\rm{FPE}}_n\left(\bm  w \right)\geq {E\left[ {{\rho _\tau }\left( {{\varepsilon }} \right)} \right]}-o_p(1)$; (ii) $\mathop {\sup }\limits_{\bm w\in \mathbb{W}} \left|{\Lambda _{n1}}\left(\bm  w \right) \right|=o_p(1)$; (iii) $\mathop {\sup }\limits_{\bm w\in \mathbb{W}} \left|{\Lambda _{n2}}\left(\bm  w \right)\right| =o_p(1)$; (iv) $\mathop {\sup }\limits_{\bm w\in \mathbb{W}} \left|{\Lambda _{n3}}\left(\bm  w \right)\right| =o_p(1)$; (v) $\mathop {\sup }\limits_{\bm w\in \mathbb{W}} \left|{\Lambda _{n4}}\left( \bm w \right)\right| =o_p(1)$ and (vi) $ \left|{\Lambda _{n5}}\right| =o_p(1)$. (vi) follows by the weak law of large numbers so we only show (i)--(v) below.

We first show (i). Let $u\left(\bm  w \right) = {Q_\tau }\left( {Y|\bm X} \right) - \sum\limits_{s = 1}^p {{w_s}} {\mu _{\tau ,s}}\left( \bm X \right)$, $u_i\left(\bm  w \right) = {Q_\tau }\left( {Y_i|\bm X_i} \right) - \sum\limits_{s = 1}^p {{w_s}} {\mu _{\tau ,s}}\left(\bm X_i \right)$, $\bm\Pi_{s}={{\left( {1,\bm X_{ \setminus s}^\top} \right)}^\top}$ and $\bm\Pi_{i,s}={{\left( {1,\bm X_{i \setminus s}^\top} \right)}^\top}$. Then by Knight's identity \eqref{A.1}, we have
\begin{align}
&{\rm{FPE}}_n\left(\bm  w \right) - E\left[ {{\rho _\tau }\left( {\varepsilon  + u\left(\bm  w \right)} \right)} \right] \nonumber\\
  =& E\left[ {\left\{ {{\rho _\tau }\left( {\varepsilon  + u\left( \bm w \right) - \sum\limits_{s = 1}^p {{w_s}} \left( {{{\hat \mu }_{\tau ,s}}\left(\bm X \right) - {\mu _{\tau ,s}}\left(\bm X \right)} \right)} \right) - {\rho _\tau }\left( {\varepsilon  + u\left( \bm w \right)} \right)} \right\}\left| {{\mathcal{D}_n}} \right.} \right] \nonumber\\
=&E\left\{ {\sum\limits_{s = 1}^p {{w_s}} \left( {{{\hat \mu }_{\tau ,s}}\left(\bm X \right) - {\mu _{\tau ,s}}\left( \bm X \right)} \right)\left[ {I\left( {\varepsilon  + u\left(\bm w \right) \le 0} \right) - \tau } \right]}\left| {{\mathcal{D}_n}} \right. \right\}\nonumber\\
  &+ E\left[ {\int_0^{\sum\limits_{s = 1}^p {{w_s}} \left( {{{\hat \mu }_{\tau ,s}}\left(\bm X \right) - {\mu _{\tau ,s}}\left( \bm X \right)} \right) } {\left[ {I\left( {\varepsilon  + u\left( \bm w \right) \le t} \right) - I\left( {\varepsilon  + u\left( \bm w \right) \le 0} \right)} \right]dt\left| {{\mathcal{D}_n}} \right.} } \right]\nonumber\\
\buildrel \Delta \over =&I_1+I_2.\nonumber
 \tag{A.5}\label{A.5}
\end{align}
By Taylor expansion, Jensen inequality, conditions (C4)--(C5) and Theorem \ref{th.1}, we have
\begin{align}
&I_2\nonumber\\
= & E_{\bm X_i}\left[ {\int_0^{\sum\limits_{s = 1}^p  {{w_s}} \left( {{{\hat \mu }_{\tau ,s}}\left(\bm X_i \right) - {\mu _{\tau ,s}}\left(\bm X_i\right)} \right) } {\left[ {F\left( {t - u_i\left( \bm w \right)\left| \bm X_i \right.} \right) - F\left( { - u_i\left(\bm  w \right)\left| \bm X_i \right.} \right)} \right]dt} } \right] \nonumber\\
  = &E_{\bm X_i}\left[ {\int_0^{\sum\limits_{s = 1}^p  {{w_s}} \left( {{{\hat \mu }_{\tau ,s}}\left(\bm X_i \right) - {\mu _{\tau ,s}}\left( \bm X_i\right)} \right) } {f\left( { - u_i\left( \bm w \right)\left| \bm X_i \right.} \right)t dt} } \right]\left\{ {1 + {o_p}\left( 1 \right)} \right\} \nonumber\\
  = &\frac{1}{2}E_{\bm X_i}\left\{ {f\left( { - u_i\left( \bm w \right)\left| \bm X_i \right.} \right){{\left[ {\sum\limits_{s = 1}^p  {{w_s}} \left( {{{\hat \mu }_{\tau ,s}}\left(\bm X_i \right) - {\mu _{\tau ,s}}\left( \bm X_i\right)} \right) } \right]}^2}} \right\}\left\{ {1 + {o_p}\left( 1 \right)} \right\}\nonumber\\
  \le &\frac{1}{2}E_{\bm X_i}\left\{ {f\left( { - u_i\left(\bm  w \right)\left| \bm X_i \right.} \right)\sum\limits_{s = 1}^p {{w_s}} {{\left[   {{{\hat \mu }_{\tau ,s}}\left(\bm X_i \right) - {\mu _{\tau ,s}}\left( \bm X_i\right)}  \right]}^2}} \right\}\left\{ {1 + {o_p}\left( 1 \right)} \right\} \nonumber\\
 =&\frac{1}{2}{E_{{\bm X_i}}}\left\{ {f\left( { - {u_i}\left( \bm w \right)\left| {{\bm X_i}} \right.} \right)\sum\limits_{s = 1}^p {{w_s}} {{\left[ {{\bm{\hat \theta }_{\tau ,s}}\left(X_{is} \right) - {\bm\theta _{\tau ,s}}\left( X_{is} \right)} \right]}^ \top }{\bm\Pi _{i,s}}\bm\Pi _{i,s}^ \top \left[ {{\bm{\hat \theta }_{\tau ,s}}\left( X_{is} \right) - {\bm\theta _{\tau ,s}}\left( X_{is} \right)} \right]} \right\} \nonumber\\
  &\times \left\{ {1 + {o_p}\left( 1 \right)} \right\}\nonumber \\
  \le& \frac{1}{2} c_f\mathop {\max }\limits_{1 \le s \le p}\mathop {\sup }\limits_{x_s \in \mathcal{C}_s} {\lambda _{\max }}\left( \bm \Sigma_s \right){\left\| {{\bm{\hat \theta }_{\tau ,s}}\left( {{x_s}} \right) - {\bm\theta _{\tau ,s}}\left( {{x_s}} \right)} \right\|^2}\left\{ {1 + {o_p}\left( 1 \right)} \right\} \nonumber\\
  =&{O_p}\left( {\log (1/\underline{h})/(n\underline{h})} +\bar{h}^4 \right)\left\{ {1 + {o_p}\left( 1 \right)} \right\} \nonumber\\
  = &{o_p}\left( 1 \right),
 \tag{A.6}\label{A.6}
\end{align}
under the conditions $\bar{h}\rightarrow 0$ and $\log (1/\underline{h})/(n\underline{h})\rightarrow 0$, as $n \rightarrow \infty$.

By the triangle inequality, the fact $\bm A^\top \bm B \bm A\leq \lambda_{max}(\bm B)\bm A^\top  \bm A$ for any real symmetric matrix $\bm B$, condition (C5), Theorem \ref{th.1} and $ \left|  {I\left( {\varepsilon  + u\left( \bm w \right) \le 0} \right) - \tau }\right| \leq 1$, we have
\begin{align}
 \left| {{I_1}} \right| \le &\sum\limits_{s = 1}^p {{w_s}}E \left| {\left( {{{\hat \mu }_{\tau ,s}}\left(\bm X \right) - {\mu _{\tau ,s}}\left(\bm X \right)} \right)} \right| \nonumber\\
  = &\sum\limits_{s = 1}^p {{w_s}} E\left\{\left[ {{\bm{\hat \theta }_{\tau ,s}}\left( {{X_s}} \right) - {\bm\theta _{\tau ,s}}\left( {{X_s}} \right)} \right]^ \top \left( {{\bm\Pi _s}\bm\Pi _s^ \top } \right)\left[ {{\bm{\hat \theta }_{\tau ,s}}\left( {{X_s}} \right) - {\bm\theta _{\tau ,s}}\left( {{X_s}} \right)} \right]\right\}^{1/2} \nonumber\\
\le& \mathop {\max }\limits_{1 \le s \le p}\mathop {\sup }\limits_{x_s \in \mathcal{C}_s}  \left\{{\lambda _{\max }}\left( \bm \Sigma_s \right)\right\}^{1/2}{\left\| {{\bm{\hat \theta }_{\tau ,s}}\left( {{x_s}} \right) - {\bm\theta _{\tau ,s}}\left( {{x_s}} \right)} \right\|} \nonumber\\
=&{O_p}\left( {\sqrt {\log (1/\underline{h})/(n\underline{h})} }+\bar{h}^2 \right)\nonumber\\
  = &{o_p}\left( 1 \right).
 \tag{A.7}\label{A.7}
\end{align}
 Therefore, combine with \eqref{A.5}--\eqref{A.7}, we have ${\rm{FPE}}_n\left(\bm  w \right) - E\left[ {{\rho _\tau }\left( {\varepsilon  + u\left(\bm  w \right)} \right)} \right] ={o_p}\left( 1 \right)$.

Let $U(t)={{\rho _\tau }\left( {\varepsilon  + t} \right)}-{{\rho _\tau }\left( {\varepsilon  } \right)}$,  where $t\in \mathbb{R}$. It is well known
that $U(t)$ has a global minimum at $t=0$. This implies that $\mathop {\min }\limits_{\bm w\in \mathbb{W}} E\left[ {{\rho _\tau }\left( {\varepsilon  + u\left(\bm  w \right)} \right)} \right]\geq E\left[ {{\rho _\tau }\left( {\varepsilon  } \right)} \right]$. Consequently, we have
$\mathop {\min }\limits_{\bm w\in \mathbb{W}}{\rm{FPE}}_n\left(\bm  w \right)= \mathop {\min }\limits_{\bm w\in \mathbb{W}} E\left[ {{\rho _\tau }\left( {\varepsilon  + u\left(\bm  w \right)} \right)} \right]-{o_p}\left( 1 \right) \geq E\left[ {{\rho _\tau }\left( {\varepsilon  } \right)} \right]-{o_p}\left( 1 \right)$.

(ii) We decompose ${\Lambda _{n1}}\left( \bm w \right)$ as follows
\begin{align}
 {\Lambda _{n1}}\left(\bm  w \right)  = &\frac{1}{n}\sum\limits_{i = 1}^n {\left[ {{Q_\tau }\left( {{Y_i}|{\bm X_i}} \right) - \sum\limits_{s = 1}^p {{w_s}} {\mu _{\tau ,s}}\left( {{\bm X_i}} \right)} \right]{\psi _\tau }\left( {{\varepsilon _i}} \right)} \nonumber\\
 - &\frac{1}{n}\sum\limits_{i = 1}^n {\sum\limits_{s = 1}^p {{w_s}} \left[ {\hat \mu _{\tau ,s}^{ \setminus i}\left( {{\bm X_i}} \right) - {\mu _{\tau ,s}}\left( {{\bm X_i}} \right)} \right]{\psi _\tau }\left( {{\varepsilon _i}} \right)}   \nonumber\\
 \buildrel \Delta \over = &{\Lambda _{n1,1}}\left(\bm  w \right) + {\Lambda _{n1,2}}\left(\bm  w \right) .\nonumber
\end{align}
In view of the fact that $E({\Lambda _{n1,1}}\left(\bm  w \right))=\bm 0$ and $Var[{\Lambda _{n1,1}}\left(\bm  w \right)]=O(1/n)$, we have ${\Lambda _{n1,1}}\left( \bm w \right))=o_p(1)$ for each $\bm w\in \mathbb{W}$. Since $p$ and $q$ are finite, one can apply the Glivenko-Cantelli theorem (e.g., Theorem 2.4.1 in \cite{VW96}) to conclude $\mathop {\sup }\limits_{\bm w\in \mathbb{W}} \left| {\Lambda _{n1,1}}\left( \bm w \right) \right|=o_p(1)$.

By the triangle inequality and $|{\psi _\tau }\left( {{\varepsilon _i}} \right)|\leq 1$, we have
\begin{align}
\mathop {\sup }\limits_{w \in \mathbb{W} } \left|{\Lambda _{n1,2}}\left(\bm  w \right)  \right|
 \le &\mathop {\sup }\limits_{w \in \mathbb{W} }\frac{1}{n}\sum\limits_{i = 1}^n\sum\limits_{s = 1}^p {{w_s} {\left| \bm\Pi _{i,s}^ \top \left( {\bm{\hat \theta} _{\tau ,s}^{ \setminus i}\left( {{X_{is}}} \right) - {\bm\theta _{\tau ,s}}\left( {{X_{is}}} \right)} \right){\psi _\tau }\left( {{\varepsilon _i}} \right) \right|} }   \nonumber\\
  \le&  \frac{1}{n} \sum\limits_{i = 1}^n\mathop {\max }\limits_{1 \le s \le p} \mathop {\sup }\limits_{x_s \in \mathcal{C}_s} \left\| {\bm{\hat \theta} _{\tau ,s}^{ \setminus i}\left( {{x_{s}}} \right) - {\bm\theta _{\tau ,s}}\left( {{x_{s}}} \right)} \right\| {\left\| {\bm\Pi _{i,s}} \right\|}   \nonumber\\
  = &{O_p}\left( {\sqrt {\log (1/\underline{h})/(n\underline{h})} } +\bar{h}^2\right){O_p}\left( 1 \right)  \nonumber\\
  = &{o_p}\left( 1 \right) . \nonumber
\end{align}
Consequently $\mathop {\sup }\limits_{w \in \mathbb{W} } \left|{\Lambda _{n1,2}}\left(\bm  w \right)  \right| = {o_p}\left( 1 \right)$ and $\mathop {\sup }\limits_{w \in \mathbb{W} } \left|{\Lambda _{n1}}\left(\bm  w \right)  \right| = {o_p}\left( 1 \right)$.

(iii) Observe that
\begin{align}
 {\Lambda _{n2}}\left( \bm w \right) =&\frac{1}{n}\sum\limits_{i = 1}^n \int_0^{\sum\limits_{s = 1}^p {{w_s}} \mu _{\tau ,s}\left( {{\bm X_i}} \right) - {Q_\tau }\left( {{Y_i}|{\bm X_i}} \right) }  {\left\{ {\left[ {I\left( {{\varepsilon _i} \le t} \right) - I\left( {{\varepsilon _i} \le 0} \right)} \right]} \right.}\nonumber\\
  &- \left. {\left[ {F\left( {t\left| {{\bm X_i}} \right.} \right) - F\left( {0\left| {{\bm X_i}} \right.} \right)} \right]} \right\}dt\nonumber\\
  &+ \frac{1}{n}\sum\limits_{i = 1}^n \int_{\sum\limits_{s = 1}^p {{w_s}}  \mu _{\tau ,s}\left( {{\bm X_i}} \right) - {Q_\tau }\left( {{Y_i}|{\bm X_i}} \right) }^{\sum\limits_{s = 1}^p {{w_s}} \hat \mu _{\tau ,s}^{ \setminus i}\left( {{\bm X_i}} \right) - {Q_\tau }\left( {{Y_i}|{\bm X_i}} \right) }
{\left\{ {\left[ {I\left( {{\varepsilon _i} \le t} \right) - I\left( {{\varepsilon _i} \le 0} \right)} \right]} \right.}\nonumber\\
   &- \left. {\left[ {F\left( {t\left| {{\bm X_i}} \right.} \right) - F\left( {0\left| {{\bm X_i}} \right.} \right)} \right]} \right\}dt\nonumber\\
\triangleq &{\Lambda _{n2,1}}\left(\bm  w \right) + {\Lambda _{n2,2}}\left( \bm w \right) . \nonumber
\end{align}
In view of the fact that $\left| {\left[ {I\left( {{\varepsilon _i} \le t} \right) - I\left( {{\varepsilon _i} \le 0} \right)} \right] - \left[ {F\left( {t\left| {{\bm X_i}} \right.} \right) - F\left( {0\left| {{\bm X_i}} \right.} \right)} \right]} \right| \le 2$, we have
\begin{align}
 \left| {{\Lambda _{n2,2}}\left( \bm w \right)} \right| \le &\frac{2}{n}\sum\limits_{i = 1}^n {\left| {\sum\limits_{s = 1}^p {{w_s}\bm\Pi _{i,s}^ \top \left( {\bm{\hat \theta} _{\tau ,s}^{ \setminus i}\left( {{X_{is}}} \right) - {\bm\theta _{\tau ,s}}\left( {{X_{is}}} \right)} \right)} } \right|}   \nonumber\\
  \le &\frac{2}{n}\sum\limits_{i = 1}^n {\sum\limits_{s = 1}^p {{w_s}} \left| \bm\Pi _{i,s}^ \top \left( {\bm{\hat \theta }_{\tau ,s}^{ \setminus i}\left( {{X_{is}}} \right) - {\bm\theta _{\tau ,s}}\left( {{X_{is}}} \right)} \right) \right|}  \nonumber\\
  \le &\frac{2}{n}\sum\limits_{i = 1}^n {\sum\limits_{s = 1}^p {{w_s}\left\| {\bm \Pi _{i,s}} \right\|\left\| {\bm{\hat \theta} _{\tau ,s}^{ \setminus i}\left( {{X_{is}}} \right) - {\bm\theta _{\tau ,s}}\left( {{X_{is}}} \right)} \right\|} }   \nonumber\\
  \le&  \frac{2}{n}\sum\limits_{i = 1}^n\mathop {\max }\limits_{1 \le s \le p} \mathop {\sup }\limits_{x_s \in \mathcal{C}_s} \left\| {\bm{\hat \theta} _{\tau ,s}^{ \setminus i}\left( {{x_{s}}} \right) - {\bm\theta _{\tau ,s}}\left( {{x_{s}}} \right)}  \right\| {\left\| {\bm\Pi _{i,s}} \right\|}   \nonumber\\
  = &{O_p}\left( {\sqrt {\log (1/\underline{h})/(n\underline{h})} } +\bar{h}^2\right){O_p}\left( 1 \right)  \nonumber\\
  = &o_p(1). \nonumber
   \tag{A.8}\label{A.8}
\end{align}
Observing that $E[{\Lambda _{n2,1}}\left( \bm w \right)]=\bm 0$ and $Var[{\Lambda _{n2,1}}\left( \bm w \right)]=O(1/n)$, we have ${\Lambda _{n2,1}}\left(\bm  w \right))=o_p(1)$ for each $\bm w\in \mathbb{W}$. Analogous to
the proof of $\Lambda _{n1,1}\left(\bm  w \right)$, we can show that $\mathop {\sup }\limits_{\bm w\in \mathbb{W}} \left|{\Lambda _{n2,1}}\left(\bm  w \right)\right| =o_p(1)$. Therefore, we have $\mathop {\sup }\limits_{\bm w\in \mathbb{W}} \left|{\Lambda _{n2}}\left( \bm w \right)\right| =o_p(1)$.

(iv) Observe that
\begin{align}
 {\Lambda _{n3}}\left(\bm  w \right) =& \frac{1}{n}\sum\limits_{i = 1}^n {\left\{ {\int_0^{\sum\limits_{s = 1}^p {{w_s}} \mu _{\tau ,s}\left( {{\bm X_i}} \right) - {Q_\tau }\left( {{Y_i}|{\bm X_i}} \right) } {\left[ {F\left( {t\left| {{\bm X_i}} \right.} \right) - F\left( {0\left| {{\bm X_i}} \right.} \right)} \right]dt} } \right.}  \nonumber\\
 &\left. { - {E_{{\bm X_i}}}\left[ {\int_0^{\sum\limits_{s = 1}^p {{w_s}}  \mu _{\tau ,s}\left( {{\bm X_i}} \right) - {Q_\tau }\left( {{Y_i}|{\bm X_i}} \right) } {\left[ {F\left( {t\left| {{\bm X_i}} \right.} \right) - F\left( {0\left| {{\bm X_i}} \right.} \right)} \right]dt} } \right]} \right\} \nonumber\\
 +&\frac{1}{n}\sum\limits_{i = 1}^n {\left\{ {\int_{\sum\limits_{s = 1}^p {{w_s}}  \mu _{\tau ,s}\left( {{\bm X_i}} \right) - {Q_\tau }\left( {{Y_i}|{\bm X_i}} \right) }^{\sum\limits_{s = 1}^p {{w_s}} \hat \mu _{\tau ,s}^{ \setminus i}\left( {{\bm X_i}} \right) - {Q_\tau }\left( {{Y_i}|{\bm X_i}} \right)} {\left[ {F\left( {t\left| {{\bm X_i}} \right.} \right) - F\left( {0\left| {{\bm X_i}} \right.} \right)} \right]dt} } \right.}   \nonumber\\
 &\left. { - {E_{{\bm X_i}}}\left[ {\int_{\sum\limits_{s = 1}^p {{w_s}} \mu _{\tau ,s}\left( {{\bm X_i}} \right) - {Q_\tau }\left( {{Y_i}|{\bm X_i}} \right) }^{\sum\limits_{s = 1}^p {{w_s}} \hat \mu _{\tau ,s}^{ \setminus i}\left( {{\bm X_i}} \right) - {Q_\tau }\left( {{Y_i}|{\bm X_i}} \right)} {\left[ {F\left( {t\left| {{\bm X_i}} \right.} \right) - F\left( {0\left| {{\bm X_i}} \right.} \right)} \right]dt} } \right]} \right\} \nonumber\\
  \triangleq & {\Lambda _{n3,1}}\left(\bm  w \right) + {\Lambda _{n3,2}}\left(\bm  w \right) . \nonumber
\end{align}
In view of the fact that $\left| {   {F\left( {t\left| {{\bm X_i}} \right.} \right) - F\left( {0\left| {{\bm X_i}} \right.} \right)} } \right| \le 1$, we have
\begin{align}
 \left| {{\Lambda _{n3,2}}\left(\bm  w \right)} \right| \le& \frac{1}{n}\sum\limits_{i = 1}^n {\left| {\sum\limits_{s = 1}^p {{w_s}\bm \Pi _{i,s}^{\top}\left( {\bm{\hat \theta} _{\tau ,s}^{ \setminus i}\left( {{X_{is}}} \right) - {\bm\theta _{\tau ,s}}\left( {{X_{is}}} \right)} \right)} } \right|}  \nonumber\\
 &+ \frac{1}{n}\sum\limits_{i = 1}^n {{E_{{\bm X_i}}}\left| {\sum\limits_{s = 1}^p {{w_s}\bm \Pi _{i,s}^{\top}\left( {\bm{\hat \theta} _{\tau ,s}^{ \setminus i}\left( {{X_{is}}} \right) - {\bm\theta _{\tau ,s}}\left( {{X_{is}}} \right)} \right)} } \right|}  \nonumber\\
 \triangleq &{{\Lambda _{n3,21}}\left(\bm  w \right)}+{{\Lambda _{n3,22}}\left(\bm  w \right)}.\nonumber
 \end{align}
 According to \eqref{A.8}, we have $\mathop {\sup }\limits_{\bm w\in \mathbb{W}} \left|{\Lambda _{n3,21}}\left( \bm w \right)\right| =o_p(1)$. For ${{\Lambda _{n3,22}}\left(\bm  w \right)}$, by
the triangle and Cauchy-Schwarz inequalities, the fact $\bm A^\top \bm B \bm A\leq \lambda_{\max}(\bm B)\bm A^\top  \bm A$ for any real symmetric matrix $\bm B$, and Theorem 1, we
have
\begin{align}
 &\mathop {\sup }\limits_{\bm w \in \mathbb{W}} \left| {{\Lambda _{n3,22}}\left(\bm  w \right)} \right| \nonumber\\
  \le &\mathop {\sup }\limits_{\bm w \in \mathbb{W} } \frac{1}{n}\sum\limits_{i = 1}^n {\sum\limits_{s = 1}^p {{w_s}} {E_{{\bm X_i}}}\left| {\bm\Pi _{i,s}^{\top}\left( {\bm{\hat \theta }_{\tau ,s}^{ \setminus i}\left( {{X_{is}}} \right) - {\bm\theta _{\tau ,s}}\left( {{X_{is}}} \right)}\right)} \right|}  \nonumber\\
  \le &\mathop {\sup }\limits_{\bm w \in \mathbb{W}} \frac{1}{n}\sum\limits_{i = 1}^n {\sum\limits_{s = 1}^p {{w_s}} {E_{{\bm X_i}}}{{\left\{ {{{\left({\bm{\hat \theta} _{\tau ,s}^{ \setminus i}\left( {{X_{is}}} \right) - {\bm\theta _{\tau ,s}}\left( {{X_{is}}} \right)} \right)}^\top}  \left(\bm\Pi _{i,s}\bm\Pi _{i,s}^{\top}\right)\left( {\bm{\hat \theta} _{\tau ,s}^{ \setminus i}\left( {{X_{is}}} \right) - {\bm\theta _{\tau ,s}}\left( {{X_{is}}} \right)}\right)} \right\}}^{1/2}}}  \nonumber\\
  \le &\frac{1}{n}\sum\limits_{i = 1}^n  \mathop {\max }\limits_{1 \le s \le p} \mathop {\sup }\limits_{x_s \in \mathcal{C}_s} \left\| {\bm{\hat \theta} _{\tau ,s}^{ \setminus i}\left( {{x_{s}}} \right) - {\bm\theta _{\tau ,s}}\left( {{x_{s}}} \right)}\right\| \lambda _{\max }^{1/2}\left( {{\bm\Sigma _s}} \right) \nonumber\\
  \le &{O_p}\left( {\sqrt {\log (1/\underline{h})/(n\underline{h})} } +\bar{h}^2\right){O_p}\left( 1 \right) \nonumber\\
  = &{o_p}\left( 1 \right). \nonumber
\end{align}
Consequently, $\mathop {\sup }\limits_{\bm w\in \mathbb{W}} \left|{\Lambda _{n3,2}}\left( \bm w \right) \right|=o_p(1)$. The proof that
 $\mathop {\sup }\limits_{\bm w\in \mathbb{W}} {\Lambda _{n3,1}}\left( \bm w \right) =o_p(1)$ is analogous to that of $\mathop {\sup }\limits_{\bm w\in \mathbb{W}} \left|{\Lambda _{n1,1}}\left(\bm  w \right)\right| =o_p(1)$ and thus omitted.

(v) For ${\Lambda _{n4}}\left(\bm w \right) $, noting that $\left| {   {F\left( {t\left| {{\bm X_i}} \right.} \right) - F\left( {0\left| {{\bm X_i}} \right.} \right)} } \right| \le 1$ and by the study of ${\Lambda _{n3,22}}\left(\bm w \right)$, we have
\[ \mathop {\sup }\limits_{\bm w\in \mathbb{W}} \left|{\Lambda _{n4}}\left(\bm w \right) \right|\le \mathop {\sup }\limits_{\bm w\in \mathbb{W}}\frac{1}{n}\sum\limits_{i = 1}^n {{E_{{\bm X_i}}}\left| \sum\limits_{s = 1}^p w_s {\bm\Pi _{i,s}^{\top}\left(  {\bm{\hat \theta} _{\tau ,s}^{ \setminus i}\left( {{X_{is}}} \right) - {\bm{\hat\theta} _{\tau ,s}}\left( {{X_{is}}} \right)} \right) } \right|} = {o_p}\left( 1 \right).\]
This completes the proof of the theorem.  $\square$\\

\section*{Acknowledgements}
This research is supported by National Natural Science Foundation of China
(Grant Numbers 11931014). Chaohui Guo's research is supported by the Science and Technology Research Program of Chongqing Municipal Education Commission (Grant No. KJQN202100526, KJQN201900511) and Chongqing University Innovation Research Group Project: Nonlinear Optimization Method and Its Application(Grant No. CXQT20014).

\bibliography{reference}
\bibliographystyle{apalike}

\tabcolsep=3pt
\begin{table}\scriptsize
\caption{\leftline{The means and standard deviations (sd) of the estimated model weights for example 1.}}
\label{table1}
\begin{tabular}{cccccccccccccccccccc} \noalign{\smallskip}\hline
\multirow{2}{*}{case 1}
&&\multicolumn{5}{c}{$n=200, p=5$}
&&\multicolumn{10}{c}{$n=400, p=10$}
\\
\cline{3-7}
\cline{9-18}
&&\multicolumn{1}{c}{$\hat w_1$}&\multicolumn{1}{c}{$\hat w_2$}&\multicolumn{1}{c}{$\hat w_3$}&\multicolumn{1}{c}{$\hat w_4$}&\multicolumn{1}{c}{$\hat w_5$}
&&\multicolumn{1}{c}{$\hat w_1$}&\multicolumn{1}{c}{$\hat w_2$}&\multicolumn{1}{c}{$\hat w_3$}&\multicolumn{1}{c}{$\hat w_4$}&\multicolumn{1}{c}{$\hat w_5$}
&\multicolumn{1}{c}{$\hat w_6$}&\multicolumn{1}{c}{$\hat w_7$}&\multicolumn{1}{c}{$\hat w_8$}&\multicolumn{1}{c}{$\hat w_9$}&\multicolumn{1}{c}{$\hat w_{10}$}
\\
\hline
$\tau=0.1$&mean& 0.390 &0.462& 0.020& 0.107& 0.020&&0.419& 0.458& 0.008& 0.104& 0.004& 0.002& 0.001& 0.002& 0.000& 0.002\\
&sd&0.165& 0.163& 0.060& 0.105& 0.071& &0.110& 0.114& 0.028& 0.079& 0.018 &0.014& 0.011& 0.014 &0.004 &0.013\\
$\tau=0.2$&mean&  0.451& 0.486& 0.006& 0.053& 0.004& &0.459& 0.491& 0.001& 0.047& 0.001 &0.000& 0.000& 0.000& 0.000& 0.001
\\
&sd&0.143& 0.142& 0.020& 0.058& 0.016&& 0.100& 0.094& 0.010 &0.045& 0.004& 0.001& 0.000& 0.004& 0.001& 0.006
\\
$\tau=0.3$&mean& 0.476& 0.494& 0.003& 0.025& 0.003&&0.488& 0.488& 0.001& 0.022& 0.000& 0.000 &0.000 &0.000& 0.000 & 0.000 \\
&sd&0.125& 0.124& 0.012& 0.038& 0.017&& 0.092& 0.092& 0.008& 0.032& 0.002& 0.000& 0.000& 0.004 &0.000 &0.001
\\
$\tau=0.4$&mean& 0.484& 0.496& 0.001& 0.017& 0.002&& 0.500& 0.491& 0.001& 0.007& 0.000& 0.000& 0.000& 0.000& 0.000& 0.000 \\
&sd&0.118& 0.119& 0.011& 0.034& 0.012&&0.088& 0.089 &0.008& 0.015 &0.002 &0.005& 0.001& 0.000 &0.000& 0.001
\\
$\tau=0.5$&mean& 0.498& 0.488& 0.001& 0.012& 0.001& &0.505 &0.491 &0.000& 0.004& 0.000& 0.000 &0.000& 0.000& 0.000 & 0.000
  \\
&sd&0.124& 0.123& 0.010& 0.027& 0.005&&0.094 &0.093& 0.002& 0.010& 0.000& 0.003& 0.000 &0.000& 0.000 &0.000
\\
$\tau=0.6$&mean&0.498& 0.491& 0.002& 0.008& 0.001& &0.514& 0.483 &0.000 &0.002& 0.000& 0.000& 0.000& 0.000& 0.000 & 0.000
\\
&sd&0.124& 0.123& 0.009& 0.022& 0.008&& 0.097& 0.096 &0.001& 0.006& 0.001& 0.001 &0.000& 0.000& 0.001& 0.000
\\
$\tau=0.7$&mean&0.518& 0.473& 0.001& 0.006& 0.000&&0.525& 0.474& 0.000& 0.001& 0.000& 0.000 &0.000 &0.000& 0.000 & 0.000 \\
&sd&0.141& 0.140& 0.008& 0.019& 0.003&& 0.100 &0.099 &0.001& 0.006& 0.001 &0.000 &0.000& 0.000& 0.000 &0.000
\\
$\tau=0.8$&mean& 0.532& 0.458& 0.001& 0.007& 0.001&& 0.534& 0.462& 0.000& 0.003& 0.000& 0.000& 0.000& 0.000& 0.000 & 0.000\\
&sd&0.142& 0.141& 0.012& 0.018& 0.008& &0.102 &0.100& 0.000& 0.012& 0.003 &0.000& 0.001& 0.000 &0.000 &0.000
\\
$\tau=0.9$&mean&0.546& 0.437& 0.003& 0.010& 0.005& &0.535& 0.457& 0.000& 0.004& 0.001& 0.000& 0.001& 0.000& 0.000 & 0.000
  \\
&sd&0.154& 0.152& 0.012& 0.025& 0.018&&0.119& 0.117& 0.004& 0.013 &0.005& 0.004& 0.003& 0.000& 0.003& 0.004
\\
\hline
\multirow{2}{*}{case 2}
&&\multicolumn{5}{c}{ $n=200, p=5$}
&&\multicolumn{10}{c}{ $n=400, p=10$}
\\
\cline{3-7}
\cline{9-18}
&&\multicolumn{1}{c}{$\hat w_1$}&\multicolumn{1}{c}{$\hat w_2$}&\multicolumn{1}{c}{$\hat w_3$}&\multicolumn{1}{c}{$\hat w_4$}&\multicolumn{1}{c}{$\hat w_5$}
&&\multicolumn{1}{c}{$\hat w_1$}&\multicolumn{1}{c}{$\hat w_2$}&\multicolumn{1}{c}{$\hat w_3$}&\multicolumn{1}{c}{$\hat w_4$}&\multicolumn{1}{c}{$\hat w_5$}&\multicolumn{1}{c}{$\hat w_6$}&\multicolumn{1}{c}{$\hat w_7$}&\multicolumn{1}{c}{$\hat w_8$}&\multicolumn{1}{c}{$\hat w_9$}&\multicolumn{1}{c}{$\hat w_{10}$}
\\
\hline
$\tau=0.1$&mean&0.386& 0.423& 0.050& 0.113& 0.028&& 0.383& 0.438&0.028& 0.116& 0.013& 0.006& 0.004 &0.007& 0.004 & 0.002
\\
&sd&0.189& 0.205& 0.101& 0.136& 0.065&& 0.134 &0.146& 0.063 &0.099& 0.037& 0.024 &0.023 &0.026& 0.019& 0.011
\\
$\tau=0.2$&mean&0.447& 0.475& 0.016& 0.055& 0.006&&0.448& 0.488& 0.009& 0.049& 0.002& 0.001& 0.000& 0.001& 0.000& 0.001 \\
&sd&0.172& 0.171& 0.047& 0.073& 0.022&&0.115& 0.123& 0.038& 0.054 &0.010& 0.014& 0.002& 0.006 &0.003 &0.009
\\
$\tau=0.3$&mean&0.484& 0.477& 0.008& 0.028& 0.002&&0.470& 0.499& 0.003& 0.025& 0.001& 0.000& 0.000&0.001 &0.000& 0.000
 \\
&sd&0.145& 0.148& 0.026& 0.052& 0.012&&0.113& 0.114& 0.013 &0.035 &0.007& 0.003& 0.003 &0.009& 0.004& 0.000
\\
$\tau=0.4$&mean& 0.490& 0.485& 0.004& 0.019& 0.002&&0.499 &0.486 &0.002& 0.011& 0.000& 0.000& 0.000& 0.001 &0.000 & 0.000
  \\
&sd&0.152& 0.157& 0.016& 0.039& 0.007&& 0.101& 0.107& 0.010& 0.026 &0.003& 0.002& 0.003& 0.005& 0.003 &0.001\\
$\tau=0.5$&mean& 0.501& 0.479& 0.003& 0.014& 0.003&& 0.507& 0.486& 0.001& 0.005 &0.000& 0.000& 0.000& 0.000 &0.000& 0.000 \\
&sd& 0.148& 0.151& 0.015& 0.033& 0.015&&0.108& 0.107& 0.004& 0.017&0.002& 0.000& 0.001& 0.002 &0.002& 0.000\\
$\tau=0.6$&mean&0.514& 0.472& 0.003& 0.009& 0.002&& 0.512& 0.483 &0.001& 0.005& 0.000& 0.000& 0.000 &0.000& 0.000& 0.000
\\
&sd& 0.144& 0.145& 0.011& 0.023& 0.009&& 0.103& 0.104& 0.005& 0.016& 0.002& 0.000& 0.000& 0.001& 0.001& 0.000
\\
$\tau=0.7$&mean&  0.517& 0.468& 0.002& 0.011& 0.002&& 0.513& 0.482& 0.001& 0.003& 0.000& 0.000& 0.000& 0.000 & 0.000 &0.000
\\
&sd& 0.158& 0.158& 0.010& 0.026& 0.008&& 0.112& 0.112& 0.006& 0.013& 0.002& 0.002& 0.001& 0.003 &0.006& 0.000
\\
$\tau=0.8$&mean&0.528& 0.455& 0.004& 0.008& 0.005&& 0.526& 0.466& 0.000 &0.005 &0.001 &0.000& 0.000& 0.001& 0.000 & 0.000\\
&sd&0.176 &0.175& 0.017& 0.021 &0.018&& 0.123& 0.124& 0.002& 0.019& 0.005 &0.003& 0.003& 0.005& 0.000& 0.004
\\
$\tau=0.9$&mean& 0.516& 0.444& 0.008 &0.020& 0.012 && 0.523& 0.453 &0.003& 0.009& 0.002& 0.002& 0.003& 0.002 & 0.002& 0.002
\\
&sd&0.198& 0.203& 0.027& 0.041& 0.033&& 0.146& 0.147& 0.014 &0.024 &0.011& 0.008& 0.014& 0.010& 0.009& 0.010
\\
\hline
\multirow{2}{*}{case 3}
&&\multicolumn{5}{c}{ $n=200, p=5$}
&&\multicolumn{10}{c}{ $n=400, p=10$}
\\
\cline{3-7}
\cline{9-18}
&&\multicolumn{1}{c}{$\hat w_1$}&\multicolumn{1}{c}{$\hat w_2$}&\multicolumn{1}{c}{$\hat w_3$}&\multicolumn{1}{c}{$\hat w_4$}&\multicolumn{1}{c}{$\hat w_5$}
&&\multicolumn{1}{c}{$\hat w_1$}&\multicolumn{1}{c}{$\hat w_2$}&\multicolumn{1}{c}{$\hat w_3$}&\multicolumn{1}{c}{$\hat w_4$}&\multicolumn{1}{c}{$\hat w_5$}&\multicolumn{1}{c}{$\hat w_6$}&\multicolumn{1}{c}{$\hat w_7$}&\multicolumn{1}{c}{$\hat w_8$}&\multicolumn{1}{c}{$\hat w_9$}&\multicolumn{1}{c}{$\hat w_{10}$}
\\
\hline
$\tau=0.1$&mean& 0.393& 0.430& 0.030& 0.121& 0.026&&0.394& 0.445& 0.014 &0.125& 0.010& 0.003& 0.001 &0.003& 0.003 & 0.002\\
&sd&0.189& 0.196& 0.069& 0.141& 0.063&& 0.119 &0.137 &0.043 &0.094 &0.033 &0.016& 0.007 &0.016 &0.018 &0.013\\
$\tau=0.2$&mean&0.450& 0.481& 0.014& 0.048& 0.007&&0.450 &0.490 &0.003 &0.054 &0.002& 0.000 &0.000 &0.000 &0.000& 0.000 \\
&sd&0.150& 0.150& 0.042& 0.068& 0.030&& 0.102& 0.114 &0.014 &0.058 &0.008 &0.002 &0.000 &0.000& 0.005 &0.005\\
$\tau=0.3$&mean& 0.472& 0.495& 0.006& 0.024& 0.003& &0.480 &0.496& 0.001 &0.021 &0.001 &0.000 &0.000& 0.000 &0.000 &0.000\\
&sd& 0.144& 0.137& 0.020& 0.042& 0.015&& 0.098& 0.104& 0.004 &0.034 &0.007 &0.003 &0.002 &0.000& 0.000 &0.000\\
$\tau=0.4$&mean& 0.491& 0.490& 0.003& 0.014& 0.002&& 0.494 &0.492& 0.000& 0.012& 0.001& 0.000& 0.000& 0.000& 0.000& 0.000 \\
&sd&0.136& 0.135& 0.013& 0.031& 0.011&&0.093& 0.099& 0.003& 0.025 &0.006 &0.001& 0.000& 0.000 &0.000& 0.000\\
$\tau=0.5$&mean& 0.493& 0.493& 0.002& 0.010& 0.002&&0.496& 0.495& 0.001& 0.007& 0.001& 0.000& 0.000& 0.000& 0.000 & 0.000 \\
&sd&0.143 &0.137& 0.011& 0.024& 0.011&&0.094& 0.098& 0.005& 0.018 &0.004& 0.000 &0.003& 0.003& 0.000& 0.003\\
$\tau=0.6$&mean&0.512& 0.478& 0.002& 0.007& 0.002&&0.510& 0.485& 0.000 &0.004 &0.000& 0.000 &0.000& 0.000& 0.000 & 0.000\\
&sd&0.141& 0.139& 0.011& 0.018& 0.010&&0.096 &0.098& 0.003 &0.013& 0.001& 0.003& 0.000& 0.000& 0.000 &0.001\\
$\tau=0.7$&mean&0.516& 0.475& 0.001& 0.007& 0.001&&0.512 &0.485 &0.000 &0.003 &0.000 &0.000 &0.000 &0.000 &0.000 &0.000\\
&sd&0.163& 0.161& 0.007& 0.019& 0.006&&0.102& 0.104& 0.000& 0.011& 0.003& 0.000& 0.000& 0.000& 0.000 &0.002\\
$\tau=0.8$&mean& 0.522& 0.465& 0.002& 0.008& 0.003&&0.524 &0.472 &0.000 &0.003 &0.000& 0.000 &0.000 &0.000 &0.000 & 0.000\\
&sd&0.171& 0.172& 0.008& 0.019& 0.011&&0.117& 0.119& 0.003& 0.010& 0.003& 0.002 &0.002& 0.001& 0.002& 0.000\\
$\tau=0.9$&mean&0.526& 0.449& 0.007& 0.013& 0.005&&0.520& 0.466 &0.001& 0.006& 0.002 &0.001 &0.001 &0.001 &0.001 &0.001\\
&sd&0.206& 0.206& 0.024& 0.032& 0.018&& 0.140& 0.144 &0.003 &0.014& 0.010& 0.007 &0.008& 0.007 &0.007 &0.006\\
\hline
\end{tabular}
\end{table}

\tabcolsep=3pt
\begin{table}\scriptsize
\caption{\leftline{The means and standard deviations (sd) of the estimated model weights for example 2.}}
\label{table2}
\begin{tabular}{cccccccccccccccccccccccccc} \noalign{\smallskip}\hline
\multirow{2}{*}{case 4}
&&\multicolumn{5}{c}{ $n=200, p=5$}
&&\multicolumn{10}{c}{ $n=400, p=10$}
\\
\cline{3-7}
\cline{9-18}
&&\multicolumn{1}{c}{$\hat w_1$}&\multicolumn{1}{c}{$\hat w_2$}&\multicolumn{1}{c}{$\hat w_3$}&\multicolumn{1}{c}{$\hat w_4$}&\multicolumn{1}{c}{$\hat w_5$}
&&\multicolumn{1}{c}{$\hat w_1$}&\multicolumn{1}{c}{$\hat w_2$}&\multicolumn{1}{c}{$\hat w_3$}&\multicolumn{1}{c}{$\hat w_4$}&\multicolumn{1}{c}{$\hat w_5$}
&\multicolumn{1}{c}{$\hat w_6$}&\multicolumn{1}{c}{$\hat w_7$}&\multicolumn{1}{c}{$\hat w_8$}&\multicolumn{1}{c}{$\hat w_9$}&\multicolumn{1}{c}{$\hat w_{10}$}
\\
\hline
$\tau=0.1$&mean&0.474& 0.240 &0.278 &0.005 &0.004&&0.482 &0.236& 0.280 &0.001 &0.000& 0.000 &0.000& 0.000& 0.000& 0.000 \\
&sd&0.183& 0.238& 0.235& 0.030& 0.029&&0.110& 0.179 &0.172 &0.006& 0.001& 0.001 &0.000 &0.001& 0.002& 0.000\\
$\tau=0.2$&mean& 0.451& 0.253& 0.293& 0.003& 0.000&& 0.463 &0.245 &0.291 &0.000 &0.000 &0.000 &0.000 &0.000 &0.000& 0.000\\
&sd&0.130 &0.222& 0.217& 0.028& 0.000&&0.084& 0.170 &0.167 &0.003 &0.000 &0.004& 0.000& 0.000& 0.000& 0.000\\
$\tau=0.3$&mean&0.397& 0.267& 0.335& 0.001 &0.000&&0.407 &0.274 &0.318 &0.001 &0.000 &0.000 &0.000& 0.000 &0.000 &0.000 \\
&sd&0.106 &0.238 &0.244& 0.007 &0.000&&0.075& 0.203& 0.196& 0.007 &0.000& 0.000& 0.000& 0.000& 0.000& 0.000\\
$\tau=0.4$&mean&0.347& 0.295& 0.357& 0.001 &0.000&&0.350 &0.295 &0.355 &0.000 &0.000& 0.000 &0.000& 0.000& 0.000& 0.000 \\
&sd&0.091& 0.242& 0.237& 0.017& 0.000&&0.068 &0.222 &0.219 &0.000& 0.000 &0.000& 0.000 &0.000& 0.000 &0.000\\
$\tau=0.5$&mean& 0.282 &0.319 &0.398& 0.001& 0.000&& 0.288 &0.351& 0.361 &0.000 &0.000& 0.000& 0.000& 0.000& 0.000 &0.000\\
&sd&0.084& 0.265& 0.263 &0.012& 0.000&&0.062 &0.215 &0.209 &0.002& 0.000& 0.000 &0.000 &0.000 &0.000 &0.000\\
$\tau=0.6$&mean& 0.210& 0.355& 0.433& 0.002& 0.001&&0.229 &0.326 &0.445& 0.000& 0.000& 0.000&0.000 &0.000& 0.000 & 0.000\\
&sd&0.080& 0.286 &0.282& 0.012& 0.010&&0.060& 0.247 &0.243 &0.001 &0.000& 0.000& 0.000& 0.000 &0.000 &0.000\\
$\tau=0.7$&mean& 0.159 &0.369& 0.465& 0.006& 0.000&& 0.174 &0.368 &0.456 &0.001 &0.000 &0.000 &0.000 &0.000 &0.000& 0.000\\
&sd&0.082 &0.304& 0.308 &0.025& 0.005&&0.059 &0.278& 0.268& 0.010 &0.003& 0.003& 0.000 &0.000& 0.000& 0.002\\
$\tau=0.8$&mean& 0.117& 0.404 &0.462& 0.012& 0.005&&0.126 &0.412 &0.449 &0.005 &0.003 &0.001 &0.001 &0.001& 0.001& 0.001\\
&sd&0.086& 0.309& 0.317& 0.038& 0.022&&0.071& 0.266& 0.265& 0.019& 0.013 &0.005 &0.007& 0.008 &0.010& 0.006\\
$\tau=0.9$&mean& 0.100& 0.421& 0.408& 0.041 &0.030&& 0.101 &0.419 &0.401 &0.017& 0.015& 0.013& 0.011 &0.008 &0.007 &0.007\\
&sd&0.111& 0.293& 0.292& 0.078& 0.065&&0.097 &0.256 &0.259& 0.042& 0.044 &0.037& 0.041& 0.029& 0.028& 0.019\\
\hline
\multirow{2}{*}{case 5}
&&\multicolumn{5}{c}{ $n=200, p=5$}
&&\multicolumn{10}{c}{ $n=400, p=10$}
\\
\cline{3-7}
\cline{9-18}
&&\multicolumn{1}{c}{$\hat w_1$}&\multicolumn{1}{c}{$\hat w_2$}&\multicolumn{1}{c}{$\hat w_3$}&\multicolumn{1}{c}{$\hat w_4$}&\multicolumn{1}{c}{$\hat w_5$}
&&\multicolumn{1}{c}{$\hat w_1$}&\multicolumn{1}{c}{$\hat w_2$}&\multicolumn{1}{c}{$\hat w_3$}&\multicolumn{1}{c}{$\hat w_4$}&\multicolumn{1}{c}{$\hat w_5$}
&\multicolumn{1}{c}{$\hat w_6$}&\multicolumn{1}{c}{$\hat w_7$}&\multicolumn{1}{c}{$\hat w_8$}&\multicolumn{1}{c}{$\hat w_9$}&\multicolumn{1}{c}{$\hat w_{10}$}
\\
\hline
$\tau=0.1$&mean&0.452& 0.248& 0.292& 0.005& 0.004&&0.455& 0.243 &0.298& 0.002 &0.001& 0.000 &0.000 &0.000 &0.001 & 0.000\\
&sd&0.151& 0.227 &0.238& 0.026& 0.026&&0.122 &0.193& 0.195& 0.019& 0.010& 0.002& 0.006& 0.000& 0.006 &0.000\\
$\tau=0.2$&mean& 0.431 &0.274& 0.293& 0.002& 0.000&  &  0.439 &0.245 &0.315& 0.000& 0.000 &0.000 &0.000 &0.000 &0.000 & 0.000\\
&sd&0.116& 0.228& 0.235&0.015 &0.002&  &0.080& 0.177& 0.179 &0.005& 0.000& 0.000 &0.000 &0.000 &0.000& 0.000\\
$\tau=0.3$&mean&0.392& 0.308& 0.299& 0.001& 0.000&  &0.398 &0.250& 0.352 &0.000& 0.000 &0.000 &0.000& 0.000 &0.000 &0.000 \\
&sd&0.093& 0.228 &0.227 &0.011& 0.000&&0.068& 0.192& 0.199 &0.000 &0.000 &0.000& 0.000 &0.000 &0.000& 0.000\\
$\tau=0.4$&mean&  0.337& 0.301 &0.360& 0.002& 0.000&& 0.341& 0.303 &0.355 &0.000 &0.000 &0.000 &0.000& 0.000 &0.000 &0.000 \\
&sd&0.098 &0.263& 0.256 &0.014& 0.000&&0.064& 0.208& 0.206 &0.000 &0.000& 0.000 &0.000& 0.000& 0.000& 0.000\\
$\tau=0.5$&mean& 0.279& 0.310& 0.409& 0.001 &0.000&&0.289 &0.336 &0.374 &0.000 &0.000 &0.000 &0.000 &0.000 &0.000 &0.000 \\
&sd& 0.086 &0.268 &0.270 &0.009 &0.000&&0.058& 0.224 &0.222&0.002 &0.000 &0.000& 0.000 &0.000& 0.000 &0.000\\
$\tau=0.6$&mean&0.228 &0.337& 0.433 &0.001& 0.000&&0.228 &0.332 &0.440 &0.000 &0.000& 0.000& 0.000 &0.000 &0.000 &0.000\\
&sd& 0.077 &0.275 &0.274 &0.011 &0.000& & 0.057 &0.234& 0.238 &0.003 &0.000 &0.000 &0.000 &0.000 &0.000& 0.000\\
$\tau=0.7$&mean& 0.173 &0.368& 0.457& 0.002& 0.001& &0.179 &0.358 &0.461 &0.001 &0.000 &0.000& 0.000 &0.000& 0.000& 0.000\\
&sd&0.078 &0.298 &0.289 &0.013& 0.006&&0.052 &0.254& 0.254 &0.004 &0.000& 0.000& 0.000 &0.002 &0.000& 0.000\\
$\tau=0.8$&mean& 0.134 &0.369 &0.489 &0.007 &0.001&&0.137& 0.424& 0.436& 0.002& 0.001& 0.000& 0.000 &0.000& 0.000 & 0.000\\
&sd&0.077 &0.290 &0.298 &0.022 &0.007&&0.052& 0.263 &0.266 &0.012& 0.009 &0.001 &0.001& 0.004& 0.002& 0.003\\
$\tau=0.9$&mean& 0.119& 0.422& 0.424& 0.024 &0.010& & 0.106& 0.430& 0.432& 0.008 &0.006 &0.004& 0.003 &0.004 &0.004& 0.003\\
&sd&0.100& 0.290 &0.300 &0.057 &0.035&&0.068 &0.250& 0.253 &0.026 &0.028& 0.019& 0.013 &0.023 &0.018& 0.015\\
\hline
\multirow{2}{*}{case 6}
&&\multicolumn{5}{c}{ $n=200, p=5$}
&&\multicolumn{10}{c}{ $n=400, p=10$}
\\
\cline{3-7}
\cline{9-18}
&&\multicolumn{1}{c}{$\hat w_1$}&\multicolumn{1}{c}{$\hat w_2$}&\multicolumn{1}{c}{$\hat w_3$}&\multicolumn{1}{c}{$\hat w_4$}&\multicolumn{1}{c}{$\hat w_5$}
&&\multicolumn{1}{c}{$\hat w_1$}&\multicolumn{1}{c}{$\hat w_2$}&\multicolumn{1}{c}{$\hat w_3$}&\multicolumn{1}{c}{$\hat w_4$}&\multicolumn{1}{c}{$\hat w_5$}
&\multicolumn{1}{c}{$\hat w_6$}&\multicolumn{1}{c}{$\hat w_7$}&\multicolumn{1}{c}{$\hat w_8$}&\multicolumn{1}{c}{$\hat w_9$}&\multicolumn{1}{c}{$\hat w_{10}$}
\\
\hline
$\tau=0.1$&mean&0.372 &0.282& 0.331& 0.011& 0.004&&0.395& 0.276 &0.317& 0.007& 0.001 &0.001 &0.001 &0.002& 0.000 & 0.001\\
&sd& 0.160& 0.249& 0.251& 0.049 &0.035&&0.116 &0.199& 0.192& 0.037& 0.007& 0.009& 0.008 &0.018& 0.000& 0.006\\
$\tau=0.2$&mean&  0.359& 0.315 &0.321 &0.005 &0.000& & 0.369& 0.299 &0.328& 0.003 &0.000& 0.000& 0.000& 0.000 & 0.000& 0.000 \\
&sd& 0.122& 0.240& 0.257& 0.032& 0.000& &0.079 &0.190 &0.192 &0.018 &0.000& 0.000& 0.000 &0.000 &0.000 &0.000\\
$\tau=0.3$&mean&0.326 &0.342 &0.327& 0.005 &0.000&  & 0.313 &0.328 &0.357& 0.002 &0.000 &0.000 &0.000 &0.000 &0.000 &0.000 \\
&sd&0.107& 0.249& 0.243& 0.033 &0.000&&0.070& 0.215 &0.212 &0.014& 0.000& 0.000 &0.000 &0.000& 0.000& 0.000\\
$\tau=0.4$&mean& 0.272 &0.397 &0.325 &0.006 &0.000& &0.262 &0.380 &0.355 &0.003 &0.000& 0.000 &0.000 &0.000 &0.000 &0.000  \\
&sd&0.098 &0.277& 0.280& 0.035& 0.000&&0.064 &0.226& 0.228& 0.020 &0.000 &0.000 &0.000& 0.000& 0.000&0.000\\
$\tau=0.5$&mean& 0.214 &0.455 &0.325 &0.005 &0.000& & 0.220 &0.430& 0.348 &0.002& 0.000 &0.000 &0.000& 0.000 &0.000&0.000 \\
&sd&0.089 &0.271& 0.272& 0.031 &0.000& & 0.057 &0.220& 0.217& 0.015& 0.000& 0.000 &0.000& 0.000&0.000 &0.000\\
$\tau=0.6$&mean&0.168& 0.467 &0.356& 0.008 &0.001&& 0.179& 0.433& 0.386 &0.003 &0.000 &0.000 &0.000& 0.000 &0.000&0.000\\
&sd&0.082 &0.278& 0.272& 0.033 &0.008& & 0.056& 0.226 &0.230& 0.015 &0.000 &0.000& 0.000& 0.000 &0.000& 0.000\\
$\tau=0.7$&mean&0.127 &0.478 &0.386 &0.008 &0.001&  & 0.139 &0.453& 0.402 &0.005 &0.000 &0.000& 0.000 &0.000&0.000 &0.000\\
&sd&0.080 &0.277& 0.280 &0.032& 0.008&&0.055 &0.233& 0.238 &0.022& 0.000 &0.000 &0.000& 0.000 &0.000& 0.002\\
$\tau=0.8$&mean&0.099 &0.479 &0.403& 0.015 &0.004&& 0.106 &0.493 &0.389 &0.009 &0.000 &0.000 &0.001 &0.001 &0.000 &0.001\\
&sd&0.079 &0.281& 0.283& 0.041& 0.019&&0.060& 0.217 &0.219 &0.028 &0.001& 0.002& 0.008& 0.006 &0.000 &0.005\\
$\tau=0.9$&mean& 0.088 &0.469 &0.390 &0.038 &0.014& &0.085 &0.469 &0.393 &0.018 &0.005 &0.005& 0.006& 0.010 &0.004 &0.006\\
&sd&0.095& 0.285& 0.282 &0.066 &0.035&& 0.078& 0.215 &0.222 &0.042 &0.024& 0.020 &0.022 &0.033& 0.018 &0.022\\
\hline
\end{tabular}
\end{table}

\tabcolsep=5.5pt
\begin{table}\scriptsize
\caption{\leftline{The means and standard deviations (sd) of the estimated model weights for example 3.}}
\label{table3}
\begin{tabular}{ccccccccccccccccccccccccccccc} \noalign{\smallskip}\hline
\multirow{2}{*}{}
&&\multicolumn{6}{c}{ case 1}
&&\multicolumn{6}{c}{ case 2}
\\
\cline{3-8}
\cline{10-15}
&&\multicolumn{1}{c}{$\hat w_1$}&\multicolumn{1}{c}{$\hat w_2$}&\multicolumn{1}{c}{$\hat w_3$}&\multicolumn{1}{c}{$\hat w_4$}&\multicolumn{1}{c}{$\hat w_5$}&\multicolumn{1}{c}{$\hat w_6$}
&&\multicolumn{1}{c}{$\hat w_1$}&\multicolumn{1}{c}{$\hat w_2$}&\multicolumn{1}{c}{$\hat w_3$}&\multicolumn{1}{c}{$\hat w_4$}&\multicolumn{1}{c}{$\hat w_5$}
&\multicolumn{1}{c}{$\hat w_6$}
\\
\hline
$\tau=0.1$&mean& 0.396& 0.113& 0.066& 0.365& 0.042 &0.018&&0.312 &0.150 &0.103& 0.286& 0.089& 0.059\\
&sd&0.180 &0.151 &0.106& 0.155 &0.092 &0.054&&0.197 &0.171& 0.146 &0.179 &0.132 &0.112\\
$\tau=0.2$&mean&0.429 &0.081 &0.033 &0.444& 0.011 &0.002&  &0.373 &0.116 &0.071 &0.380 &0.046& 0.014 \\
&sd&0.172 &0.119& 0.075 &0.137 &0.035& 0.011&  &0.192 &0.146 &0.125& 0.169& 0.094 &0.051\\
$\tau=0.3$&mean&0.402 &0.084 &0.024 &0.482 &0.006 &0.001&  &0.369& 0.109 &0.057& 0.440& 0.022 &0.004 \\
&sd&0.154 &0.126 &0.050& 0.126 &0.024 &0.008&&0.170& 0.135& 0.101 &0.157 &0.058 &0.018\\
$\tau=0.4$&mean& 0.352 &0.106 &0.022& 0.514& 0.006 &0.001&&0.325 &0.135 &0.049 &0.473 &0.016 &0.002  \\
&sd&0.148 &0.134 &0.049 &0.131& 0.028& 0.009&& 0.165& 0.153 &0.092 &0.153& 0.052& 0.014\\
$\tau=0.5$&mean& 0.307 &0.145 &0.026 &0.515 &0.007 &0.001&& 0.279& 0.160& 0.057& 0.490 &0.013& 0.001\\
&sd& 0.145& 0.152 &0.058 &0.130 &0.029 &0.005&&0.151 &0.155 &0.102 &0.150& 0.046& 0.004\\
$\tau=0.6$&mean&0.248& 0.172& 0.039& 0.531 &0.009& 0.001&& 0.220& 0.184& 0.065 &0.518& 0.013 &0.000\\
&sd&  0.145 &0.143 &0.075 &0.141 &0.034 &0.012&& 0.145& 0.170 &0.104& 0.164 &0.039& 0.001\\
$\tau=0.7$&mean& 0.198& 0.183 &0.063 &0.540 &0.014 &0.002& &0.171 &0.196& 0.087& 0.520& 0.025 &0.001\\
&sd&0.133& 0.146 &0.097& 0.164 &0.048 &0.016&&0.131 &0.160& 0.125& 0.176& 0.066 &0.008\\
$\tau=0.8$&mean&0.141 &0.193& 0.106 &0.529 &0.029 &0.003&&0.123& 0.196 &0.131 &0.485 &0.055 &0.010\\
&sd&0.116 &0.145 &0.126 &0.180 &0.074 &0.019&&0.121 &0.172 &0.152 &0.202& 0.099 &0.049\\
$\tau=0.9$&mean& 0.078 &0.164& 0.184& 0.477 &0.086 &0.011& & 0.089 &0.146& 0.201 &0.415 &0.121 &0.028\\
&sd&0.097 &0.145 &0.158 &0.210 &0.123 &0.036&&0.101 &0.159 &0.185& 0.212 &0.137 &0.074\\
\hline
\multirow{2}{*}{}
&&\multicolumn{6}{c}{ case 3}
&&\multicolumn{6}{c}{ case 4}
\\
\cline{3-8}
\cline{10-15}
&&\multicolumn{1}{c}{$\hat w_1$}&\multicolumn{1}{c}{$\hat w_2$}&\multicolumn{1}{c}{$\hat w_3$}&\multicolumn{1}{c}{$\hat w_4$}&\multicolumn{1}{c}{$\hat w_5$}&\multicolumn{1}{c}{$\hat w_6$}
&&\multicolumn{1}{c}{$\hat w_1$}&\multicolumn{1}{c}{$\hat w_2$}&\multicolumn{1}{c}{$\hat w_3$}&\multicolumn{1}{c}{$\hat w_4$}&\multicolumn{1}{c}{$\hat w_5$}
&\multicolumn{1}{c}{$\hat w_6$}
\\
\hline
$\tau=0.1$&mean&0.355& 0.117& 0.096 &0.324& 0.079& 0.028  &&0.373 &0.163& 0.038 &0.415& 0.009& 0.002\\
&sd& 0.202& 0.155& 0.136& 0.187 &0.131 &0.069  &&0.194& 0.166& 0.080& 0.143& 0.029 &0.013 \\
$\tau=0.2$&mean&0.380 &0.113& 0.044& 0.422 &0.034& 0.006  &&0.374& 0.149& 0.018& 0.457& 0.002& 0.000 \\
&sd&0.190 &0.142& 0.083& 0.163& 0.069& 0.027  &&0.176& 0.155& 0.048& 0.133& 0.014& 0.004\\
$\tau=0.3$&mean& 0.365& 0.108 &0.042& 0.467 &0.016& 0.002  &&0.318& 0.175& 0.021& 0.480& 0.005& 0.001\\
&sd& 0.173& 0.135 &0.077& 0.149 &0.046 &0.010   &&0.164 &0.151 &0.052& 0.138& 0.029& 0.013\\
$\tau=0.4$&mean&0.311& 0.146& 0.037& 0.495& 0.010& 0.002   &&0.265& 0.206& 0.029& 0.497 &0.003& 0.000  \\
&sd& 0.165& 0.152 &0.072& 0.147 &0.046 &0.012  &&0.150& 0.150& 0.064& 0.136& 0.022& 0.000\\
$\tau=0.5$&mean&   0.270 &0.178& 0.038 &0.504 &0.010 &0.000 &&0.200& 0.231 &0.049& 0.513 &0.008& 0.000 \\
&sd& 0.162 &0.153 &0.080 &0.148& 0.039& 0.000  && 0.125& 0.137& 0.085& 0.140& 0.034 &0.004\\
$\tau=0.6$&mean& 0.226 &0.197& 0.045 &0.522 &0.009& 0.000  &&0.160& 0.251& 0.065& 0.514& 0.011& 0.000\\
&sd& 0.158 &0.146& 0.086& 0.151 &0.031& 0.007  && 0.120 &0.147 &0.106& 0.153 &0.036& 0.002\\
$\tau=0.7$&mean&0.173 &0.212 &0.073 &0.517 &0.023 &0.002   &&0.111 &0.243& 0.100 &0.518& 0.026& 0.001\\
&sd&0.143 &0.155 &0.112 &0.171& 0.059& 0.014   &&0.108 &0.156 &0.128& 0.175& 0.072 &0.012\\
$\tau=0.8$&mean& 0.126 &0.204 &0.104& 0.520& 0.042& 0.003 && 0.089& 0.216& 0.151& 0.468 &0.068& 0.007\\
&sd& 0.122 &0.172 &0.142 &0.198 &0.093& 0.022  &&0.098 &0.165 &0.164 &0.197 &0.098 &0.028\\
$\tau=0.9$&mean& 0.091 &0.160& 0.198& 0.420& 0.113 &0.019  &&0.074 &0.166& 0.228& 0.340 &0.147& 0.044\\
&sd& 0.110 &0.162 &0.176 &0.231& 0.125& 0.055  && 0.104 &0.169 &0.196 &0.221& 0.162 &0.094\\
\hline
\multirow{2}{*}{}
&&\multicolumn{6}{c}{ case 5}
&&\multicolumn{6}{c}{ case 6}
\\
\cline{3-8}
\cline{10-15}
&&\multicolumn{1}{c}{$\hat w_1$}&\multicolumn{1}{c}{$\hat w_2$}&\multicolumn{1}{c}{$\hat w_3$}&\multicolumn{1}{c}{$\hat w_4$}&\multicolumn{1}{c}{$\hat w_5$}&\multicolumn{1}{c}{$\hat w_6$}
&&\multicolumn{1}{c}{$\hat w_1$}&\multicolumn{1}{c}{$\hat w_2$}&\multicolumn{1}{c}{$\hat w_3$}&\multicolumn{1}{c}{$\hat w_4$}&\multicolumn{1}{c}{$\hat w_5$}
&\multicolumn{1}{c}{$\hat w_6$}
\\
\hline
$\tau=0.1$&mean&0.337& 0.198& 0.047& 0.405& 0.012& 0.002&&0.285& 0.280& 0.044& 0.368& 0.018& 0.005\\
&sd&0.187& 0.176& 0.078& 0.139 &0.035& 0.010&&0.157 &0.185 &0.086& 0.148 &0.051 &0.023\\
$\tau=0.2$&mean&0.333& 0.197& 0.019 &0.449 &0.003& 0.000  && 0.258& 0.300& 0.019 &0.417& 0.007& 0.000\\
&sd&0.183& 0.163& 0.048& 0.130& 0.016& 0.000&  &0.156 &0.174& 0.054 &0.143 &0.031 &0.001\\
$\tau=0.3$&mean&0.282& 0.215& 0.019& 0.481& 0.003 &0.001&  & 0.219 &0.329 &0.017 &0.429& 0.005& 0.000\\
&sd& 0.167& 0.160& 0.047 &0.131& 0.014 &0.008&&0.141 &0.149 &0.053 &0.146 &0.028 &0.000\\
$\tau=0.4$&mean& 0.223 &0.239& 0.022 &0.513 &0.001& 0.000&& 0.176 &0.349 &0.026 &0.444& 0.004 &0.001 \\
&sd&0.152& 0.157 &0.055& 0.124& 0.012& 0.000&&0.130 &0.160& 0.068 &0.158 &0.021 &0.010\\
$\tau=0.5$&mean&0.177& 0.268& 0.028& 0.524& 0.003 &0.000& & 0.151 &0.346 &0.034& 0.464& 0.005 &0.000\\
&sd&0.139 &0.154 &0.059 &0.140 &0.016 &0.003& & 0.126 &0.161 &0.075 &0.152 &0.028 &0.000\\
$\tau=0.6$&mean&0.134 &0.287 &0.044 &0.528 &0.006& 0.000&& 0.122& 0.341& 0.055& 0.474& 0.008& 0.000\\
&sd&0.128 &0.150 &0.074 &0.146 &0.022 &0.005& &0.120 &0.162 &0.098 &0.150 &0.034 &0.003\\
$\tau=0.7$&mean& 0.111 &0.269& 0.078& 0.532& 0.009& 0.000 &&0.097& 0.330& 0.085& 0.475 &0.013& 0.000\\
&sd& 0.112 &0.157 &0.102 &0.153 &0.034& 0.004&& 0.110 &0.170 &0.120 &0.160 &0.037& 0.002\\
$\tau=0.8$&mean&0.081 &0.229 &0.129 &0.516 &0.041& 0.005&&0.072 &0.280& 0.149 &0.456 &0.041 &0.001\\
&sd&0.092& 0.153& 0.134 &0.181 &0.076& 0.032&&0.109 &0.173 &0.158 &0.191 &0.073 &0.007\\
$\tau=0.9$&mean&0.065& 0.164& 0.227& 0.426& 0.096& 0.022&  & 0.057 &0.218& 0.213& 0.379 &0.104& 0.029\\
&sd&0.102 &0.157 &0.183& 0.219& 0.126& 0.064&&0.091 &0.171 &0.161 &0.209 &0.130& 0.072\\
\hline
\end{tabular}
\end{table}

\tabcolsep=5.5pt
\begin{table}\scriptsize
\caption{\leftline{The means and standard deviations (sd) of the estimated model weights for example 4.}}
\label{table4}
\begin{tabular}{ccccccccccccccccccccccccccccc} \noalign{\smallskip}\hline
\multirow{2}{*}{}
&&\multicolumn{6}{c}{ case 1}
&&\multicolumn{6}{c}{ case 2}
\\
\cline{3-8}
\cline{10-15}
&&\multicolumn{1}{c}{$\hat w_1$}&\multicolumn{1}{c}{$\hat w_2$}&\multicolumn{1}{c}{$\hat w_3$}&\multicolumn{1}{c}{$\hat w_4$}&\multicolumn{1}{c}{$\hat w_5$}&\multicolumn{1}{c}{$\hat w_6$}
&&\multicolumn{1}{c}{$\hat w_1$}&\multicolumn{1}{c}{$\hat w_2$}&\multicolumn{1}{c}{$\hat w_3$}&\multicolumn{1}{c}{$\hat w_4$}&\multicolumn{1}{c}{$\hat w_5$}
&\multicolumn{1}{c}{$\hat w_6$}
\\
\hline
$\tau=0.1$&mean&0.106 &0.220& 0.234 &0.135& 0.170 &0.135&& 0.094&0.212 &0.224 &0.179& 0.182& 0.108\\
&sd&0.127 &0.173 &0.166 &0.143 &0.143 &0.128&& 0.124 &0.186& 0.185& 0.172& 0.148& 0.120\\
$\tau=0.2$&mean& 0.054& 0.184& 0.198& 0.084& 0.272& 0.207& & 0.068& 0.186 &0.195 &0.108 &0.270 &0.173\\
&sd& 0.084 &0.136 &0.156 &0.109 &0.181 &0.175& &0.102 &0.149 &0.167& 0.127& 0.168 &0.154\\
$\tau=0.3$&mean&0.042 &0.157 &0.153 &0.051 &0.328 &0.270&  &0.036& 0.160 &0.158 &0.080 &0.325& 0.241 \\
&sd&0.071 &0.109 &0.132 &0.077 &0.202 &0.200&&0.065 &0.130& 0.143& 0.099& 0.195& 0.190\\
$\tau=0.4$&mean&0.038 &0.119& 0.121 &0.044& 0.356 &0.322& &0.037 &0.117 &0.134& 0.063 &0.366 &0.284  \\
&sd& 0.062 &0.103 &0.106 &0.060 &0.222 &0.218&&0.063 &0.103 &0.129& 0.090 &0.199 &0.198\\
$\tau=0.5$&mean&0.036 &0.102& 0.094 &0.036 &0.357 &0.376& & 0.046& 0.108& 0.095 &0.053 &0.387 &0.311\\
&sd& 0.062 &0.098 &0.096& 0.055& 0.233 &0.228&& 0.071 &0.099 &0.113& 0.079& 0.216 &0.208\\
$\tau=0.6$&mean&0.039 &0.088& 0.073& 0.041 &0.384 &0.374&&0.044& 0.094 &0.083& 0.043 &0.390& 0.346\\
&sd& 0.068 &0.090 &0.083 &0.064 &0.232 &0.217&&0.074 &0.096 &0.105 &0.076 &0.219 &0.215\\
$\tau=0.7$&mean&0.044& 0.087& 0.054& 0.036& 0.381& 0.398&  &0.049 &0.089& 0.074& 0.046 &0.392& 0.351\\
&sd&0.070& 0.089 &0.070 &0.063 &0.230 &0.231&&0.078 &0.097& 0.096 &0.080& 0.234 &0.226\\
$\tau=0.8$&mean&0.051 &0.075& 0.053& 0.044 &0.364 &0.413&&0.066 &0.088 &0.056 &0.064 &0.358& 0.369\\
&sd&0.077 &0.082& 0.075 &0.066 &0.241 &0.244&& 0.099& 0.096 &0.088& 0.087& 0.233 &0.238\\
$\tau=0.9$&mean& 0.078& 0.083 &0.056 &0.050& 0.352 &0.382& &0.083& 0.101 &0.077& 0.085& 0.349& 0.306\\
&sd&0.096 &0.105& 0.081& 0.075 &0.198 &0.204&&0.125& 0.109& 0.106 &0.106& 0.212& 0.210\\
\hline
\multirow{2}{*}{}
&&\multicolumn{6}{c}{ case 3}
&&\multicolumn{6}{c}{ case 4}
\\
\cline{3-8}
\cline{10-15}
&&\multicolumn{1}{c}{$\hat w_1$}&\multicolumn{1}{c}{$\hat w_2$}&\multicolumn{1}{c}{$\hat w_3$}&\multicolumn{1}{c}{$\hat w_4$}&\multicolumn{1}{c}{$\hat w_5$}&\multicolumn{1}{c}{$\hat w_6$}
&&\multicolumn{1}{c}{$\hat w_1$}&\multicolumn{1}{c}{$\hat w_2$}&\multicolumn{1}{c}{$\hat w_3$}&\multicolumn{1}{c}{$\hat w_4$}&\multicolumn{1}{c}{$\hat w_5$}
&\multicolumn{1}{c}{$\hat w_6$}
\\
\hline
$\tau=0.1$&mean&0.094& 0.200& 0.235 &0.172 &0.172&0.127&&0.097 &0.245 &0.197& 0.134& 0.212& 0.115\\
&sd& 0.121& 0.173 &0.185& 0.168 &0.155& 0.126&& 0.125 &0.168 &0.174 &0.157 &0.157& 0.134\\
$\tau=0.2$&mean&   0.061& 0.172& 0.194 &0.114& 0.269 &0.190&&0.049 &0.204 &0.143 &0.060& 0.347 &0.198 \\
&sd&0.086& 0.146& 0.155& 0.134& 0.167& 0.148&  &0.090 &0.143& 0.127 &0.092 &0.174& 0.168\\
$\tau=0.3$&mean&0.037 &0.152 &0.148& 0.066 &0.330& 0.267&  &  0.035 &0.135 &0.099 &0.036& 0.386 &0.309\\
&sd&0.064 &0.122 &0.130 &0.093 &0.192& 0.183&&0.061 &0.110& 0.092 &0.063& 0.215& 0.206\\
$\tau=0.4$&mean&0.034 &0.120 &0.113& 0.047 &0.360 &0.325& &0.032 &0.109 &0.077& 0.028 &0.423 &0.331  \\
&sd&0.063 &0.106 &0.106 &0.075 &0.199& 0.192&& 0.054 &0.092 &0.075& 0.050 &0.230 &0.220\\
$\tau=0.5$&mean&0.030 &0.102 &0.084& 0.044 &0.394 &0.345& & 0.041& 0.096 &0.063 &0.024 &0.424& 0.352\\
&sd& 0.056& 0.094& 0.091 &0.069 &0.224 &0.217& &0.062 &0.090 &0.071& 0.045 &0.236 &0.228\\
$\tau=0.6$&mean&0.032 &0.082 &0.074 &0.040 &0.407 &0.364&&0.049& 0.090 &0.054& 0.032 &0.394 &0.381\\
&sd& 0.059& 0.087 &0.087 &0.067& 0.224& 0.212& &0.065& 0.089 &0.074 &0.054 &0.235 &0.223\\
$\tau=0.7$&mean& 0.038 &0.082& 0.068 &0.037 &0.409 &0.366& &0.064 &0.088 &0.056 &0.044 &0.385 &0.363\\
&sd&0.064 &0.092& 0.084&0.058 &0.220 &0.214&&0.085& 0.103 &0.080 &0.069 &0.245 &0.229\\
$\tau=0.8$&mean&0.050& 0.083& 0.056 &0.047& 0.410 &0.352&&0.080& 0.102& 0.075 &0.059& 0.338 &0.345\\
&sd&0.074 &0.095 &0.083 &0.078 &0.220 &0.216&&0.113 &0.112& 0.118& 0.095& 0.227 &0.228\\
$\tau=0.9$&mean&0.076 &0.098& 0.064 &0.072 &0.359 &0.331&  &0.135 &0.136 &0.081& 0.101 &0.273 &0.273\\
&sd& 0.103 &0.112& 0.102 &0.092& 0.210 &0.206&&0.152 &0.153& 0.123 &0.136 &0.210& 0.214\\
\hline
\multirow{2}{*}{}
&&\multicolumn{6}{c}{ case 5}
&&\multicolumn{6}{c}{ case 6}
\\
\cline{3-8}
\cline{10-15}
&&\multicolumn{1}{c}{$\hat w_1$}&\multicolumn{1}{c}{$\hat w_2$}&\multicolumn{1}{c}{$\hat w_3$}&\multicolumn{1}{c}{$\hat w_4$}&\multicolumn{1}{c}{$\hat w_5$}&\multicolumn{1}{c}{$\hat w_6$}
&&\multicolumn{1}{c}{$\hat w_1$}&\multicolumn{1}{c}{$\hat w_2$}&\multicolumn{1}{c}{$\hat w_3$}&\multicolumn{1}{c}{$\hat w_4$}&\multicolumn{1}{c}{$\hat w_5$}
&\multicolumn{1}{c}{$\hat w_6$}
\\
\hline
$\tau=0.1$&mean&0.094 &0.266 &0.196& 0.137 &0.211 &0.096&&0.099& 0.261 &0.183 &0.140& 0.225 &0.092\\
&sd&0.124& 0.174 &0.158 &0.149& 0.152 &0.121&&0.132& 0.184 &0.166& 0.157 &0.151 &0.117\\
$\tau=0.2$&mean&0.052& 0.225 &0.144& 0.062& 0.338 &0.180&  & 0.071 &0.221& 0.132& 0.061 &0.330 &0.185\\
&sd&0.083 &0.134 &0.125 &0.090& 0.180 &0.159&  &0.104 &0.135 &0.139 &0.088& 0.169 &0.163\\
$\tau=0.3$&mean& 0.044 &0.162 &0.094& 0.034& 0.385 &0.282& &0.061& 0.179& 0.086& 0.038& 0.356& 0.280 \\
&sd&0.068 &0.108& 0.096 &0.058 &0.210 &0.205&& 0.083& 0.115& 0.096 &0.064 &0.194 &0.191\\
$\tau=0.4$&mean& 0.039 &0.130 &0.072 &0.026 &0.395 &0.338&&  0.056 &0.160 &0.055 &0.030 &0.396 &0.302\\
&sd&0.065 &0.093 &0.084& 0.045 &0.224 &0.220&&0.074 &0.106 &0.075& 0.055 &0.212& 0.201\\
$\tau=0.5$&mean&0.041 &0.110& 0.058& 0.021& 0.420& 0.350& & 0.072& 0.132 &0.041 &0.025 &0.374 &0.356\\
&sd& 0.064 &0.089 &0.070& 0.038 &0.224 &0.226& &0.095 &0.107& 0.064 &0.048 &0.212 &0.212\\
$\tau=0.6$&mean&0.052 &0.097& 0.048& 0.028& 0.409& 0.367&&0.086 &0.114& 0.037 &0.024& 0.348 &0.392\\
&sd&0.072& 0.094 &0.069 &0.050 &0.220 &0.219& &0.092 &0.102 &0.057& 0.049 &0.220& 0.220\\
$\tau=0.7$&mean& 0.058& 0.099& 0.050 &0.025 &0.406 &0.362& &0.095& 0.109 &0.036 &0.030 &0.363& 0.367\\
&sd&0.081 &0.098 &0.078& 0.047 &0.230 &0.225&& 0.107 &0.109 &0.065 &0.057& 0.215 &0.214\\
$\tau=0.8$&mean&0.083 &0.098& 0.056 &0.039& 0.353 &0.371&&0.118 &0.114& 0.038& 0.043& 0.362& 0.326\\
&sd&0.097 &0.107 &0.087 &0.070& 0.216 &0.221&&0.129 &0.115& 0.068&0.072& 0.209& 0.213\\
$\tau=0.9$&mean&  0.116 &0.109& 0.065& 0.067 &0.310& 0.333& &0.142 &0.126 &0.063 &0.066& 0.300 &0.303\\
&sd&0.119 &0.127& 0.106 &0.102 &0.223 &0.212&&0.142& 0.140 &0.093 &0.101 &0.201 &0.195\\
\hline
\end{tabular}
\end{table}

\tabcolsep=6pt
\begin{table}\scriptsize
\caption{{Mean-FPEs at different quantiles with $\tau$ ranging from 0.1 to 0.9 for the Boston housing data.}}
\label{tablereal}
\begin{tabular}{ccccccccccccc} \noalign{\smallskip}\hline
&\multicolumn{1}{c}{Method}
&\multicolumn{1}{c}{$\tau=0.1$}
&\multicolumn{1}{c}{$\tau=0.2$}
&\multicolumn{1}{c}{$\tau=0.3$}
&\multicolumn{1}{c}{$\tau=0.4$}
&\multicolumn{1}{c}{$\tau=0.5$}
&\multicolumn{1}{c}{$\tau=0.6$}
&\multicolumn{1}{c}{$\tau=0.7$}
&\multicolumn{1}{c}{$\tau=0.8$}
&\multicolumn{1}{c}{$\tau=0.9$}
\\
\hline
$n_{test}=50$&{\sf LQR}&0.590 &0.967& 1.251 &1.466 &1.587& 1.657& 1.619 &1.423 &1.067\\
&{\sf PLQR}& 0.651 &1.034 &1.359& 1.557 &1.681 &1.708 &1.702 &1.521& 1.159\\
&{\sf LQMA}&0.595& 0.973& 1.254 &1.470& 1.591 &1.655 &1.640& 1.445 &1.090\\
&{\sf VCQR$_1$}&0.579& 0.925 &1.174 &1.378& 1.522& 1.616 &1.548& 1.353& 0.937\\
&{\sf VCQR$_2$}&0.555& 0.820 &1.047 &1.244 &1.315 &1.398& 1.336 &1.134 &0.819\\
&{\sf VCQR$_3$}&0.574& 0.827& 1.023& 1.147 &1.241& 1.245 &1.249& 1.065 &0.727\\
&{\sf VCQR$_4$}&0.556 &0.855 &1.080& 1.222& 1.293& 1.316& 1.256& 1.072 &0.869\\
&{\sf VCQR$_5$}&0.572& 0.883 &1.130 &1.315 &1.459 &1.528 &1.457 &1.289 &0.877\\
&{\sf VCQR$_6$}&0.562 &0.884 &1.143& 1.331 &1.425& 1.429& 1.366& 1.136& 0.812\\
&{\sf VCQR$_7$}&0.498 &0.806 &1.058& 1.224 &1.304 &1.333 &1.307 &1.125 &0.828\\
&{\sf VCQR$_8$}&0.532& 0.836& 1.055& 1.207 &1.265& 1.318& 1.262 &1.087 &0.803\\
&{\sf VCQR$_9$}&0.665& 1.049 &1.313 &1.477 &1.586 &1.626 &1.595 &1.474 &1.059\\
&{\sf VCQR$_{10}$}&0.658& 0.926 &1.117 &1.195& 1.271& 1.282 &1.235& 1.091 &0.755\\
&{\sf AQR}&2.387 &3.347& 3.823 &4.136 &4.260 &4.306 &4.033& 3.377 &2.174\\
&{\sf VCQMA1}&0.472& 0.751 &0.952& 1.096 &1.185 &1.216 &1.165& 0.986 &0.666\\
&{\sf VCQMA2}&0.513 &0.846& 1.052& 1.203& 1.287 &1.303 &1.291 &1.139 &0.752\\
&{\sf JVCQMA}& \textbf{0.469} & \textbf{0.749}& \textbf{0.938}&  \textbf{1.065}&  \textbf{1.134}&  \textbf{1.157} & \textbf{1.090}&  \textbf{0.908} & \textbf{0.626}\\
\hline
$n_{test}=100$&{\sf LQR}&0.595 &0.971 &1.256 &1.476& 1.603& 1.675 &1.639& 1.435 &1.079\\
&{\sf PLQR}& 0.649 &1.050& 1.364 &1.571& 1.685 &1.730 &1.714& 1.528& 1.154\\
&{\sf LQMA}&0.599& 0.975& 1.259 &1.480& 1.610& 1.675 &1.654 &1.453 &1.093\\
&{\sf VCQR$_1$}&0.585 &0.919& 1.170 &1.378& 1.520& 1.591& 1.541 &1.334 &0.930\\
&{\sf VCQR$_2$}&0.569 &0.840 &1.066& 1.256 &1.358 &1.418 &1.359 &1.176& 0.881\\
&{\sf VCQR$_3$}&0.594& 0.845 &1.034& 1.172 &1.254& 1.286& 1.273& 1.107& 0.773\\
&{\sf VCQR$_4$}& 0.560 &0.858 &1.082 &1.229 &1.312 &1.333 &1.270 &1.104& 0.887\\
&{\sf VCQR$_5$}&0.570& 0.883 &1.126& 1.319 &1.459& 1.534 &1.469& 1.293& 0.902\\
&{\sf VCQR$_6$}&0.570 &0.893 &1.154 &1.337& 1.443& 1.468 &1.401 &1.175 &0.851\\
&{\sf VCQR$_7$}&0.505& 0.809& 1.057 &1.220& 1.309 &1.346& 1.296 &1.120 &0.822\\
&{\sf VCQR$_8$}& 0.552 &0.845 &1.082 &1.233& 1.304 &1.351& 1.302 &1.136 &0.893\\
&{\sf VCQR$_9$}& 0.670& 1.048& 1.317 &1.488& 1.597 &1.633 &1.605 &1.493& 1.093\\
&{\sf VCQR$_{10}$}&0.664 &0.938 &1.120& 1.204 &1.267 &1.271 &1.215 &1.079& 0.768\\
&{\sf AQR}&2.288 &3.268& 3.684& 3.989& 4.124& 4.166 &3.906 &3.282& 2.174\\
&{\sf VCQMA1}&0.468 &0.746 &0.955& 1.108& 1.200& 1.232& 1.180 &1.007& 0.679\\
&{\sf VCQMA2}&0.527& 0.844& 1.066& 1.218& 1.311& 1.342& 1.281 &1.152& 0.825\\
&{\sf JVCQMA}& \textbf{0.467}& \textbf{0.742}&  \textbf{0.939} &\textbf{1.073} &\textbf{1.143} &\textbf{1.169} & \textbf{1.098} & \textbf{0.942}& \textbf{0.671}\\
\hline
$n_{test}=200$&{\sf LQR}&0.609& 0.991& 1.280 &1.494 &1.624& 1.699& 1.668 &1.466 &1.105\\
&{\sf PLQR}&0.663& 1.065 &1.370& 1.574& 1.693 &1.748 &1.731 &1.542 &1.163\\
&{\sf LQMA}&0.613 &0.995& 1.284& 1.502& 1.635 &1.710& 1.682& 1.490& 1.110\\
&{\sf VCQR$_1$}&0.605& 0.947& 1.218& 1.441& 1.587& 1.643& 1.583& 1.375& 1.010\\
&{\sf VCQR$_2$}&0.649& 0.931& 1.174& 1.365& 1.477& 1.539& 1.530& 1.477& 1.084\\
&{\sf VCQR$_3$}&0.669& 0.919& 1.103& 1.244& 1.329& 1.369& 1.342& 1.191& 0.888\\
&{\sf VCQR$_4$}&0.594& 0.896& 1.116& 1.269& 1.355& 1.373& 1.305& 1.153& 0.951\\
&{\sf VCQR$_5$}&0.593& 0.911& 1.159& 1.355& 1.487& 1.554& 1.508& 1.323& 0.945\\
&{\sf VCQR$_6$}&0.599& 0.937& 1.194& 1.386& 1.511& 1.552 &1.488 &1.285& 0.939\\
&{\sf VCQR$_7$}&0.540& 0.845 &1.089 &1.261& 1.353& 1.392& 1.336 &1.156 &0.862\\
&{\sf VCQR$_8$}&0.646 &0.926 &1.173& 1.323& 1.401 &1.429 &1.369 &1.211 &1.012\\
&{\sf VCQR$_9$}&0.737& 1.109 &1.375& 1.547 &1.659& 1.706& 1.699 &1.576& 1.226\\
&{\sf VCQR$_{10}$}&0.715 &0.982 &1.157 &1.259 &1.311 &1.317& 1.255 &1.119 &0.828\\
&{\sf AQR}& 2.247& 3.243& 3.691& 3.950& 4.085& 4.140& 3.869 &3.254& 2.238\\
&{\sf VCQMA1}& \textbf{0.488} &0.773 &0.984 &1.138 &1.240& 1.278 &1.234& 1.081& 0.788\\
&{\sf VCQMA2}&0.575& 0.888& 1.123 &1.251 &1.347 &1.365 &1.298& 1.154 &0.883\\
&{\sf JVCQMA}&0.496 &\textbf{0.772} &\textbf{0.974} &\textbf{1.108}&  \textbf{1.185}&  \textbf{1.200}& \textbf{1.139} &\textbf{0.993} &0\textbf{.748}\\
\hline
\end{tabular}
\end{table}

\end{document}